\documentclass[journal, onecolumn]{IEEEtran}
\usepackage{amsthm} 
\usepackage[tbtags]{amsmath}
\usepackage{algorithmic}
\usepackage{algorithm}
\usepackage{amssymb}
\allowdisplaybreaks[4]
\usepackage{amsfonts}
\usepackage[caption=false,font=footnotesize,labelfont=rm,textfont=rm]{subfig}
\usepackage{textcomp}
\usepackage{url}
\usepackage{verbatim}
\usepackage{graphicx}
\usepackage{epstopdf}
\usepackage{makeidx}
\usepackage{cite}
\usepackage{enumerate}
\usepackage{bbm}
\usepackage{array}
\usepackage{color}
\usepackage{bm}
\usepackage{float}
\usepackage[colorlinks,linkcolor=blue,anchorcolor=blue,citecolor=blue]{hyperref}

\theoremstyle{definition} 
\newtheorem{theorem}{Theorem}
\newtheorem{corollary}{Corollary}

\newtheorem{lemma}{Lemma}
\newtheorem{remark}{Remark}
\newtheorem{definition}{Definition}
\newtheorem{proposition}{Proposition}

\IEEEoverridecommandlockouts

\makeatletter
\renewcommand{\maketag@@@}[1]{\hbox{\m@th\normalsize\normalfont#1}}%

\newcommand{\Rmnum}[1]{\expandafter\@slowromancap\romannumeral#1@}
\newcommand{\Ico}[2]{[#1, #2)} 
\newcommand{\Ioc}[2]{(#1, #2]} 
\makeatother

\begin{document}

\title{Rate-Distortion-Perception Tradeoff for Vector Gaussian Sources with Limited Common Randomness: Exact Characterization and Approximate Universality}

\author{
Wentao~Wang,
Shuo~Shao,~\IEEEmembership{Member,~IEEE,}
Ruibin~Li,
Jingjing~Qian,
Jun~Chen,~\IEEEmembership{Senior~Member,~IEEE,}
and~Bin~Xia,~\IEEEmembership{Senior~Member,~IEEE}
\thanks{Wentao Wang and Bin Xia are with State Key Laboratory of Photonics and
Communications, Shanghai Key Laboratory of Digital Media Processing and Transmission,
School of Integrated Circuits (School of Information and Electronic Engineering),
Shanghai Jiao Tong University, Shanghai, 200240, China (email: \{wt\_wang, bxia\}@sjtu.edu.cn).}
\thanks{Shuo Shao is with the School of Information and Electronic Engineering, East China 
Normal University, Shanghai 200241, China (e-mail: sshao@ee.ecnu.edu.cn).}
\thanks{Ruibin Li and Jun Chen are with the Department of Electrical and Computer
Engineering, McMaster University, Hamilton, ON L8S 4K1, Canada 
(e-mail: \{lir144, chenjun\}@mcmaster.ca).}
\thanks{Jingjing Qian is with the Research Institute of Science and Technology 
Innovation, Civil Aviation University of China, Tianjin 300300, China 
(e-mail: jjqian@cauc.edu.ca).}
}




\maketitle

\begin{abstract}
  The distortion-rate-common randomness-perception (DRCP) theory characterizes
  how limited common randomness (CR) improves reconstruction fidelity under
  perceptual constraints, thereby providing an information-theoretic foundation
  for generative compression. In this paper, we characterize the DRCP tradeoff
  for vector Gaussian sources with limited CR under the squared-error distortion
  measure and the squared Wasserstein-2 perception measure. Specifically, we
  first establish a generalized reverse waterfilling characterization of globally
  optimal allocations. Second, we derive an analytical upper bound on the DRCP
  function through sequential greedy allocation of rate and CR. For fixed rate
  and CR budgets, the resulting representation is independent of the perception
  threshold. Finally, we establish its approximate universality through uniform
  performance guarantees: the achievable distortion exceeds the optimal distortion
  by at most $2\rho^*\approx0.2897$ times the total source variance, whereas the
  optimal-to-greedy distortion ratio is at least $\eta^*\approx0.7517$.
\end{abstract}


\section{Introduction}
\IEEEPARstart{G}{enerative} compression seeks reconstructions that preserve both
sample fidelity and statistical characteristics of the source, an
objective illustrated by image compression with generative models
\cite{agustsson2019generative}. Classical rate-distortion (RD) theory quantifies
the communication cost of reconstruction fidelity \cite{shannon1959coding},
but a small distortion may not imply perceptually realistic outputs.
Blau and Michaeli formalized this tension through the perception-distortion
tradeoff \cite{blau2018perception} and its rate-constrained extension
\cite{blau2019rethinking}. The resulting
rate-distortion-perception (RDP) framework supplements the distortion constraint
with a discrepancy between source and reconstruction distributions, providing
an information-theoretic model for the cost of distributional realism.

\subsection{Motivation and Challenges}
The RDP tradeoff depends not only on the communication rate but also on the randomness available to the encoder and decoder. Common randomness (CR) allows them to coordinate their stochastic operations and can improve reconstruction fidelity under a perception constraint. In practice, however, this shared resource may be limited. It is therefore important to understand how communication and CR should be allocated to achieve the best distortion–perception tradeoff.

For vector Gaussian sources, this allocation problem remains coupled even after diagonalizing the source into independent eigenmodes. Communication and CR must be allocated across these modes under shared resource budgets, while the perception constraint links their contributions to the overall distortion. Consequently, classical reverse waterfilling does not directly determine which modes should receive resources or how their allocations should change with the required perceptual quality. Resolving this coupling is necessary to characterize the optimal tradeoff and quantify the value of limited CR.

A related question is whether the representation must be redesigned for each level of perceptual quality. For fixed communication and CR budgets, it would be desirable to use a single representation across different perception constraints, with only the decoder adapting to the desired perceptual quality. This motivates the study of approximate universality: whether such a representation can achieve distortion uniformly close to the optimum across all perception levels. Addressing this question requires both a suitable representation and explicit guarantees on its performance loss.


\subsection{Related Work}
The RDP framework builds on earlier results in source coding with distribution constraints. Saldi et al. established a coding theorem for randomized quantization with a prescribed output distribution \cite{saldi2015randomized}, and subsequently characterized the tradeoff between communication rate and CR under an output distribution constraint \cite{saldi2015output}, encompassing the near-perfect realism setting studied by Wagner \cite{wagner2022role}. These results provide a foundation for treating shared randomness as a rate-limited resource. Within the RDP framework, Theis and Wagner proved an operational coding theorem using stochastic variable-length codes \cite{theis2021coding}, while Chen et al. clarified the role of randomness and the distinction between different notions of perceptual quality \cite{chen2022on}. In particular, matching individual sample distributions and matching the joint distribution of a reconstruction block impose different operational requirements, under which CR can play different roles.


Another line of work studies the geometry and evaluation of RDP tradeoffs.
Freirich et al. characterized the distortion-perception tradeoff in the quadratic Wasserstein space \cite{freirich2021theory}; related studies considered binary sources under Hamming distortion and total variation perception \cite{qian2022binary}, as well as finite channels under arbitrary distortion measures and Wasserstein-1 perception measures induced by arbitrary metrics \cite{freirich2024characterization}. Universal representations
\cite{zhang2021universal} and optimally controllable perceptual compression
\cite{yan2022optimally} further relate different perception levels through
decoder design. Gaussian sources are particularly useful for analyzing these
tradeoffs: classical vector Gaussian RD theory admits reverse waterfilling
over the source eigenmodes \cite{cover1999elements}. For Gaussian RDP problems,
Serra et al. developed computational characterizations under jointly Gaussian
reconstruction \cite{serra2024computation}, and Qian et al. derived a generalized reverse
waterfilling solution, including the squared Wasserstein-$2$ perception measure
\cite{qian2024rate}. These results characterize the information RDP function,
which corresponds to the unlimited-CR limit of the sequence-level formulation
considered in the present work. Conditional distribution perception yields another distinct
formulation, for which Salehkalaibar et al. established coding theorems and a
Gaussian vector characterization \cite{salehkalaibar2024conditional}.

For the formulation requiring i.i.d. reconstructions, Xie et al. derived bounds on the Gaussian RDP tradeoff with limited CR \cite{xie2025output}, and subsequently improved the lower bound and demonstrated limitations of existing bounds through entropy-constrained scalar quantization \cite{xie2026gaussian}. In contrast, the present work imposes a perception constraint on the entire reconstruction block without requiring an i.i.d. output law. Under squared-error distortion and squared Wasserstein-$2$ perception, Qu et al. established a single-letter characterization for this setting with limited CR and explicitly evaluated it for scalar Gaussian sources \cite{qu2025rate}. For vector Gaussian sources, their characterization reduces the coding problem to a finite-dimensional optimization over two eigenmode allocation vectors. This provides the starting point for the present work: determining the globally optimal allocation and constructing a tractable representation with provable approximation guarantees across all perception thresholds.


\subsection{Main Contributions}
We study the distortion-rate-CR-perception (DRCP) function $D(R,C,P)$ for a
vector Gaussian source. Our main contributions are as follows.
\begin{itemize}
  \item \emph{Optimal allocation.} We identify the exact perception threshold
        separating the classical RD regime from the regime with a positive
        perception penalty. An equivalent trajectory optimization reformulation
        then yields a generalized reverse waterfilling characterization of the
        globally optimal allocations. This characterization specifies the eigenmode
        activation condition and optimal allocation through shared
        multipliers, which recovers the perfect-realism and zero-CR special cases.
  \item \emph{Analytical greedy bound.} We construct an achievable distortion
        bound $\tilde D(R,C,P)$ by first allocating communication rate using
        classical reverse waterfilling and then optimizing the CR
        allocation. The resulting allocation pair depends on $R$ and $C$, but not on $P$,
        allowing the same representation to serve different perception thresholds through
        decoder adaptation. This sequential procedure requires only scalar budget equations
        and is exact when CR is absent or the perception constraint is inactive.
  \item \emph{Approximate universality with sharp guarantees.} We establish that
        $\tilde D-D\leq2\rho^*\operatorname{tr}(\Gamma)$, where
        $2\rho^*\approx0.2897$, and $D/\tilde D\geq\eta^*\approx0.7517$.
        Both constants are independent of the source distribution, communication rate,
        CR budget, and perception threshold, and
        are asymptotically sharp over the problem class. These bounds establish approximate
        universality of the proposed representation with respect to the perception threshold,
        with guarantees that hold uniformly over the source dimension and covariance,
        communication rate, and CR budget.
\end{itemize}

The remainder of this work is organized as follows.
Section~\Rmnum{2} formulates the problem and recalls the single-letter results.
Section~\Rmnum{3} characterizes the optimal allocations.
Section~\Rmnum{4} develops the greedy allocation and establishes its approximate universality.
Section~\Rmnum{5} concludes the paper.

Notation: We use $\mathbb{R}^L$ to denote the $L$-dimensional real vector space,
and $\mathbb{R}_{+}^L$ to denote its non-negative orthant.
The set of all probability distributions defined over
$\mathbb{R}^L$ is written as $\mathcal{P}(\mathbb{R}^L)$.
For any $a \in \mathbb{R}$, let $(a)_+ \triangleq \max\{a, 0\}$, and
$\log^+(a) \triangleq \max\{\log (a), 0\}$ assuming $a>0$.
We use $\operatorname{diag} (a_1,a_2,\ldots,a_L)$ to refer to a diagonal
matrix with the diagonal entries being $a_1, a_2, \ldots, a_L$, and
$\operatorname{tr}(\cdot)$ to denote the trace of a matrix.
Unless otherwise specified, let $\mathbf{a}=(a_1,a_2,\ldots,a_L)$ be
a vector in $\mathbb{R}_{+}^L$.
The Gaussian distribution with zero mean and covariance matrix $\Gamma$ is
denoted by $\mathcal{N}(0, \Gamma)$.
Throughout the paper, $\log(\cdot)$ and $\ln(\cdot)$ denote logarithms to bases
2 and $e$, respectively.

\section{Problem Formulation}

Let ${\{X(t)\}}_{t=1}^{\infty}$ be a sequence of independent and identically distributed
random variables with marginal distribution $p_X$ over $\mathbb{R}^L$.
We assume that $p_X$ is square-integrable, i.e., $\mathbb{E}[{\|X\|}^2] < \infty$.
The notation $X^n$ and $\hat{X}^n$ represent the
length-$n$ blocks of the sequence and its reconstruction, respectively.
Let $d$: $\mathcal{X} \times \hat{\mathcal{X}} \mapsto \mathbb{R}_0^+ $
be a single-letter distortion measure with
\begin{equation}
  d(X^n,\hat{X}^n) \triangleq \frac{1}{n} \sum\limits_{t=1}^{n} d(X(t),\hat{X}(t)).
\end{equation}
To better control the perception quality of the reconstruction at the
sequence level, we use $\frac{1}{n} \phi (p_{X^n}, p_{\hat{X}^n})$ to
denote the perception loss function.

Consider a length-$n$ perception-aware lossy source coding system in which the
encoder and decoder have access to a shared random seed $K$, uniformly distributed
over $\mathcal K$ and independent of the source sequence. For a $(2^{nR},n)$
code, the stochastic encoder
\begin{equation}
  f^{(n)}:\mathcal X^n \times \mathcal{K} \rightarrow \mathcal{J},
\end{equation}
generates the message $J=f^{(n)}(X^n,K)\in\mathcal J$ according to the conditional
distribution $P_{J|X^nK}$. Given $(J,K)$, the stochastic decoder
\begin{equation}
  g^{(n)}:\mathcal J\times\mathcal K\rightarrow\hat{\mathcal X}^n,
\end{equation}
produces the reconstruction $\hat X^n=g^{(n)}(J,K)$ according to $P_{\hat X^n|JK}$.
Consequently, the induced joint distribution factorizes as
$p_{X^n J K \hat{X}^n} = p_{X^n} p_K p_{J|X^n K} p_{\hat{X}^n|J K}$.

The compression rate of the code is defined as $R = \frac{1}{n} \log |\mathcal{J}|$.
We refer to $R$ as the transmission rate in terms of bits per source vector.
The main contribution of this work is a complete
solution for the vector Gaussian sources with the squared error serving as
the distortion measure, i.e.,
\begin{equation}
  d(x, \hat{x}) = {\|x - \hat{x}\|}^2 < \infty,
\end{equation}
and the squared Wasserstein-2 distance serving as the perception measure, i.e.,
\begin{equation}\label{eq:phi_def}
  \phi (p_X, p_{\hat{X}}) = W_2^2(p_X, p_{\hat{X}}) \triangleq
  \!\inf_{\mu \in \Pi (p_X\!, p_{\hat{X}})}\! \mathbb{E}_{\mu}[\|X \!-\! \hat{X}\|^2],
\end{equation}
where $\Pi(p_X, p_{\hat{X}})$ is the set of all joint distributions with marginals
$p_X$ and $p_{\hat{X}}$.

\begin{definition}
  A tuple $(R,D,C,P)$ is achievable
  if for every $\epsilon > 0$, there exists $(f^{(n)}, g^{(n)})$ such that
  \begin{align}
    \frac{1}{n} \log |\mathcal{J}|                  & \leq R + \epsilon, \\
    \frac{1}{n} \log |\mathcal{K}|                  & \leq C + \epsilon, \\
    \frac{1}{n} \mathbb{E}[{\|X^n - \hat{X}^n\|}^2] & \leq D + \epsilon, \\
    \frac{1}{n} W_2^2(p_{X^n}, p_{\hat{X}^n})       & \leq P + \epsilon,
  \end{align}
  for every sufficiently large $n$. For a fixed rate budget $R$, CR limit $C$ and
  perception threshold $P$, the minimum distortion $D$ for which tuple $(R,D,C,P)$ is
  achievable is denoted by $D(R,C,P)$.
\end{definition}

Now consider the vector Gaussian source setting, where
$\mathbf{X} \triangleq {(X_1,X_2,\ldots,X_L)}^T \sim \mathcal{N}(0,\Gamma)$.
Owing to the invariance of the Euclidean distance under unitary transformations,
we can, without loss of generality,
diagonalize the covariance matrix via eigenvalue decomposition and assume
$\Gamma = \operatorname{diag} (\gamma_1, \gamma_2, \ldots, \gamma_L)$. In the
subsequent analysis, we concentrate on the non-degenerate case where $\Gamma \succ 0$.
Under this assumption, $X_1,X_2,\ldots,X_L$ are mutually independent.

According to~\cite{qu2025rate}, the DRCP function $D(R,C,P)$ for the vector Gaussian
sources is given as follows.
\begin{proposition}[The DRCP Function for Vector Gaussian Sources]\label{prop:original_problem}
  For $p_X=\mathcal{N}(0,\Gamma)$,
  \begin{align}
    D(R, C, P) \triangleq
    \min_{\mathbf{r}, \mathbf{r}' \in \mathbb{R}_+^L}
     & \sum_{\ell=1}^L \gamma_\ell 2^{-2r_\ell} \nonumber                                     \\
     & \hspace{-9em} + \Bigg[ \Bigg( \sqrt{\sum_{\ell=1}^L \gamma_\ell \big( 2 - 2^{-2r_\ell}
    - 2 \psi(r_\ell,r'_\ell) \big)} - \sqrt{P} \Bigg)_{\!\!+} \Bigg]^2,\label{eq:prob5_1}     \\
    \text{s.t.}\quad
     & \sum_{\ell=1}^L r_\ell \le R,\label{eq:prob5_2}                                        \\
     & \sum_{\ell=1}^L r'_\ell \le R + C,\label{eq:prob5_3}
  \end{align}
  where $\psi(r_\ell, r'_\ell) \triangleq \sqrt{(1 - 2^{-2r_\ell})(1 - 2^{-2r'_\ell})}$.
\end{proposition}

\begin{lemma}[Convexity of the DRCP Problem]\label{lem:convexity_original_rates}
  The optimization problem~\eqref{eq:prob5_1}-\eqref{eq:prob5_3} is convex
  in $(\mathbf r,\mathbf r')$ on $\mathbb R_+^L\times\mathbb R_+^L$.
\end{lemma}
\begin{IEEEproof}
  See Appendix~\ref{app:convexity_original_rates}.
\end{IEEEproof}

The DRCP function $D(R,C,P)$ can be achieved by incorporating a pair of stochastic
encoder and decoder that share a common random seed with rate $C$.
Based on a set of codebooks generated according to the optimal distribution,
the encoder employs a random coding scheme with the help of a random seed,
while the decoder performs an optimal linear
interpolation between the MMSE estimate and a perceptually
perfect generative reconstruction for each component. By optimally allocating rate
and CR resources across the orthogonal components, the proposed
scheme asymptotically achieves the fundamental DRCP tradeoff in the overall
high-dimensional space. Denote the optimal allocation vectors by
$\hat{\mathbf{r}}$ and $\hat{\mathbf{r}}'$, respectively.
Our goal in this paper is to solve the optimization problem defined in
Prop.~\ref{prop:original_problem} and evaluate $D(R, C, P)$, along
with its analytical upper bound.

\section{Exact Characterization and Optimal Allocation}
Although the optimization problem in Prop.~\ref{prop:original_problem} is
convex, its objective couples the eigenmode allocations through the perception
penalty. Consequently, classical reverse waterfilling alone does not
characterize the optimal allocation. To address this, we propose a trajectory
optimization method to derive the optimality conditions and obtain a generalized
reverse waterfilling characterization. We first identify the perception threshold
separating the classical RD regime from the regime with a positive perception penalty.

\subsection{Threshold for the Perception-Limited Regime}

In this part, we give a necessary and sufficient condition for the perceptual
constraint to be active, which can be used to remove the $(\cdot)_+$ operation
in the objective~\eqref{eq:prob5_1} and simplify the analysis.
To establish the theoretical foundation of our proposed active condition, we
first define two baseline rate allocation strategies that serve as critical reference points
for the subsequent analysis.
\begin{definition}[Classical Reverse Waterfilling Solution]\label{def:r_tilde}
  For any $R>0$, we define a rate tuple $\tilde{\mathbf{r}}$
  whose $\ell$-th component $\tilde{r}_\ell$ is
  \begin{equation}\label{eq:def_r_tilde}
    \tilde{r}_\ell \triangleq \frac{1}{2} \log^+ \left( \frac{\gamma_\ell}{\alpha} \right),
  \end{equation}
  for $\ell = 1, 2, \dots, L$, where $\alpha$
  is the unique positive scalar satisfying the rate constraint that
  \begin{equation}\label{eq:alpha_condition}
    \sum_{\ell=1}^L \frac{1}{2} \log^+ \left( \frac{\gamma_\ell}{\alpha} \right) = R.
  \end{equation}
  When $R=0$, define $\tilde{\mathbf{r}} \triangleq \mathbf{0}$.
\end{definition}

Notice that this rate tuple $\tilde{\mathbf{r}}$ is the optimal rate allocation
of the classical RD problem for the vector Gaussian sources, which is formulated as
\begin{align}
  \underline{D}(R) \triangleq
   & \min_{\mathbf{r} \in \mathbb{R}_+^L} \sum_{\ell=1}^L \gamma_\ell 2^{-2r_\ell},\label{eq:opt1} \\
   & \text{s.t.} \sum_{\ell=1}^L r_\ell \le R.\label{eq:opt1_constraint}
\end{align}
The minimizer to this problem admits a well-known reverse waterfilling
interpretation~\cite{cover1999elements}.
It is worth noting that $\tilde{\mathbf{r}}$ is separable
across components and exhibits no coupling among different components except
through the global water level $\alpha$.

Based on $\tilde{\mathbf{r}}$, we further define another rate
tuple $\tilde{\mathbf{r}}'$.

\begin{definition}[Secondary Sum-Rate Allocation]
  \label{def:r_prime}
  For any given $R>0$, $C\geq 0$, and $\tilde{\mathbf{r}}$ given
  in Def.~\ref{def:r_tilde}, we define a rate tuple $\tilde{\mathbf{r}}'$ whose
  $\ell$-th component $\tilde{r}'_\ell$ is
  \begin{equation}\label{eq:def_r_prime}
    \tilde{r}'_\ell \triangleq \frac{1}{2} \log
    \frac{1 + \sqrt{1 + \beta \gamma_\ell^2(1 - 2^{-2\tilde{r}_\ell})}}{2},
  \end{equation}
  for $\ell = 1, 2, \dots, L$, where $\beta \in \mathbb{R}_{0}^{+}$ is a unique
  scalar satisfying the sum rate constraint that
  \begin{equation}\label{eq:lambda_condition}
    \sum_{\ell=1}^L \tilde{r}'_\ell = R + C.
  \end{equation}
  When $R=0$, define $\tilde{\mathbf{r}}' \triangleq \mathbf{0}$.
\end{definition}

We can see that the rate tuple $\tilde{\mathbf{r}}'$ is the optimal rate allocation
for another optimization problem when $R$, $C$ and $\tilde{\mathbf{r}}$ are given.
\begin{lemma}\label{lem:optimal_of_r_prime}
  For a given $\tilde{\mathbf{r}}$ defined in Def.~\ref{def:r_tilde}, the optimal
  solution to the following maximization problem is given by $\tilde{\mathbf{r}}'$, i.e.,
  \begin{align}
    \tilde{\mathbf{r}}' \in \arg \max_{\mathbf{r}' \in\mathbb{R}_+^L}
     & \sum_{\ell=1}^L \gamma_\ell \psi(\tilde{r}_\ell, r'_\ell), \label{eq:opt2} \\
    \text{s.t.} \enspace
     & \sum_{\ell=1}^L r_\ell' \le R + C. \label{eq:opt2_constraint}
  \end{align}
\end{lemma}
This lemma can be verified by analyzing the KKT conditions, where the detailed proof
can be found in Appendix~\ref{app:optimal_of_r_prime}.
Now we establish the relationship between these two baseline allocations.
\begin{corollary}[Properties of Baseline Allocations]\label{cor:properties_greedy}
  The inequality $\tilde{r}_\ell \le \tilde{r}'_\ell$ holds
  for all $\ell = 1, 2, \ldots, L$.
\end{corollary}
\begin{IEEEproof}
  See Appendix~\ref{app:properties_greedy}.
\end{IEEEproof}

Def.~\ref{def:r_tilde} and Def.~\ref{def:r_prime} together give us two rate
tuples $\tilde{\mathbf{r}}$ and $\tilde{\mathbf{r}}'$, which are a set of feasible
solutions of Prop.~\ref{prop:original_problem}. They can be regarded as a solution
by the greedy strategy. Specifically, rather than jointly optimizing $\mathbf{r}$
and $\mathbf{r}'$, we decouple the original optimization problem into two sequential
stages, yielding the stage-wise optimal solution $\tilde{\mathbf{r}}$ and
$\tilde{\mathbf{r}}'$ for each. Initially, disregarding
the limitations of CR and perception constraints, we allocate
the total transmission rate budget $R$ via $\tilde{\mathbf{r}}$ to minimize
the classical RD distortion $\underline{D}(R)$. Building upon $\tilde{\mathbf{r}}$
and Cor.~\ref{cor:properties_greedy}, we further allocate the limited CR
$C$ via $\tilde{\mathbf{r}}'-\tilde{\mathbf{r}}$ to minimize
the total distortion penalty incurred by satisfying the perception constraint.
This scheme naturally yields an upper bound on $D(R,C,P)$, as detailed
in Section~\ref{sec:greedy_solution}.

Based on the baseline rate allocations $\tilde{\mathbf{r}}$ and $\tilde{\mathbf{r}}'$,
we now investigate the original problem~\eqref{eq:prob5_1}-\eqref{eq:prob5_3} incorporating
the perception limit $P$.
The structural behavior of the optimal solution $\hat{\mathbf{r}}$ and $\hat{\mathbf{r}}'$
is fundamentally governed by the strictness of the perception constraint, which is
completely determined in the following theorem.
\begin{theorem}[Active Condition of Perception Constraint]\label{thm:P_active_condition}
  The DRCP function $D(R,C,P)$ can be formulated as
  \begin{equation}
    D(R, C, P) =
    \begin{cases}
      \underline{D}(R),    & P \ge \tilde{P}, \\
      \overline{D}(R,C,P), & P < \tilde{P},
    \end{cases}
  \end{equation}
  where $\tilde{P}$ is the critical perception threshold defined as
  \begin{equation}\label{eq:P_tilde_explicit}
    \tilde{P} \triangleq \sum_{\ell=1}^L \gamma_\ell
    \left( 2 - 2^{-2\tilde{r}_\ell} - 2\psi(\tilde{r}_\ell, \tilde{r}'_\ell) \right),
  \end{equation}
  and $\overline{D}(R,C,P)$ is an upper bound of $D(R,C,P)$ by simply removing
  the non-negative operator $(\cdot)_+$ in the original objective function~\eqref{eq:prob5_1}.
\end{theorem}
\begin{IEEEproof}
  See Appendix~\ref{app:P_active_condition}.
\end{IEEEproof}

The above theorem shows that $D(R,C,P)$ is a piecewise function of $P$ with a breakpoint
at $\tilde{P}$. The entire perception region can be divided into two regimes by $\tilde{P}$,
which are referred to as the \emph{inactive perception regime} $\Ico{\tilde{P}}{\infty}$
and the \emph{active perception regime} $\Ico{0}{\tilde{P}}$, respectively.
Specifically, when $P \ge \tilde{P}$, the perception constraint is inherently
satisfied by the baseline distortion-optimal allocation $\tilde{\mathbf{r}}$ and
$\tilde{\mathbf{r}}'$. Thus, the perception penalty term $[(\cdot)_{+}]^2=0$,
and the objective function $D(R,C,P)$ reduces to the classical RD
function $\underline{D}(R)$. Hence, the optimal value of $D(R,C,P)$ can be
trivially found under this circumstance. Meanwhile, when $P$ lies in the active
perception region $\Ico{0}{\tilde{P}}$, the perception constraint is strictly
active, thereby we can
remove the non-negative operator $(\cdot)_+$, and the value of  $D(R,C,P)$
equals its upper bound $\overline{D}(R,C,P)$. However, it is still not an
easy problem to solve $\overline{D}(R,C,P)$, and we will focus on that instead
in the remainder of the paper.

\subsection{Optimal Solution}

In this subsection, we are going to characterize the optimal solution of $D(R,C,P)$.
We will first show the results for the inactive perception regime
as a corollary, and then focus on the more challenging part of the active
perception regime to establish an exact parametric characterization.
Further, we investigate some extreme cases of $D(R,C,P)$ where $R,C,P$ tend
to some particular value, and provide some insights into the optimal
solution by analyzing its structural properties.

When $R=0$, it is trivial to obtain that $\hat{\mathbf{r}} = \hat{\mathbf{r}}' = \mathbf{0}$
is a set of optimal solutions for any $C,P \geq 0$, which yields
\begin{equation}
  D(0,C,P) = \operatorname{tr}(\Gamma)+
  \left[\Big(\sqrt{\operatorname{tr}(\Gamma)}-\sqrt{P}\Big)_{+}\right]^2.
\end{equation}
Thus, we will focus on the case of $R>0$ in the subsequent analysis of this paper.

\subsubsection{The case of \texorpdfstring{$P \in \Ico{\tilde{P}}{\infty}$}{P in [Ptilde, infty)}}

The inactive perception constraint means that
the system possesses sufficient degrees of freedom such that the optimal
$\hat{\mathbf{r}}'$ is no longer strictly unique (except for $P = \tilde{P}$).
To maintain mathematical tractability and facilitate deterministic system design, we can
uniquely identify a practical solution by the following corollary.
\begin{corollary}[Optimal Allocation for Inactive Perception Regime]\label{cor:solution_for_inactive_P}
  When $P \in \Ico{\tilde{P}}{\infty}$, the DRCP function is
  \begin{equation}\label{eq:inactive_P}
    D(R,C,P) = \sum_{\ell=1}^L \min\{\alpha, \gamma_\ell\},
  \end{equation}
  where $\alpha$ is the classical water level determined by
  Eq.~\eqref{eq:alpha_condition}. One optimal allocation can be characterized as
  $(\hat{\mathbf{r}}, \hat{\mathbf{r}}') = (\tilde{\mathbf{r}}, \tilde{\mathbf{r}}')$.
\end{corollary}
\begin{IEEEproof}
  Recall that the baseline allocation $\tilde{\mathbf{r}}$ defined in Def.~\ref{def:r_tilde}
  is the unique allocation that minimizes $\underline{D}(R)$, thus substituting
  $\tilde{\mathbf{r}}$ into Eq.~\eqref{eq:opt1} yields
  \begin{equation}
    \underline{D}(R) = \sum_{\ell=1}^L \gamma_\ell 2^{-2\tilde{r}_\ell}
    = \sum_{\ell=1}^L \min\{\alpha, \gamma_\ell\}.
  \end{equation}
  As discussed in Appendix~\ref{app:P_active_condition}, in the inactive perception regime
  where $P \geq \tilde{P}$, the $D(R,C,P)$ is precisely equal to $\underline{D}(R)$, and the optimal
  solution must satisfy $\hat{r}_\ell = \tilde{r}_\ell$ for all $\ell = 1, 2, \dots, L$,
  thus Eq.~\eqref{eq:inactive_P} holds.
  For the secondary allocation $\tilde{\mathbf{r}}'$ given in Def.~\ref{def:r_prime},
  it is also a set of feasible points of $D(R,C,P)$. Therefore, we can select a canonical
  optimal allocation $\hat{\mathbf{r}}' = \tilde{\mathbf{r}}'$, which completes the proof.
\end{IEEEproof}

The closed-form solution of $D(R,C,P)$ in the inactive perception regime
has been fully characterized by Cor.~\ref{cor:solution_for_inactive_P},
thus we will primarily focus on the regime of $P \in \Ico{0}{\tilde{P}}$
in the following analysis.

\subsubsection{The case of \texorpdfstring{$P \in \Ico{0}{\tilde{P}}$}{P in [0, Ptilde)}}

Building upon the formulation given in Thm.~\ref{thm:P_active_condition}, we first introduce
a lemma to provide an overview of $D(R,C,P)$ within the active perception regime.
This lemma demonstrates that the total perception penalty can be divided
into component-wise contributions, which is essential for analyzing the optimal rate allocation.

\begin{lemma}[Component-wise Decoupling of the Objective]\label{lem:D_reformulation}
  In the active perception regime $P \in \Ico{0}{\tilde{P}}$, the DRCP function
  $D(R,C,P)$ can be decomposed into a sum of component-wise effective distortions, given by:
  \begin{equation}\label{eq:D_RCP_definition_2}
    D(R,C,P) = \sum_{\ell=1}^L \left[ D_\ell + \left(\sqrt{W_\ell(\hat{r}_\ell, \hat{r}'_\ell)} - \sqrt{P_\ell}\right)^2 \right],
  \end{equation}
  where the optimal component-wise distortion $D_\ell$, perception penalty
  $W_\ell(\hat{r}_\ell, \hat{r}'_\ell)$, and the allocated perception limit $P_\ell$ for the
  $\ell$-th component are respectively defined as:
  \begin{align}
     & D_\ell \triangleq \gamma_\ell 2^{-2\hat{r}_\ell},                        \\
     & W_\ell(\hat{r}_\ell, \hat{r}'_\ell) \triangleq \gamma_\ell
    \left( 2 - 2^{-2\hat{r}_\ell} - 2\psi(\hat{r}_\ell, \hat{r}'_\ell) \right), \\
     & P_\ell \triangleq
    \frac{W_\ell(\hat{r}_\ell, \hat{r}'_\ell)}{\sum_{\ell'=1}^L W_{\ell'}(\hat{r}_{\ell'}, \hat{r}'_{\ell'})} P. \label{eq:P_l}
  \end{align}
\end{lemma}

\begin{IEEEproof}
  By substituting the definition of the proportionally allocated
  perception limit $P_\ell$ given in Eq.~\eqref{eq:P_l} into the quadratic term on the right-hand
  side of Eq.~\eqref{eq:D_RCP_definition_2}, we obtain:
  \begin{align}
     & \quad \sum_{\ell=1}^L \left( \sqrt{W_\ell(\hat{r}_\ell, \hat{r}'_\ell)} - \sqrt{P_\ell} \right)^2 \nonumber                          \\
     & = \sum_{\ell=1}^L \left( \sqrt{W_\ell(\hat{r}_\ell, \hat{r}'_\ell)}
    - \sqrt{\frac{W_\ell(\hat{r}_\ell, \hat{r}'_\ell)P}{\sum_{\ell'=1}^L W_{\ell'}(\hat{r}_{\ell'}, \hat{r}'_{\ell'})}} \right)^2 \nonumber \\
     & = \sum_{\ell=1}^L \left[W_\ell(\hat{r}_\ell, \hat{r}'_\ell)
    \left( 1 - \sqrt{\frac{P}{\sum_{\ell'=1}^L W_{\ell'}(\hat{r}_{\ell'}, \hat{r}'_{\ell'})}} \right)^2\right] \nonumber                    \\
     & = \left( 1 - \sqrt{\frac{P}{\sum_{\ell'=1}^L W_{\ell'}(\hat{r}_{\ell'}, \hat{r}'_{\ell'})}} \right)^2
    \cdot \sum_{\ell=1}^L W_\ell(\hat{r}_\ell, \hat{r}'_\ell) \nonumber                                                                     \\
     & = \left( \sqrt{\sum_{\ell=1}^L W_\ell(\hat{r}_\ell, \hat{r}'_\ell)} - \sqrt{P} \right)^2. \label{eq:component-wise}
  \end{align}
  Recognizing that $\sum_{\ell=1}^L W_\ell(\hat{r}_\ell, \hat{r}'_\ell)$ is precisely the underlying
  summation within the perception penalty of the original objective function~\eqref{eq:prob5_1},
  thus adding the total distortion $\sum_{\ell=1}^L D_\ell$ to both sides of
  Eq.~\eqref{eq:component-wise} perfectly recovers the exact formulation of $D(R,C,P)$
  in the active perception regime.
\end{IEEEproof}

\begin{remark}
  The structural decomposition presented in Lem.~\ref{lem:D_reformulation} provides
  insights into how the system internally manages the total distortion and perception quality.
  The composite term $D_\ell + (\sqrt{W_\ell(\hat{r}_\ell, \hat{r}'_\ell)} - \sqrt{P_\ell})^2$ and $P_\ell$
  can be interpreted as the effective distortion loss and the assigned perception limit for the
  $\ell$-th source component, respectively.
\end{remark}

Notably, the assigned distortion loss and perceptual limit for the $\ell$-th
component inherently vary with the total perception penalty
$\sum_{\ell=1}^{L} W_\ell(\hat{r}_\ell, \hat{r}'_\ell)$, which renders the global coupling effect
prominent in the regime $P \in \Ico{0}{\tilde{P}}$. This dynamic, state-dependent
assignment stands in contrast to the static, uncoupled allocation
observed in classical waterfilling paradigms, explicitly highlighting the
structural complexity introduced by the perception constraint. Addressing
this coupling effect constitutes a primary analytical challenge.

To overcome this, we propose a tailored trajectory optimization method.
Firstly, we introduce an auxiliary variable $T$ to bound the total perception
penalty $\sum_{\ell=1}^{L} W_\ell(r_\ell, r'_\ell)$, thereby obtaining a relaxed problem that
is convex for any given $T$. Then, we search for the optimal $T$ that
minimizes the total distortion. Here, $T$ can be interpreted as a trajectory
variable that characterizes the total perception penalty, as the $L$-dimensional
optimal allocations $\hat{\mathbf{r}}$ and
$\hat{\mathbf{r}}'$ can be optimized along the trajectory value of $T$ to minimize
the total distortion.

\begin{definition}[Optimization Problem for Fixed Total Perception Penalty]\label{def:problem_re3}
  With a fixed total perception penalty $T \ge P$, we define an optimization
  problem as follows:
  \begin{align}
    D(R,C
     & ,P,T)=\min_{\mathbf{d},\mathbf{d'}\in {\Ioc{0}{1}}^L}\sum_{\ell=1}^L \gamma_\ell {d_\ell}
    + {\Big(\sqrt{T}\!-\!\sqrt{P} \Big)}^2, \label{eq:problem_re3_1}                                                \\
    \text{s.t.} \enspace
     & \sum_{\ell=1}^{L}\gamma_\ell\Big(2-d_\ell-2\sqrt{(1-d_\ell)(1-d_\ell')}\Big) \leq T,\label{eq:problem_re3_2} \\
     & \sum_{\ell=1}^L -\frac{1}{2}\log d_\ell \leq R, \label{eq:problem_re3_3}                                     \\
     & \sum_{\ell=1}^L -\frac{1}{2}\log d_\ell' \leq R+C. \label{eq:problem_re3_4}
  \end{align}
  For completeness, we set $D(R,C,P,T)=+\infty$ whenever the above problem is infeasible.
\end{definition}

Now, we give the following lemma to demonstrate the convexity property of the above
optimization problem.
\begin{lemma}[Convexity Property]\label{lem:convexity_of_fixed_T}
  The optimization problem \eqref{eq:problem_re3_1}-\eqref{eq:problem_re3_4} is jointly
  convex in $(\mathbf{d},\mathbf{d'},T)$.
\end{lemma}
This lemma holds trivially by verifying the convexity of the objective function~\eqref{eq:problem_re3_1}
and the constraints~\eqref{eq:problem_re3_2}-\eqref{eq:problem_re3_4}, so we omit the
detailed proof here for brevity. With the convexity property established, we can now
show the relationship between the original problem~\eqref{eq:prob5_1}-\eqref{eq:prob5_3}
and the auxiliary optimization problem~\eqref{eq:problem_re3_1}-\eqref{eq:problem_re3_4}.

\begin{lemma}[Problem Equivalence]\label{lem:problem_equivalence}
  The original problem~\eqref{eq:prob5_1}-\eqref{eq:prob5_3}
  of $D(R,C,P)$ equals minimizing $D(R,C,P,T)$ over $T \ge P$, i.e.,
  \begin{equation}
    D(R,C,P) = \min_{T \ge P} D(R,C,P,T).
  \end{equation}
\end{lemma}
\begin{IEEEproof}
  See Appendix~\ref{app:problem_equivalence}.
\end{IEEEproof}

Lem.~\ref{lem:problem_equivalence} implies that the optimal solution to
the original problem defined in Eqs.~\eqref{eq:prob5_1}-\eqref{eq:prob5_3} can be retrieved
by minimizing $D(R,C,P,T)$ over the domain $T \ge P$.
For any fixed $T\ge P$ for which the problem is feasible, the optimal allocations for
$\mathbf{d}(T)$ and $\mathbf{d'}(T)$ can be derived analytically via
the method of Lagrange multipliers, which is guaranteed by the convexity
established in Lem.~\ref{lem:convexity_of_fixed_T}.
Motivated by this observation, we propose a trajectory optimization framework to jointly
optimize $T$, $\mathbf{d}$, and $\mathbf{d'}$, thereby yielding the
optimal rate allocations $\hat{\mathbf{r}}$ and $\hat{\mathbf{r}}'$.

To characterize the optimal solution in the active perception regime, we
first introduce some auxiliary parameters.

\begin{definition}[Perception-Coupled Auxiliary Parameters]\label{def:w_l}
  For a given set of multipliers $(\mu, \nu, \nu')$ with $0<\mu\le1$ and
  $\nu,\nu'>0$, define
  \begin{equation}\label{eq:varpi_def}
    \varpi \triangleq \frac{\sqrt{\nu'^2(1-\mu)^2+4\mu^2\nu\nu'}-\nu'(1-\mu)}{2\mu^2}.
  \end{equation}
  For each component with $\gamma_\ell>\varpi$, let
  $\hat w_\ell$ be the unique real root in $(0,\frac{\mu^2\gamma_\ell}{\nu'})$ of
  \begin{align}
     & w_\ell^3 + \left(1-\mu+\frac{\nu'-\nu}{\gamma_\ell}\right)w_\ell^2
    + \left(\frac{(1-\mu)\nu'}{\gamma_\ell}-\mu^2\right)w_\ell \nonumber  \\
     & \quad - \mu^2(1-\mu) = 0.\label{eq:w_l}
  \end{align}
  For each component with $\gamma_\ell\le\varpi$, set
  $\hat w_\ell\triangleq\frac{\mu^2\gamma_\ell}{\nu'}$.
\end{definition}

With the auxiliary parameter $\hat{w}_\ell$ defined, we are now positioned to present
the optimal joint allocation strategy for the active perception regime.

\begin{theorem}[Optimal Joint Allocation for Active Perception Regime]\label{thm:optimal_solution}
  In the active perception regime where $P \in \Ico{0}{\tilde{P}}$, the optimal
  solution $(\hat{\mathbf{r}}, \hat{\mathbf{r}}')$ of the
  DRCP function $D(R,C,P)$ is given by
  \begin{align}
     & \hat{r}_\ell = \frac{1}{2} \log^{+} \left(\frac{\gamma_\ell \hat{w}_\ell^2}
    {\gamma_\ell \hat{w}_\ell^2+\nu'\hat{w}_\ell - \mu^2\gamma_\ell}\right), \label{eq:theorem2_r} \\
     & \hat{r}'_\ell = \frac{1}{2} \log^{+}
    \left(\frac{\mu^2\gamma_\ell}{\nu'\hat{w}_\ell}\right), \label{eq:theorem2_r_prime}
  \end{align}
  Furthermore, the optimal multipliers $\mu$, $\nu$, and $\nu'$ are determined by:
  \begin{align}
     & (1-\mu)^2\sum_{\ell=1}^{L} \left[ \gamma_\ell \left( 1 - \frac{\mu}{\hat{w}_\ell} \right)^2
    + \frac{2\nu'}{\mu} - \frac{\nu'}{\hat{w}_\ell} \right] = P, \label{eq:lag_para_opt_1}                 \\
     & \sum_{\ell=1}^L \frac{1}{2} \log^{+} \left(\frac{\gamma_\ell \hat{w}_\ell^2}
    {\gamma_\ell \hat{w}_\ell^2+\nu'\hat{w}_\ell - \mu^2\gamma_\ell}\right) = R, \label{eq:lag_para_opt_2} \\
     & \sum_{\ell=1}^L \frac{1}{2} \log^{+}
    \left(\frac{\mu^2\gamma_\ell}{\nu'\hat{w}_\ell}\right) = R+C. \label{eq:lag_para_opt_3}
  \end{align}
\end{theorem}
\begin{IEEEproof}
  See Appendix~\ref{app:optimal_solution}.
\end{IEEEproof}

Thm.~\ref{thm:optimal_solution} provides a parametric expression for the optimal
solution $(\hat{\mathbf{r}}, \hat{\mathbf{r}}')$ in the active perception regime where
$P \in \Ico{0}{\tilde{P}}$.
Similar to the classical water level $\alpha$ in Def.~\ref{def:r_tilde},
the critical cutoff threshold for this extended reverse
waterfilling solution is $\varpi$ in Def.~\ref{def:w_l}.
If the variance of the $\ell$-th component $\gamma_\ell$ is less than the cutoff
threshold $\varpi$, i.e., $\gamma_\ell < \varpi$, the optimal solution will allocate
zero rate and shared randomness to this component, namely $\hat{r}_\ell = \hat{r}'_\ell = 0$.

\begin{remark}\label{remark:infty_C}
  Notice that when $C\to \infty$ and $P<\tilde{P}$, the cutoff threshold $\varpi$ will
  approach to zero. In this case, all source components will be active regardless of their
  variances, which is the phenomenon that aligns with the findings in~\cite{qian2024rate}.
\end{remark}

Having established the exact analytical form for the optimal dual resources
allocation in Thm.~\ref{thm:optimal_solution}, we now turn
our attention to the intrinsic structural properties of this optimal solution.

\begin{corollary}[Property of the Optimal Solution]\label{cor:properties_opt}
  For any active perception constraint $P \in \Ico{0}{\tilde{P}}$, the secondary
  allocation universally dominates the baseline allocation across
  all components, i.e.,
  \begin{equation}
    \hat{r}_\ell \leq \hat{r}'_\ell, \qquad \ell=1,\dots,L.
  \end{equation}
\end{corollary}
\begin{IEEEproof}
  See Appendix~\ref{app:properties}.
\end{IEEEproof}


It is worth noting that the analysis method of this optimal solution can be used
to analyze a wide range of extreme scenarios, such as the cases of perfect
perception ($P=0$) given in the following corollary.

\begin{corollary}[Optimal Joint Allocation with Perfect Perception Constraint]\label{cor:optimal_solution_perfect_P}
  Under the assumptions of Thm.~\ref{thm:optimal_solution}, the optimal allocations
  for perfect perceptual quality $P=0$ are
  \begin{align}
     & \hat{r}_\ell = \frac{1}{2}\log^{+}\left(\frac{\sqrt{(\nu_1-\nu_1')^2 + 4\gamma_{\ell}^2}
    +(\nu_1-\nu_1')}{2\nu_1}\right), \label{eq:opt_perfectP_r}                                     \\
     & \hat{r}_{\ell}' = \frac{1}{2}\log^{+}\left(\frac{\sqrt{(\nu_1-\nu_1')^2 + 4\gamma_{\ell}^2}
        - (\nu_1 - \nu_1')}{2\nu_1'}\right), \label{eq:opt_perfectP_r_prime}
  \end{align}
  where $\nu_1,\nu_1'>0$ are the unique multipliers satisfying
  \begin{align}
    \sum_{\ell=1}^L\hat r_\ell  & =R, \label{eq:lam1}   \\
    \sum_{\ell=1}^L\hat r_\ell' & =R+C. \label{eq:lam2}
  \end{align}
  Their common cutoff threshold is $\varpi_0=\sqrt{\nu_1\nu_1'}$, which is also
  the unique solution of
  \begin{equation}\label{eq:varpi0_condition}
    \sum_{\ell=1}^L\log^+ \left(\frac{\gamma_\ell}{\varpi_0}\right)=2R+C.
  \end{equation}
  Thus both rates are positive if $\gamma_\ell>\varpi_0$ and zero otherwise.
\end{corollary}
\begin{IEEEproof}
  See Appendix~\ref{app:optimal_solution_perfect_P}.
\end{IEEEproof}

In particular, for the high rate regime where $\varpi_0 < \min_\ell\gamma_\ell$, all
components are active and the cutoff threshold $\varpi_0$ becomes
\begin{equation}
  \varpi_0= 2^{\frac{1}{L}\left(\sum_{\ell=1}^L\log\gamma_\ell-2R-C\right)}.
\end{equation}

We now proceed to demonstrate that the results of
Cor.~\ref{cor:optimal_solution_perfect_P} recover known results
in extreme operational scenarios.

\begin{corollary}[{\cite[Theorem~1]{yan2022optimally}}]\label{cor:opt_perfectP_noC}
  When $C=0$, $P=0$, and $R>0$, the function $D(R,0,0)$ is given by
  \begin{equation}
    D(R,0,0) = 2 \sum_{\ell=1}^L \min\{\alpha, \gamma_\ell\},
  \end{equation}
  and the corresponding optimal dual allocations are given by
  \begin{equation}
    \hat{r}_\ell = \hat{r}_\ell' = \frac{1}{2}\log^{+}\left(\frac{\gamma_\ell}{\alpha}\right),
  \end{equation}
  for all $\ell=1,2,\dots,L$, where $\alpha$ is the classical
  reverse waterfilling level determined by Eq.~\eqref{eq:alpha_condition}.
\end{corollary}
\begin{IEEEproof}
  For $R>0$ and $C=0$, Cor.~\ref{cor:properties_opt} and the active rate
  constraints~\eqref{eq:lam1}-\eqref{eq:lam2} give $\hat r_\ell'-\hat r_\ell\ge0$
  for $\ell=1,\dots,L$ and
  $\sum_{\ell=1}^L(\hat r_\ell'-\hat r_\ell)=(R+C)-R=0$.
  Hence $\hat{\mathbf r}'=\hat{\mathbf r}$.
  With $P=0$, the objective in~\eqref{eq:prob5_1} reduces to
  $2\sum_{\ell=1}^L\gamma_\ell2^{-2r_\ell}$, whose minimum is
  attained by the classical reverse waterfilling allocation in
  Def.~\ref{def:r_tilde}. This completes the proof.
\end{IEEEproof}

\begin{corollary}[{\cite[Corollary~1]{qian2024rate}}]\label{cor:opt_perfectP_infC}
  When $C \to \infty$ and $P=0$, the function $D(R,\infty,0)$ is given by
  \begin{equation}
    D(R, \infty, 0) = \sum_{\ell=1}^{L} \left( 2\gamma_{\ell}
    - 2\sqrt{\nu_2^2 + \gamma_{\ell}^{2}} + 2\nu_2 \right),
  \end{equation}
  and the corresponding optimal allocations are given by
  \begin{align}
     & \hat{r}_\ell = \frac{1}{2}\log\frac{\sqrt{\nu_2^2 + \gamma_{\ell}^2}
    + \nu_2}{2\nu_2},\label{eq:r_hat_Cinf_P0}                               \\
     & \hat{r}_{\ell}' \to \infty,\label{eq:r_prime_Cinf_P0}
  \end{align}
  for all $\ell=1,2,\dots,L$, where $\nu_2$ is the unique multiplier chosen to satisfy
  $\sum_{\ell=1}^L \hat{r}_\ell = R$.
\end{corollary}
\begin{IEEEproof}
  When $C \to \infty$, according to the complementary slackness condition, the multiplier $\nu_1'$
  given in Cor.~\ref{cor:optimal_solution_perfect_P} must satisfy $\nu_1' \to 0$.
  Taking $\nu_1' \to 0$ into Eq.~\eqref{eq:opt_perfectP_r} and Eq.~\eqref{eq:opt_perfectP_r_prime},
  the optimal solution can be simplified to Eq.~\eqref{eq:r_hat_Cinf_P0}
  and Eq.~\eqref{eq:r_prime_Cinf_P0}, which completes the proof.
\end{IEEEproof}

\section{Greedy Allocation and Approximate Universality}\label{sec:greedy_solution}
Due to the difficulty of solving the coupled non-linear equations in
Thm.~\ref{thm:optimal_solution}, it is generally intractable to
obtain an analytical expression for the DRCP function $D(R,C,P)$.
Therefore, we propose a sub-optimal but universal ``greedy'' strategy
to simplify the coupled optimization. By sequentially determining
the rate allocations, we establish a rigorous upper bound on the DRCP function.

\begin{theorem}[Greedy Upper Bound]\label{thm:greedy_solution}
  We define $\tilde{D}(R, C, P)$ as
  \begin{equation}\label{eq:D_tilde_definition}
    \tilde{D}(R, C, P) \triangleq
    \sum_{\ell=1}^L \gamma_\ell 2^{-2 \tilde{r}_\ell}
    + {\left[{\left( \sqrt{\tilde{P}} - \sqrt{P} \right)}_{+}\right]}^2,
  \end{equation}
  where the value of $\tilde{\mathbf{r}}$, $\tilde{\mathbf{r}}'$ and $\tilde{P}$ are taken
  as Def.~\ref{def:r_tilde}, Def.~\ref{def:r_prime} and Eq.~\eqref{eq:P_tilde_explicit},
  respectively. Then, $\tilde{D}(R, C, P)$ is an upper bound of the DRCP function
  $D(R, C, P)$, namely
  \begin{equation}
    \tilde{D}(R, C, P) \ge D(R, C, P).
  \end{equation}
\end{theorem}

\begin{IEEEproof}
  This theorem can be easily proved by showing the aforementioned $\tilde{\mathbf{r}}$
  and $\tilde{\mathbf{r}}'$ are a set of feasible solutions of $D(R,C,P)$ and achieves
  the value of $\tilde{D}(R, C, P)$ .

  The feasibility of $\tilde{\mathbf{r}}$ and $\tilde{\mathbf{r}}'$ have been discussed
  in Def.~\ref{def:r_tilde} and Def.~\ref{def:r_prime}, which hold trivially.
  Meanwhile, by taking $\mathbf{r}=\tilde{\mathbf{r}}$ and $\mathbf{r}'=\tilde{\mathbf{r}}'$,
  we have
  \begin{align}
     & \quad D(R, C, P) \nonumber                                                                                      \\
     & = \min_{\mathbf{r},\mathbf{r}'}\sum_{\ell=1}^L \gamma_\ell 2^{-2 r_\ell} +
    \left[\left(\sqrt{\sum_{\ell=1}^L  W_\ell(r_\ell, r'_\ell)}-\sqrt{P} \right)_{+}\right]^2 \nonumber                \\
     & \le \sum_{\ell=1}^L \gamma_\ell 2^{-2 \tilde{r}_\ell} +
    \left[\left(\sqrt{\sum_{\ell=1}^L W_\ell(\tilde{r}_\ell, \tilde{r}'_\ell)}-\sqrt{P}\right)_{+} \right]^2 \nonumber \\
     & = \sum_{\ell=1}^L \gamma_\ell 2^{-2 \tilde{r}_\ell} + \left[ \left( \sqrt{\tilde{P}}
    - \sqrt{P} \right)_{+} \right]^2 \nonumber                                                                         \\
     & = \tilde{D}(R, C, P).
  \end{align}
  This completes the proof.
\end{IEEEproof}

For fixed $R$ and $C$, the allocation pair $(\tilde{\mathbf r},\tilde{\mathbf r}')$ is independent of $P$. In the associated coding construction, the encoder and representation are therefore fixed as $P$ varies, while the decoder adapts its interpolation rule to meet the perception constraint. Thus, $\tilde D(R,C,P)$ describes the distortion attainable by this representation across different perception thresholds. We establish its approximate universality by bounding the gap and ratio between $\tilde D(R,C,P)$ and the optimal distortion $D(R,C,P)$.


We can observe that $\tilde{D}(R,C,P)$ is tight in some special cases, such as the two
operational extremes discussed in the following corollary.
\begin{corollary}[Tightness at Operational Extremes]\label{cor:tightness_extremes}
  The greedy upper bound is tight when the perception constraint is inactive
  or no CR is available. That is,
  \begin{align}
    D(R,0,P)
     & =\underline D(R)+\Bigl[\bigl(\sqrt{\underline D(R)}-\sqrt P\bigr)_+\Bigr]^2 \nonumber \\
     & =\tilde D(R,0,P),
  \end{align}
  and $\tilde{D}(R,C,P')=D(R,C,P')$ for any $P'\geq \tilde{P}$.
\end{corollary}
This corollary can be proved trivially by
comparing the optimal solutions $(\hat{\mathbf{r}},\hat{\mathbf{r}}')$ in these
two extreme cases with $(\tilde{\mathbf{r}},\tilde{\mathbf{r}}')$ given in
Def.~\ref{def:r_tilde} and Def.~\ref{def:r_prime}, which are omitted here for brevity.

\subsection{Difference between Greedy and Optimal Solutions}
To characterize the performance gap within the general cases, we further define
the gap function as
\begin{equation}\label{eq:Delta_D_define}
  \Delta D(R,C,P) \triangleq \tilde{D}(R,C,P) - D(R,C,P),
\end{equation}
which is a non-negative function that quantifies the tightness of the greedy
strategy. Although we can easily obtain the value of $\tilde{D}(R,C,P)$ from
Eq.~\eqref{eq:D_tilde_definition}, directly evaluating $D(R,C,P)$ is complex
due to the coupling of the optimal solution with the Lagrange multipliers and
the cubic equation.
Consequently, directly evaluating $\Delta D(R,C,P)$ is as intractable as
deriving an exact analytical solution for the original optimization problem.
To circumvent this, we first derive an analytical upper bound on
$\Delta D(R,C,P)$ that holds across all $C$ and $P$.

\begin{theorem}[Analytical Bound for the Gap with Given $R$ and $\Gamma$]\label{thm:gap_upper_bound}
  For any given thresholds $R>0$, $C, P \geq 0$ and source covariance matrix
  $\Gamma = \operatorname{diag} (\gamma_1, \gamma_2, \ldots, \gamma_L)$, the gap
  function is upper bounded by $\Delta D(R,C,P) \leq U(R, \Gamma)$,
  where $U(R, \Gamma)$ is defined as:
  \begin{equation}\label{eq:U_R_definition}
    U(R, \Gamma) \triangleq 2 \sum_{\ell=1}^{L} \left(\sqrt{\gamma_{\ell}^2 \!+\! \nu_2^2}
    - \sqrt{\gamma_{\ell}(\gamma_{\ell} \!-\! \alpha)_{+}} - \nu_2 \right),
  \end{equation}
  for $R>0$, and $U(0,\Gamma) \triangleq 0$.
  Here, $\alpha$ is the classical reverse waterfilling level determined in
  Def.~\ref{def:r_tilde} that satisfies Eq.~\eqref{eq:alpha_condition},
  and $\nu_2$ is a unique positive scalar satisfying
  \begin{equation}\label{eq:opt_D_multiplier}
    \sum_{\ell=1}^{L} \log
    \left( 1 + \sqrt{1 + \frac{\gamma_{\ell}^{2}}{\nu_2^2}} \right) = 2R + L.
  \end{equation}
\end{theorem}
\begin{IEEEproof}
  See Appendix~\ref{app:gap_upper_bound}.
\end{IEEEproof}

Thm.~\ref{thm:gap_upper_bound} says that the gap $\Delta D(R,C,P)$ can be upper
bounded by $U(R, \Gamma)$, which essentially is a function of $R$ alone for a
given source $\Gamma$. Moreover, we can further find the constant upper bound
of $U(R, \Gamma)$ in the following corollary, which is also an upper bound on
$\Delta D(R,C,P)$ that holds for any given vector Gaussian source.

\begin{corollary}[Upper Bound for the Gap with Given $\Gamma$]\label{cor:property_of_U}
  For any specific $\Gamma$, the global maximum of $U(R, \Gamma)$ over $R>0$
  is attained at one of the following discrete rate points:
  \begin{equation}
    \max_{R > 0} U(R, \Gamma) = \max_{k\in \{1,2,\ldots,L\}} U(R_k, \Gamma),
  \end{equation}
  where $R_k = \sum_{\ell=1}^{L} \frac{1}{2}\log^+\left( \frac{\gamma_\ell}{\gamma_k} \right)$
  for $k=1,2,\ldots,L$. Therefore, the gap function $\Delta D(R,C,P)$ is upper bounded by
  \begin{equation}
    \Delta D(R,C,P) \le \max_{\ell\in \{1,2,\ldots,L\}} U(R_\ell, \Gamma).
  \end{equation}
\end{corollary}
\begin{IEEEproof}
  See Appendix~\ref{app:property_of_U}.
\end{IEEEproof}

\begin{remark}
  The proof of Thm.~\ref{thm:gap_upper_bound} implies that $U(R, \Gamma)$
  is asymptotically tight when $C \to \infty$ and $P=0$, and we also know that
  there is no cut-off component for the optimal solution when $C \to \infty$
  from the Remark~\ref{remark:infty_C}.
  However, the greedy strategy still cuts off some components by the classical
  water level $\alpha$. This is the main reason why the upper bound $U(R,\Gamma)$
  takes its extreme value at the transition points $R_\ell$, where the water
  level $\alpha$ coincides with the $\ell$-th component variance $\gamma_\ell$.
\end{remark}

To directly find the supremum of $\Delta D(R,C,P)$ for a given source $\Gamma$,
one would need to exhaustively evaluate all possible rate budgets $R$ over
$\Ico{0}{\infty}$. However, Cor.~\ref{cor:property_of_U} significantly
streamlines this procedure. It establishes that the continuous maximization
problem over $R \in \Ico{0}{\infty}$ is
equivalent to a discrete search over the finite set of rate points
$\{R_1, R_2, \ldots, R_L\}$. Consequently, determining the
supremum of the gap requires evaluating merely $L$ discrete points, rendering
the process strictly independent of $R$.

Furthermore, we can also derive a universal upper bound for the gap function
$\Delta D(R,C,P)$ that quantifies the largest possible performance gap between
the greedy strategy and the optimal solution among all vector Gaussian
sources with a fixed trace $\operatorname{tr}(\Gamma)$.
For $\nu,\gamma>0$, define
\begin{align}
  \varphi(\nu,\gamma)
   & \triangleq \sqrt{\gamma^2+\nu^2}-\sqrt{\gamma(\gamma-1)_+}-\nu \nonumber       \\
   & \quad-\nu\left(\ln\frac{\nu+\sqrt{\gamma^2+\nu^2}}{2\nu}-\ln^{+}\gamma\right).
  \label{eq:H_nu_gamma_def}
\end{align}
\begin{corollary}[Universal Bound for the Gap]\label{cor:universal_gap_bound}
  For any vector Gaussian source $\Gamma$ subject to a fixed trace
  $\operatorname{tr}(\Gamma) = \sum_{\ell=1}^{L}\gamma_\ell$,
  the performance gap $\Delta D(R, C, P)$ is universally bounded by
  \begin{equation}
    0 \le \Delta D(R,C,P)\le 2\rho^*\operatorname{tr}(\Gamma)\approx 0.2897\operatorname{tr}(\Gamma),
  \end{equation}
  where $\rho^*=\varphi(\bar{\nu}_2^*,1)\approx 0.1448$ is a fundamental constant, and
  $(\bar{\nu}_2^*,\tilde{\gamma}_1^*)\in (0,\infty)\times (1,\infty)$ is the unique solution of
  \begin{equation}\label{eq:universal_gap_constant}
    \frac{\partial \varphi(\bar{\nu}_2,\tilde{\gamma}_1)}{\partial \tilde{\gamma}_1}
    =\frac{\varphi(\bar{\nu}_2,\tilde{\gamma}_1)}{\tilde{\gamma}_1}=\varphi(\bar{\nu}_2,1).
  \end{equation}
  Moreover, the universal upper bound $2\rho^*\operatorname{tr}(\Gamma)$
  is asymptotically achievable.
\end{corollary}
\begin{IEEEproof}
  See Appendix~\ref{app:universal_gap_bound}.
\end{IEEEproof}

This corollary characterizes the sharp universal bound on the performance gap between
the greedy upper bound $\tilde{D}(R,C,P)$ and the optimal $D(R,C,P)$, which is
independent of the specific covariance matrix $\Gamma$, the rate budget $R$, the
CR limit $C$, and the perception threshold $P$.
That is, for an arbitrary vector Gaussian source, the absolute error incurred
by employing $\tilde{D}(R,C,P)$ as an approximation for $D(R,C,P)$ is universally
upper bounded by $2 \rho^* \operatorname{tr}(\Gamma)$.

\subsection{Ratio between Greedy and Optimal Solutions}
We further evaluate the tightness of the greedy upper bound $\tilde{D}(R,C,P)$ by analyzing
the minimum and maximum ratio between $\tilde{D}(R, C, P)$ and
the optimal solution $D(R, C, P)$ in this part. To simplify the notation, we define
the ratio function $\eta(R,C,P)$ as
\begin{equation}
  \eta(R,C,P) \triangleq \frac{D(R,C,P)}{\tilde{D}(R,C,P)}.
\end{equation}
Now we give an analytical lower bound on $\eta(R,C,P)$ that holds across
all $C$ and $P$.

\begin{theorem}[Analytical Bound for the Ratio with Given $R$ and $\Gamma$]\label{thm:analytical_ratio}
  For any given thresholds $R>0$, $C, P \geq 0$ and source covariance matrix
  $\Gamma$, the ratio function is lower bounded by $\eta (R,C,P) \geq L(R,\Gamma)$,
  where $L(R,\Gamma)$ is defined as
  \begin{equation}\label{eq:L_R_definition_ratio}
    L(R,\Gamma) \triangleq
    \frac{\sum_{\ell=1}^{L}\left(\nu_2 + \gamma_{\ell} - \sqrt{\nu_2^2 + \gamma_{\ell}^{2}}\right)}
    {\sum_{\ell=1}^{L}\left(\gamma_{\ell} - \sqrt{\gamma_{\ell}(\gamma_{\ell}-\alpha)_{+}}\right)},
  \end{equation}
  for $R>0$, and $L(0,\Gamma) \triangleq 1$.
  Here, $\alpha$ and $\nu_2$ are implicit functions of $(R,\Gamma)$ determined by
  Eq.~\eqref{eq:alpha_condition} and~\eqref{eq:opt_D_multiplier}, respectively.
\end{theorem}
\begin{IEEEproof}
  See Appendix~\ref{app:analytical_ratio}.
\end{IEEEproof}

Thm.~\ref{thm:analytical_ratio} says that the ratio $\eta(R,C,P)$ can be lower
bounded by $L(R,\Gamma)$, which essentially is a function of $R$ and $\Gamma$.
Similarly, we can further find the constant lower bound of $L(R,\Gamma)$ in the following
corollary, which is also a lower bound on $\eta(R,C,P)$ that holds for any specific
source $\Gamma$.

\begin{corollary}[Lower Bound for the Ratio with Given $\Gamma$]\label{cor:property_of_L}
  For any specific $\Gamma$, the global minimum of $L(R, \Gamma)$ over $R>0$
  is attained at one of the discrete rate points as follows
  \begin{equation}\label{eq:global_min_L_overR}
    \min_{R > 0} L(R, \Gamma) = \min_{k\in \{1,2,\ldots,L\}} L(R_k, \Gamma),
  \end{equation}
  where $R_k = \sum_{\ell=1}^{L} \frac{1}{2}\log^+\left( \frac{\gamma_\ell}{\gamma_k} \right)$
  for $k=1,2,\ldots,L$. Therefore, the ratio function $\eta(R,C,P)$ is lower bounded by
  \begin{equation}
    \eta(R,C,P) \ge \min_{\ell\in \{1,2,\ldots,L\}} L(R_\ell, \Gamma).
  \end{equation}
\end{corollary}
\begin{IEEEproof}
  See Appendix~\ref{app:property_of_L}.
\end{IEEEproof}

Moreover, we also establish a universal bound for the ratio $\eta(R,C,P)$ that is
valid for any vector Gaussian source.

\begin{corollary}[Universal Bound for the Ratio]\label{cor:universal_ratio_bound}
  The ratio $\eta(R,C,P)$ between the greedy and optimal solutions always satisfies
  \begin{equation}
    0.7517 \approx \eta^* \leq \eta(R,C,P) \leq 1.
  \end{equation}
  where $\eta^*=2\tilde{\nu}_2 \left( 1 - \ln(2\tilde{\nu}_2) \right)$ is a
  fundamental constant, and $\tilde{\nu}_2\in (\frac{1}{2}, \infty)$ is the unique
  real root of the equation
  \begin{equation}\label{eq:nu1_star}
    \nu \ln \left( 2\nu^2 + 2\nu\sqrt{1+\nu^2} \right) = \sqrt{1+\nu^2} + \nu - 1.
  \end{equation}
\end{corollary}
\begin{IEEEproof}
  See Appendix~\ref{app:universal_ratio_bound}.
\end{IEEEproof}

The corollary above establishes a universal lower bound on the ratio
between the greedy and optimal solutions, which is independent of the specific
source $\Gamma$, the rate budget $R$, the CR limit $C$, and the perception
threshold $P$.

\section{Conclusion}
In this paper, we have provided a comprehensive analytical characterization
of the DRCP tradeoff for vector Gaussian sources under limited CR. By resolving
the multi-coupling caused by the Wasserstein-2 perception constraint,
we reformulated the problem into a convex trajectory optimization
framework. This transformation allowed us to derive exact,
parametric expressions for the optimal allocation of transmission rate
and CR. As a result, we established a generalized reverse waterfilling
theory for the limited-CR scenario, along with explicit characterizations for
the special case of perfect realism. To bridge the gap between
theoretical limits and practical implementation, we further proposed a
greedy allocation strategy as an analytical upper bound. For fixed communication and CR budgets, the proposed representation is independent of the perception threshold, with only the decoder adapting to the desired perceptual quality. The sharp bounds on its performance gap and ratio establish approximate universality across all perception thresholds.
Our findings offer solid theoretical guidelines for resource management in high-dimensional
generative compression systems and pave the way for future research into
non-Gaussian sources and practical lattice-based coding designs.

{
\appendices
\section{Proof of Lemma~\ref{lem:convexity_original_rates}}\label{app:convexity_original_rates}
The convexity of the objective~\eqref{eq:prob5_1} is not
apparent from its original form, owing to the coupling of the variables
within the nested square-root and non-negative operations.
To address this, we first establish the following identity to rewrite
the objective in a more tractable form.
For any $t,P\ge0$, we have
\begin{equation}\label{eq:original_convexity_scalar_identity}
  \Bigl[\bigl(\sqrt t-\sqrt P\bigr)_+\Bigr]^2
  =\sup_{0\le\lambda<1}\left\{\lambda t-\frac{\lambda P}{1-\lambda}\right\}.
\end{equation}
Indeed, for $P>0$, the derivative of the expression in braces
with respect to (w.r.t.) $\lambda$ is $t-P/(1-\lambda)^2$. The expression is maximized at
$\lambda=0$ when $t\le P$, and at $\lambda=1-\sqrt{P/t}$ when $t>P$,
yielding the left-hand side in both cases. For $P=0$, the right-hand
side of Eq.~\eqref{eq:original_convexity_scalar_identity} becomes
$\sup_{0\le\lambda<1}\lambda t=t$, and this supremum is
approached as $\lambda\uparrow1$ when $t>0$.

For any $(\mathbf r,\mathbf r')\in\mathbb R_+^L\times\mathbb R_+^L$, let
\begin{equation}
  t =\sum_{\ell=1}^L\gamma_\ell \bigl(2-2^{-2r_\ell}-2\psi(r_\ell,r_\ell')\bigr)\overset{(a)}{\geq} 0,
\end{equation}
where step $(a)$ follows from $2\psi(r_\ell,r_\ell')\le2-2^{-2r_\ell}-2^{-2r_\ell'}$
by the arithmetic-geometric mean inequality.
Using Eq.~\eqref{eq:original_convexity_scalar_identity}, the
objective~\eqref{eq:prob5_1} can be written as
\begin{align}
   & \sum_{\ell=1}^L\gamma_\ell2^{-2r_\ell} + \Bigl[\bigl(\sqrt t-\sqrt P\bigr)_+\Bigr]^2
  = \sup_{0\le\lambda<1} \Bigg\{\sum_{\ell=1}^L\gamma_\ell \nonumber                             \\
   & \qquad \cdot \Bigl[(1-\lambda)2^{-2r_\ell}+2\lambda\bigl(1-\psi(r_\ell,r_\ell')\bigr)\Bigr]
  -\frac{\lambda P}{1-\lambda}\Bigg\},
  \label{eq:original_convexity_supremum}
\end{align}
where the expression in braces is jointly convex in $(\mathbf r,\mathbf r')$
for each fixed $\lambda\in[0,1)$, since
$2^{-2r_\ell}$ and $-\psi(r_\ell,r_\ell')$ are convex,
their coefficients are nonnegative, and the last term is constant
w.r.t. $\mathbf r$ and $\mathbf r'$.
Notice that a pointwise supremum of convex functions is also convex and
the constraints~\eqref{eq:prob5_2}-\eqref{eq:prob5_3} are affine,
the original DRCP problem is convex. This completes the proof.

\section{Proof of Lemma~\ref{lem:optimal_of_r_prime}}\label{app:optimal_of_r_prime}
To prove Lem.~\ref{lem:optimal_of_r_prime}, we first analyze the structural properties
of the optimal solution. When $R=0$, the objective in~\eqref{eq:opt2} is identically zero,
so $\tilde{\mathbf r}'=\mathbf0$ in Def.~\ref{def:r_prime} is an optimal solution.
Subsequently, we concentrate on the case $R>0$ and let
$\mathcal A\triangleq\{\ell:\tilde r_\ell>0\}$ denote the active set of components.
The objective is continuous on the compact feasible set, so a maximizer exists.
It is strictly increasing in $r'_\ell$ for $\ell\in\mathcal A$ and
independent of $r'_\ell$ for $\ell\notin\mathcal A$, thus
every maximizer uses the entire budget $R+C$ and assigns zero
rate outside $\mathcal A$.
Moreover, every maximizer assigns positive rate to every active component
in $\mathcal A$, since its right derivative at zero is infinite, whereas
the derivative at any positive rate is finite, which means transferring
a sufficiently small amount of rate from a positive component to a zero
active component would strictly improve the objective.

Based on the properties established above, we can restrict the optimization
problem to the components in $\mathcal A$. In this restricted problem, the
objective function~\eqref{eq:opt2} is strictly concave, and the sum-rate
constraint~\eqref{eq:opt2_constraint} is linear.
Consequently, the restricted problem is a convex optimization problem
with a unique maximizer, which can be characterized by the first-order KKT conditions.
The Lagrangian function of this restricted problem can be expressed as
\begin{align}
  \mathcal{L}(\mathbf{r'}, \lambda_1)
   & = \sum_{\ell\in\mathcal A} \gamma_\ell\sqrt{(1 - 2^{-2\tilde{r}_\ell})(1 - 2^{-2r'_\ell})} \nonumber \\
   & \quad - \lambda_1 \left( \sum_{\ell\in\mathcal A} r'_\ell - R - C \right),
\end{align}
with $\lambda_1 \in \mathbb{R}_0^+$ being the Lagrange multiplier.
Since all retained rates are positive, the first-order KKT stationarity
condition for the optimal solution $\tilde{\mathbf{r}}'$ is
\begin{equation}
  \frac{\gamma_\ell \sqrt{1 - 2^{-2\tilde{r}_\ell}}}{2\sqrt{1 - 2^{-2\tilde{r}'_\ell}}}
  \cdot \left( 2\ln 2 \cdot 2^{-2\tilde{r}'_\ell} \right) - \lambda_1^* = 0,
\end{equation}
for $\ell\in\mathcal A$, which implies $\lambda_1^*>0$.
Therefore, the optimal solution can be found as
\begin{equation}
  \tilde{r}'_\ell = \frac{1}{2} \log
  \frac{1 + \sqrt{1 + \beta \gamma_\ell^2(1 - 2^{-2\tilde{r}_\ell})}}{2},
\end{equation}
for $\ell = 1, 2, \dots, L$, where
$\beta \triangleq \big[ (2\ln 2)/\lambda_1^* \big]^2 \in \mathbb{R}_{+}$.
The same expression also gives $\tilde r'_\ell=0$ for $\ell\notin\mathcal A$.
By the complementary slackness condition~\cite{boyd2004convex} and
$\lambda_1^*>0$, the sum-rate constraint is satisfied with equality, i.e.,
\begin{equation}
  \sum_{\ell=1}^L \tilde{r}'_\ell = R + C.
\end{equation}
The left-hand side is continuous and strictly increasing from $0$ to
$\infty$ as $\beta$ increases from $0$ to $\infty$.
Thus, this equation determines a unique $\beta>0$, yielding precisely
the allocation in Def.~\ref{def:r_prime}. This completes the proof.

\section{Proof of Corollary~\ref{cor:properties_greedy}}\label{app:properties_greedy}

We divide the proof into two cases based on the state of the components
(inactive or active).

\subsubsection*{1) Inactive components}
According to the Def.~\ref{def:r_tilde}, for the inactive components where
$\gamma_\ell \le \alpha$, we have $\log^+ \left( \frac{\gamma_\ell}{\alpha} \right) = 0$.
Thus, $\tilde{r}_\ell = 0$ for these inactive components.
Substituting $\tilde{r}_\ell = 0$ into the expression for $\tilde{r}'_\ell$
in Eq.~\eqref{eq:def_r_prime}, we have
\begin{equation}
  \tilde{r}'_\ell = \frac{1}{2} \log \frac{1 + \sqrt{1 + \beta \gamma_\ell^2(1 - 2^{0})}}{2} = 0.
\end{equation}
Therefore, for inactive components, $\tilde{r}_\ell = \tilde{r}'_\ell = 0$, and
the inequality holds with equality.

\subsubsection*{2) Active components}
For active components, there exists $\tilde{r}_\ell > 0$, thus
$2^{2\tilde{r}_\ell} = \gamma_\ell/\alpha$.
Substitute this equality into Eq.~\eqref{eq:def_r_prime}, we can
express $\tilde{r}'_\ell$ as a function of $\beta$, i.e.,
\begin{equation}
  \tilde{r}'_\ell(\beta) = \frac{1}{2} \log
  \frac{1 + \sqrt{1 + \beta \gamma_\ell (\gamma_\ell - \alpha)}}{2}.
\end{equation}
Note that when $\beta'=\frac{4}{\alpha^2}$, there is
\begin{equation}
  \sum_{\ell=1}^L \tilde{r}'_\ell (\beta) = R+C\ge R
  = \sum_{\ell=1}^L \tilde{r}_\ell \overset{(a)}{=} \sum_{\ell=1}^L \tilde{r}_\ell'(\beta'),
\end{equation}
where step $(a)$ comes from the fact that
\begin{align}
  \tilde{r}'_\ell (\beta') & = \frac{1}{2} \log
  \frac{1 + \sqrt{1 + \frac{4}{\alpha^2} \gamma_\ell (\gamma_\ell - \alpha)}}{2} = \tilde{r}_\ell.
\end{align}
Since $\tilde{r}'_\ell(\beta)$ is a strictly monotonically increasing function
w.r.t. $\beta$, there must be $\beta \ge \beta' = \frac{4}{\alpha^2}$,
which ensures that
\begin{equation} \label{eq:inequality_target}
  \frac{1 + \sqrt{1 + \beta \gamma_\ell (\gamma_\ell - \alpha)}}{2} \ge \frac{\gamma_\ell}{\alpha}.
\end{equation}
Therefore, we have $\tilde{r}'_\ell \ge \tilde{r}_\ell$ for all active
components, which completes the proof.

\section{Proof of Theorem~\ref{thm:P_active_condition}}\label{app:P_active_condition}
To establish the desired conclusion, we first introduce an operational perception
threshold $\bar{P}$ as the breakpoint of the continuous function $D(R,C,P)$.
Then we proceed to prove that $D(R,C,P)=\underline{D}(R)$ when $P\geq \bar{P}$,
and $D(R,C,P)=\overline{D}(R,C,P)$ when $ P < \bar{P}$. Finally, we prove that
$\bar{P} = \tilde{P}$ exactly, and hence the whole proof of
Thm.~\ref{thm:P_active_condition} is complete.

\subsection{Introducing the Operational Threshold \texorpdfstring{$\bar{P}$}{Pbar}}
Since the perception penalty term $[(\cdot)_+]^2$ is inherently non-negative, relaxing
this penalty yields a trivial lower bound:
\begin{equation}
  D(R, C, P) \ge \underline{D}(R).
\end{equation}
Hence, we define $\bar{P}$ as the minimum perceptual tolerance required to achieve this lower bound, i.e.,
\begin{equation}\label{eq:property_of_P_tilde}
  \bar{P} \triangleq \min \left\{ P \ge 0 : D(R,C,P)=\underline{D}(R) \right\}.
\end{equation}

Note that $\bar{P}$ is well-defined and finite. Since $\psi(r_\ell,r_\ell')$ is
a non-negative function, we have
\begin{equation}
  \sum_{\ell=1}^L \gamma_\ell \left(2-2^{-2r_\ell}-2\psi(r_\ell,r_\ell') \right)
  \le \sum_{\ell=1}^L 2 \gamma_\ell.
\end{equation}
Thus there must exist a sufficiently large threshold $P' \le \sum_{\ell=1}^L 2 \gamma_\ell$
such that
\begin{equation}
  \sum_{\ell=1}^L \gamma_\ell \left(2-2^{-2\hat{r}_\ell}-2\psi(\hat{r}_\ell,\hat{r}'_\ell) \right)\le P',
\end{equation}
then the perception penalty term $[(\cdot)_+]^2=0$, and the DRCP function
$D(R,C,P')$ achieves the lower bound $\underline{D}(R)$.
Even in the extreme case that the rate budget $R$ is sufficiently small such
that $\tilde{r}_\ell \to 0$ for all $\ell=1,2,\ldots,L$ and
$\underline{D}(R) \to \sum_{\ell=1}^L \gamma_\ell$,
we can still find that $\bar{P} \le \sum_{\ell=1}^L \gamma_\ell$, which is finite.

\subsection{Discussing \texorpdfstring{$D(R,C,P)$}{D(R,C,P)} on either side of \texorpdfstring{$\bar{P}$}{Pbar}}
For $P\ge\bar P$, we have
\begin{equation}
  \underline D(R) \overset{(a)}{\leq} D(R,C,P)
  \overset{(b)}{\leq} D(R,C,\bar P) \overset{(c)}{=} \underline D(R),
\end{equation}
where step $(a)$ is due to $\underline{D}(R)$ is a fundamental lower bound;
step $(b)$ comes from the non-increasing property of $D(R,C,P)$ w.r.t. $P$;
and step $(c)$ is based on Eq.~\eqref{eq:property_of_P_tilde}.
Thus, we can conclude that $D(R,C,P)=\underline D(R)$ for $P\ge\bar P$.

For $P<\bar P$, we proceed to prove $D(R,C,P) = \overline{D}(R,C,P)$ by contradiction.
Suppose there exists a specific perception limit $P' \in \Ico{0}{\bar{P}}$ such that
$D(R,C,P') \neq \overline{D}(R,C,P')$, i.e., for its optimal solution
$(\hat{\mathbf{r}},\hat{\mathbf{r}}')$, we have
\begin{equation}
  \sum_{\ell=1}^L \gamma_\ell
  \left( 2 - 2^{-2\hat{r}_\ell} - 2\psi(\hat{r}_\ell, \hat{r}'_\ell) \right) < P'.
\end{equation}
By continuity, the perception penalty term $(\cdot)_+$ would remain zero under sufficiently
small feasible perturbations of $\hat{\mathbf r}$ with $\hat{\mathbf r}'$ fixed, which
means $\hat{\mathbf r}$ would minimize the classical RD objective function~\eqref{eq:opt1}
under the same constraint~\eqref{eq:opt1_constraint}. Thus $\hat{\mathbf r}=\tilde{\mathbf r}$.
This implies that $D(R,C,P)=\underline D(R)$, which contradicts the definition
of $\bar{P}$ in Eq.~\eqref{eq:property_of_P_tilde}.
Therefore, $D(R,C,P)=\overline D(R,C,P)$ for any $P<\bar P$.

\subsection{Equivalence of \texorpdfstring{$\bar{P}$}{Pbar} and \texorpdfstring{$\tilde{P}$}{Ptilde}}

In this part, we prove $\bar{P}=\tilde{P}$ by solving the optimization
problem in Eq.~\eqref{eq:property_of_P_tilde} that defines $\bar{P}$. Specifically, we denote
$\bar{\mathbf{r}}=(\bar{r}_1, \dots,\bar{r}_L)$ and $\bar{\mathbf{r}}'=(\bar{r}'_1, \dots,\bar{r}'_L)$
as the optimal rate tuple which achieves $D(R,C,\bar{P})$, then we will show that
$(\bar{\mathbf{r}},\bar{\mathbf{r}}')$ is equal to $(\tilde{\mathbf{r}},\tilde{\mathbf{r}}')$ exactly.

Eq.~\eqref{eq:property_of_P_tilde} implies that for any $P' \ge \bar{P}$, there exists a set of optimal
solutions $(\hat{\mathbf{r}},\hat{\mathbf{r}}')$ such that $D(R,C,P')=\underline{D}(R)$, i.e.,
\begin{align}
  \sum_{\ell=1}^{L} \gamma_\ell 2^{-2 \hat{r}_\ell}
  = \sum_{\ell=1}^{L} \gamma_\ell 2^{-2 \tilde{r}_\ell},\label{eq:P_tilde_condition_1}
\end{align}
and
\begin{align}\label{eq:P_tilde_condition_2}
  \sqrt{\sum_{\ell=1}^L W_\ell(\hat{r}_\ell,\hat{r}_\ell')} - \sqrt{P'} \le 0.
\end{align}
Eq.~\eqref{eq:P_tilde_condition_1} ensures that $\hat{\mathbf{r}} = \tilde{\mathbf{r}}$ for
any $P'\ge \bar{P}$. Then for the second condition~\eqref{eq:P_tilde_condition_2},
the \emph{existence} of a solution implies
\begin{equation}\label{eq:def_P_bar}
  \bar{P} = \min P' = \min_{\mathbf{r}'\in \mathbb{R}_{+}^L,\sum r_\ell' \leq R + C}
  \sum_{\ell=1}^L W_\ell(\tilde{r}_\ell,r_\ell').
\end{equation}
Hence, $\bar{\mathbf{r}}$ and $\bar{\mathbf{r}}'$ must satisfy that
$\bar{\mathbf{r}} = \hat{\mathbf{r}} = \tilde{\mathbf{r}}$ and
\begin{align}
  \bar{\mathbf{r}}' = \arg \min_{\mathbf{r}' \in \mathbb{R}_+^L}
   & \sum_{\ell=1}^L W_\ell(\tilde{r}_\ell,r_\ell'),\label{eq:P_hat_condition} \\
  \text{s.t.} \enspace
   & \sum_{\ell=1}^L r'_\ell = R + C.
\end{align}
Note that minimizing the objective function in Eq.~\eqref{eq:P_hat_condition} is
equivalent to maximizing the problem given in Lem.~\ref{lem:optimal_of_r_prime}, so
we have $\bar{\mathbf{r}}' = \tilde{\mathbf{r}}'$.
Taking $\bar{\mathbf{r}} = \tilde{\mathbf{r}}$ and $\bar{\mathbf{r}}' = \tilde{\mathbf{r}}'$
into Eq.~\eqref{eq:def_P_bar} yields $\bar{P} = \tilde{P}$, which completes the proof.

\section{Proof of Lemma~\ref{lem:problem_equivalence}}\label{app:problem_equivalence}
This lemma can be verified by equivalent reformulation of the optimization problem
defined in Def.~\ref{def:problem_re3}.
Recalling the $W_\ell(r_\ell,r_\ell')$
defined in Lem.~\ref{lem:D_reformulation}, we define two feasible sets
\begin{align}
  \mathcal{A}_1 & \triangleq \left\{(\mathbf{r},\! \mathbf{r}') \!\in\! \mathbb{R}_+^{L} \!\times\!
  \mathbb{R}_+^{L} \!:\! \sum_{\ell=1}^L r_\ell \!\leq\! R, \sum_{\ell=1}^L r_\ell' \!\leq\! R \!+\! C \right\}\!, \\
  \mathcal{A}_2 & \triangleq \left\{T \geq \max
  \bigg(P, \sum_{\ell=1}^L W_\ell(r_\ell,r_\ell')\bigg)\right\}.
\end{align}
Then, we have
\begin{align}
   & \min_{T\ge P} D(R,C,P,T) \nonumber                                                                            \\
   & \overset{(a)}{=} \min_{(\mathbf r,\mathbf r') \in \mathcal{A}_1}
  \Bigg\{\sum_{\ell=1}^L \gamma_\ell 2^{-2r_\ell} +
  \min_{T\in \mathcal{A}_2}\bigl(\sqrt T-\sqrt P\bigr)^2\Bigg\}\nonumber                                           \\
   & \overset{(b)}{=} \min_{(\mathbf r,\mathbf r') \in \mathcal{A}_1}
  \Bigg\{\sum_{\ell=1}^L \gamma_\ell 2^{-2r_\ell} + \nonumber                                                      \\
   & \qquad \Bigg[\sqrt{\max \bigg(P,\sum_{\ell=1}^L W_\ell(r_\ell,r_\ell')\bigg)}-\sqrt P\Bigg]^2\Bigg\}\nonumber \\
   & = \min_{(\mathbf r,\mathbf r') \in \mathcal{A}_1}
  \Bigg\{\sum_{\ell=1}^L \gamma_\ell 2^{-2r_\ell} +
  \Bigg(\sqrt{\sum_{\ell=1}^L W_\ell(r_\ell,r_\ell')}-\sqrt P\Bigg)_+^2\Bigg\}\nonumber                            \\
   & \overset{(c)}{=} D(R,C,P), \label{eq:problem_equivalence_direct}
\end{align}
where step~$(a)$ follows from Eqs.~\eqref{eq:problem_re3_1}-\eqref{eq:problem_re3_4}
by the substitution $d_\ell=2^{-2r_\ell}$, $d_\ell'=2^{-2r_\ell'}$ and reordering the
joint minimization; step~$(b)$ uses the fact that $(\sqrt T-\sqrt P)^2$ is non-decreasing
for $T\ge P$; and step~$(c)$ is precisely Eqs.~\eqref{eq:prob5_1}-\eqref{eq:prob5_3}.
The identity holds for every $P\ge0$, which completes the proof.

\section{Proof of Theorem~\ref{thm:optimal_solution}}\label{app:optimal_solution}
In this proof, we focus on the active perception regime $P \in \Ico{0}{\tilde P}$,
with $R>0$ and $0\le C<\infty$.
According to Lem.~\ref{lem:problem_equivalence}, we can obtain
$D(R,C,P)$ by minimizing $D(R,C,P,T)$ over all $T\ge P$.
Meanwhile, as shown in Lem.~\ref{lem:convexity_of_fixed_T}, $D(R,C,P,T)$ is a convex
optimization problem, so we can obtain the $D(R,C,P,T)$ by analyzing its KKT conditions.
Specifically, the DRCP function can be reformulated as
\begin{align}
  D(R,C
    & ,P) =\min_{T \ge P} D(R,C,P,T) \nonumber                                                               \\
  = & \min_{\mathbf{d},\mathbf{d'}\in {\Ioc{0}{1}}^L, T \ge P}\sum_{\ell=1}^L \gamma_\ell {d_\ell}
  + {\Big(\sqrt{T}-\sqrt{P} \Big)}^2, \label{eq:prob_re1}                                                    \\
  \text{s.t.} \enspace
    & \sum_{\ell=1}^{L}\gamma_\ell\Big(2-d_\ell-2\sqrt{(1-d_\ell)(1-d_\ell')}\Big)\leq T,\label{eq:prob_re2} \\
    & \sum_{\ell=1}^L -\frac{1}{2}\log d_\ell \leq R, \label{eq:prob_re3}                                    \\
    & \sum_{\ell=1}^L -\frac{1}{2}\log d_\ell' \leq R+C.\label{eq:prob_re4}
\end{align}
The associated Lagrangian function can be constructed by introducing the
non-negative multipliers $\lambda_2,\lambda_3$, and $\lambda_4$ as
\begin{align}
   & \mathcal{L}(\mathbf{d},\mathbf{d'},T,\lambda_2,\lambda_3,\lambda_4) =
  \sum_{\ell=1}^L \gamma_\ell d_\ell + {\Big(\sqrt{T}-\sqrt{P} \Big)}^2 \nonumber                   \\
   & \quad +\lambda_2 \left( \sum_{\ell=1}^L -\frac{1}{2}\log d_\ell \!-\! R \right)
  \!+\! \lambda_3 \left( \sum_{\ell=1}^L -\frac{1}{2}\log d_\ell' \!-\! R \!-\! C \right) \nonumber \\
   & \quad +\lambda_4 \left[\sum_{\ell=1}^{L}\gamma_\ell
    \Big(2 \!-\! d_\ell \!-\! 2\sqrt{(1-d_\ell)(1-d_\ell')}\Big) \!-\! T \right]\!,
  \label{eq:Lagrangian_RDCP}
\end{align}
where the domain bounds $0<d_\ell,d_\ell'\le1$ and $T\ge P$
are retained when minimizing $\mathcal L$.
Let $(\hat{\mathbf d},\hat{\mathbf d}',\hat T)$ be an optimal solution,
and denote the normalized optimal multipliers as
$\nu=\lambda_2^*/(2\ln2)$, $\nu'=\lambda_3^*/(2\ln2)$ and $\mu=\lambda_4^*$.
The following lemma establishes the properties of the optimal solution and
associated multipliers.

\begin{lemma}\label{lem:optimal_solution_properties}
  \renewcommand{\labelenumi}{(\theenumi)}
  For $R>0$, $C\geq 0$, and $0\le P<\tilde P$, the following properties hold
  for the problem~\eqref{eq:prob_re1}-\eqref{eq:prob_re4}:
  \begin{enumerate}
    \item\label{item:optimal_solution_boundary}
          Each component pair is either interior, i.e., $0<\hat d_\ell,\hat d_\ell'<1$,
          or equals $(1,1)$, and at least one pair is interior.
    \item\label{item:optimal_solution_multipliers}
          The normalized optimal multiplier triple $(\mu,\nu,\nu')$
          exists and is unique, with $\nu,\nu'>0$ and $0<\mu\le1$.
    \item\label{item:optimal_solution_active}
          All three constraints~\eqref{eq:prob_re2}-\eqref{eq:prob_re4} hold with equality
          at the optimal solution.
    \item\label{item:optimal_solution_root}
          The corner condition for $\hat d_\ell=\hat d_\ell'=1$ is given by
          $\gamma_\ell \leq \varpi$, where $\varpi$ is defined in Eq.~\eqref{eq:varpi_def}.
          For each $\gamma_\ell>\varpi$, Eq.~\eqref{eq:w_l} has
          exactly one root $\hat w_\ell \in (0,\mu^2\gamma_\ell/\nu')$.
  \end{enumerate}
\end{lemma}
\begin{IEEEproof}
  See Appendix~\ref{app:existence_uniqueness_w}.
\end{IEEEproof}

Then the KKT stationarity conditions for the interior component pairs are given by
\begin{align}
   & \frac{\sqrt{\hat{T}}-\sqrt{P}}{\sqrt{\hat{T}}} - \mu = 0,\label{eq:partial_3_1}   \\
   & \gamma_\ell (1-\mu) + \mu\gamma_\ell\sqrt{\frac{1-\hat{d}_\ell'}{1-\hat{d}_\ell}}
  - \frac{\nu}{\hat{d}_\ell} = 0,\label{eq:partial_3_2}                                \\
   & \mu \gamma_\ell \sqrt{\frac{1-\hat{d}_\ell}{1-\hat{d}_\ell'}}
  - \frac{\nu'}{\hat{d}_\ell'} = 0.\label{eq:partial_3_3}
\end{align}
From Eq.~\eqref{eq:partial_3_1}, the optimal total perception penalty
$\hat T$ for $P>0$ can be expressed as
\begin{equation}
  \hat T=\frac{P}{(1-\mu)^2}.
\end{equation}
For each interior component, substituting Eq.~\eqref{eq:partial_3_3} into
Eq.~\eqref{eq:partial_3_2} to eliminate $\hat d_\ell$ yields
\begin{align}\label{eq:d_l_prime_3}
   & \mu^4 \gamma_\ell^3 (\hat d_\ell')^3 + \mu^2 \gamma_\ell \nu'
  \big[\gamma_\ell(1-\mu)-\nu+\nu'\big](\hat d_\ell')^2 \nonumber    \\
   & \quad +\nu'^2\big[(1-\mu)\nu'-\mu^2\gamma_\ell\big]\hat d_\ell'
  -(1-\mu)\nu'^3=0.
\end{align}

By Lem.~\ref{lem:optimal_solution_properties}\eqref{item:optimal_solution_root},
$\varpi$ can be set as the critical cutoff threshold of the extended reverse waterfilling
strategy: both rates $\hat{r}_\ell$ and $\hat{r}_\ell'$ are zero for the $\ell$-th
component when $\gamma_\ell\le\varpi$.
For each component with $\gamma_\ell > \varpi$, the substitution
\begin{equation}\label{eq:d_l_prime_opt}
  \hat d_\ell'=\frac{\nu'\hat w_\ell}{\mu^2\gamma_\ell}
\end{equation}
transforms Eq.~\eqref{eq:d_l_prime_3} into Eq.~\eqref{eq:w_l}, with
$0<\hat w_\ell<\mu^2\gamma_\ell/\nu'$.
Substituting Eq.~\eqref{eq:d_l_prime_opt} back into
Eq.~\eqref{eq:partial_3_3} yields
\begin{equation}
  \hat d_\ell=\frac{\gamma_\ell\hat w_\ell^2+\nu'\hat w_\ell
    -\mu^2\gamma_\ell}{\gamma_\ell\hat w_\ell^2}.
\end{equation}
For $\gamma_\ell\le\varpi$, setting $\hat w_\ell=\mu^2\gamma_\ell/\nu'$ as in Def.~\ref{def:w_l}
recovers $\hat d_\ell=\hat d_\ell'=1$ in the same formulas. Converting $\hat d_\ell$ and
$\hat d_\ell'$ back to rate domain gives Eqs.~\eqref{eq:theorem2_r}-\eqref{eq:theorem2_r_prime}.
Finally, substituting these expressions into the active
constraints in Lem.~\ref{lem:optimal_solution_properties}\eqref{item:optimal_solution_active}
and using Eq.~\eqref{eq:partial_3_1} yields Eqs.~\eqref{eq:lag_para_opt_1}-\eqref{eq:lag_para_opt_3}.
This completes the proof.

\section{Proof of Corollary~\ref{cor:properties_opt}}\label{app:properties}
Based on the KKT conditions derived in Eqs.~\eqref{eq:partial_3_1}-\eqref{eq:partial_3_3},
we prove this corollary by showing that all nonzero differences
$\hat{r}_\ell' - \hat{r}_\ell$ have the same sign for $\ell=1,2,\ldots,L$.

Recall from Eqs.~\eqref{eq:theorem2_r} and~\eqref{eq:theorem2_r_prime} that
$\hat{\mathbf{r}}$ and $\hat{\mathbf{r}}'$ have the same active thresholds $\varpi$.
Hence, if $\hat{r}_{\ell}=0$, then $\hat{r}_{\ell}'=0$, and the desired inequality
holds immediately. It therefore suffices to consider the active component satisfying
$\hat{r}_{\ell}>0$ and $\hat{r}_{\ell}'>0$.

For each active component, the KKT stationarity conditions in Eqs.~\eqref{eq:partial_3_2} and~\eqref{eq:partial_3_3} give
\begin{align}
  \gamma_{\ell}\hat{d}_{\ell}
  \left[ 1-\mu +\mu\sqrt{\frac{1-\hat{d}_{\ell}'}{1-\hat{d}_{\ell}}} \right]
   & = \nu, \label{eq:KKT_conj1_1}  \\
  \gamma_{\ell}\hat{d}_{\ell}'\mu \sqrt{\frac{1-\hat{d}_{\ell}}{1-\hat{d}_{\ell}'}}
   & = \nu'. \label{eq:KKT_conj1_2}
\end{align}
Define $z_{\ell}\triangleq \sqrt{\frac{1-\hat{d}_{\ell}'}{1-\hat{d}_{\ell}}}$, then
we have
\begin{equation}\label{eq:d_prime_z_direct}
  \hat{d}_{\ell}' = 1-z_{\ell}^{2}(1-\hat{d}_{\ell}).
\end{equation}
Dividing Eq.~\eqref{eq:KKT_conj1_1} by Eq.~\eqref{eq:KKT_conj1_2} yields
\begin{equation}\label{eq:multiplier_ratio_z}
  \frac{\nu}{\nu'} = \frac{\hat{d}_{\ell}}{\hat{d}_{\ell}'}
  \left( \frac{1-\mu}{\mu}z_{\ell}+z_{\ell}^{2} \right).
\end{equation}
Using Eqs.~\eqref{eq:multiplier_ratio_z} and \eqref{eq:d_prime_z_direct}, we have
\begin{align}
  \frac{\nu}{\nu'}-\frac{1}{\mu}
   & \overset{(a)}{=} \left( \frac{\hat{d}_{\ell}}{\hat{d}_{\ell}'}-1 \right)
  \left( \frac{1-\mu}{\mu}z_{\ell}+z_{\ell}^{2} \right) \nonumber                                \\
   & \quad + \frac{1-\mu}{\mu}(z_{\ell}-1)+z_{\ell}^{2}-1 \nonumber                              \\
   & \overset{(b)}{=} (z_{\ell}-1) \Bigg[ \frac{(1-\hat{d}_{\ell})(z_{\ell}+1)}{\hat{d}_{\ell}'}
  \left( \frac{1-\mu}{\mu}z_{\ell} +z_{\ell}^{2} \right) \nonumber                               \\
   & \quad + \frac{1-\mu}{\mu} +z_{\ell}+1 \Bigg],
  \label{eq:direct_common_sign}
\end{align}
where step $(a)$ follows from Eq.~\eqref{eq:multiplier_ratio_z} and
$\frac{1}{\mu}=\frac{1-\mu}{\mu}+1$; and step $(b)$ follows from
Eq.~\eqref{eq:d_prime_z_direct} and $z_{\ell}^{2}-1=(z_{\ell}-1)(z_{\ell}+1)$.
Since $0<\mu\leq1$, $\hat{d}_{\ell},\hat{d}_{\ell}'\in(0,1)$, and $z_{\ell}>0$,
the bracketed term $[\cdot]$ in Eq.~\eqref{eq:direct_common_sign} is strictly positive.
Therefore,
\begin{equation}\label{eq:rate_gap_common_sign}
  \operatorname{sgn}(z_{\ell}-1) = \operatorname{sgn}
  \left( \frac{\nu}{\nu'}-\frac{1}{\mu} \right).
\end{equation}
Because the right-hand side of Eq.~\eqref{eq:rate_gap_common_sign} is
independent of $\ell$, all nonzero differences $z_{\ell}-1$ have the
same sign, which implies that all nonzero differences
$\hat{r}_{\ell}'-\hat{r}_{\ell}$ have the same sign.
Meanwhile, the active sum-rate constraints yield
\begin{align}\label{eq:total_rate_gap}
  \sum_{\ell=1}^{L} \big(\hat{r}_{\ell}'-\hat{r}_{\ell}\big) = (R+C)-R = C\geq0.
\end{align}
Hence, the common sign of $\hat{r}_{\ell}'-\hat{r}_{\ell}$ cannot be negative.
Together with the zero-rate components, we have $\hat{r}_{\ell}\leq\hat{r}_{\ell}'$
for all $\ell=1,2,\ldots,L$. This completes the proof.

\section{Proof of Corollary~\ref{cor:optimal_solution_perfect_P}}\label{app:optimal_solution_perfect_P}
We use the method described in the proof of Thm.~\ref{thm:optimal_solution} for this
special case. When $P=0$, we have $\hat T>0$
and the KKT condition~\eqref{eq:partial_3_1} becomes
\begin{equation}
  \frac{\sqrt{\hat{T}}-\sqrt{0}}{\sqrt{\hat{T}}} - \mu_0 = 0,
\end{equation}
which implies that $\mu_0=1$. For each active component with $0<\hat d_\ell,\hat d_\ell'<1$,
substituting this value into Eq.~\eqref{eq:d_l_prime_3} and
Eq.~\eqref{eq:partial_3_2}-\eqref{eq:partial_3_3} yields
\begin{equation}\label{eq:degenerate_equation}
  \nu_1' {\left(\frac{1 \!-\! \hat{d}_\ell'}{\hat{d}_\ell'}\right)}^2 + (\nu_1 \!+\! \nu_1')
  \left(\frac{1 \!-\! \hat{d}_\ell'}{\hat{d}_\ell'}\right)
  + \left(\nu_1 \!-\! \frac{\gamma_\ell^2}{\nu_1'}\right) = 0,
\end{equation}
and
\begin{equation}\label{eq:mult1}
  \gamma_\ell^2 \hat{d}_\ell \hat{d}_\ell' = \nu_1 \nu_1',
\end{equation}
respectively. As established in Appendix~\ref{app:optimal_solution}, solving
Eq.~\eqref{eq:degenerate_equation} on the active components and setting
$\hat d_\ell=\hat d_\ell'=1$ on the remaining components gives the expression
for $(\hat{\mathbf r},\hat{\mathbf r}')$ in
Eqs.~\eqref{eq:opt_perfectP_r}-\eqref{eq:opt_perfectP_r_prime}, where $\nu_1,\nu_1'>0$
are the unique multipliers satisfying Eqs.~\eqref{eq:lam1}-\eqref{eq:lam2}.
Taking $\mu_0=1$ in Eq.~\eqref{eq:varpi_def} gives the cutoff threshold
$\varpi_0=\sqrt{\nu_1\nu_1'}$ when $P=0$, thus
\begin{align}
  R + (R+C)
   & =\sum_{\ell'=1}^L\bigl(\hat r_{\ell'}+\hat r_{\ell'}'\bigr)\nonumber   \\
   & =-\frac12\sum_{\ell'=1}^L
  \log\bigl(\hat d_{\ell'}\hat d_{\ell'}'\bigr)\nonumber                    \\
   & =\sum_{\ell'=1}^L\log^+\!\left(\frac{\gamma_{\ell'}}{\varpi_0}\right),
\end{align}
where the last equality uses Eq.~\eqref{eq:mult1} on active components
and $\hat d_{\ell'}\hat d_{\ell'}'=1$ on the others.
This proves~\eqref{eq:varpi0_condition}. Its right-hand side is strictly
decreasing in $\varpi_0$ on $(0,\max_{\ell'}\gamma_{\ell'})$, with
limits $+\infty$ and $0$ at the respective endpoints. Therefore,
the cutoff threshold $\varpi_0$ is uniquely determined, and
$\hat r_\ell, \hat{r}_\ell'>0$ if and only if
$2R+C>\sum_{\ell'=1}^L\log^+\!\big(\frac{\gamma_{\ell'}}{\gamma_\ell}\big)$.
This completes the proof.

\section{Proof of Theorem~\ref{thm:gap_upper_bound}}\label{app:gap_upper_bound}
Firstly, we will prove that the supremum of $\Delta D(R,C,P)$ will be taken
at $C \to \infty$ and $P = 0$ by monotonicity. Subsequently, we let
$U(R,\Gamma) = \Delta D(R,\infty,0)$, which naturally serves as an analytical
upper bound on the $\Delta D(R,C,P)$.

The monotonicity of $\Delta D(R,C,P)$ w.r.t. $P$ and $C$ can be
shown as the following two lemmas.
\begin{lemma}[Monotonicity of $\Delta D(R,C,P)$ in $P$]\label{lem:monotonicity_P}
  For any given rate budget $R \geq 0$ and fixed CR $C \geq 0$,
  the gap function $\Delta D(R,C,P)$ is monotonically non-increasing w.r.t.
  the perception limit $P$.
\end{lemma}
\begin{IEEEproof}
  See Appendix~\ref{app:monotonicity_P}.
\end{IEEEproof}

\begin{lemma}[Monotonicity of $\Delta D(R,C,0)$ in $C$]\label{lem:monotonicity_C}
  For any given rate budget $R \geq 0$ and perfect perception quality $P = 0$, the
  gap function $\Delta D(R,C,0)$ is monotonically non-decreasing w.r.t.
  the CR $C$.
\end{lemma}
\begin{IEEEproof}
  See Appendix~\ref{app:monotonicity_C}.
\end{IEEEproof}

Lem.~\ref{lem:monotonicity_P} and Lem.~\ref{lem:monotonicity_C} ensure that
\begin{equation}
  \Delta D(R,C,P) \leq \Delta D(R,C,0) \leq \Delta D(R,\infty,0).
\end{equation}
Thus, we have
\begin{equation}
  \sup_{C,P} \Delta D(R,C,P) = \Delta D(R,\infty,0).
\end{equation}
This equation indicates that the supremum of $\Delta D(R,C,P)$
is always achieved at $\Delta D(R,\infty,0)$, thus we can let $U(R, \Gamma)=\Delta D(R,\infty,0)$
as an asymptotically tight upper bound of $\Delta D(R,C,P)$, which is a function
of the rate budget $R$ alone for a given source $\Gamma$.

According to Cor.~\ref{cor:opt_perfectP_infC}, the optimal
distortion $D(R, \infty, 0)$ is given by
\begin{equation}\label{eq:D_Cinf_P0}
  D(R, \infty, 0) = 2 \sum_{\ell=1}^{L} \left( \gamma_{\ell}
  - \sqrt{\nu_2^2 + \gamma_{\ell}^{2}} + \nu_2 \right),
\end{equation}
where $\nu_2$ is a unique positive scalar satisfying
\begin{equation}
  \sum_{\ell=1}^{L} \log
  \left( 1 + \sqrt{1 + \frac{\gamma_{\ell}^{2}}{\nu_2^2}} \right) = 2R + L.
\end{equation}
Meanwhile, the greedy distortion $\tilde{D}(R, \infty, 0)$ is given by
\begin{equation}\label{eq:D_tilde_Cinf_P0}
  \tilde{D}(R,\infty,0) = 2 \sum_{\ell=1}^{L}
  \left( \gamma_{\ell} - \sqrt{\gamma_{\ell}(\gamma_{\ell}-\alpha)_{+}} \right).
\end{equation}
Subtracting Eq.~\eqref{eq:D_Cinf_P0} from Eq.~\eqref{eq:D_tilde_Cinf_P0}
directly yields the expression of the upper bound $U(R, \Gamma)$ given
in Eq.~\eqref{eq:U_R_definition}, which completes the proof.

\section{Proof of Corollary~\ref{cor:property_of_U}}\label{app:property_of_U}
The key to this proof lies in showing that for any given $\Gamma \succ 0$, $U(R, \Gamma)$
is a piecewise function over $R$ and, unless all eigenvalues are equal, any stationary
point within a sub-interval divided by the breakpoints is a strict local minimum.
With this local minimum property, the global maximum of $U(R, \Gamma)$ can be
found by evaluating the function values at the finite set of breakpoints, which
are exactly $\{R_1, R_2, \dots, R_L\}$.

Firstly, we construct the local minimum property of $U(R, \Gamma)$ over the entire
feasible domain. Without loss of generality, assume the variance sequence is sorted
in descending order, i.e., $\gamma_1 \ge \gamma_2 \ge \dots \ge \gamma_L > 0$.
If $\gamma_1=\gamma_L$, Eqs.~\eqref{eq:alpha_condition},
\eqref{eq:opt_D_multiplier}, and~\eqref{eq:U_R_definition} give
$U(R,\Gamma)\equiv 0$, and the conclusion follows immediately.
Hence, assume $\gamma_1>\gamma_L$ and $R_{L+1} \triangleq +\infty$.
Then the continuous feasible domain of $R \in [0,\infty)$ can be decomposed as
\begin{equation}\label{eq:domain_divide}
  \Ico{0}{\infty} = \bigcup_{\ell=1}^{L} [R_{\ell}, R_{\ell+1}],
\end{equation}
where $R_{\ell} = \sum_{k=1}^{\ell} \frac{1}{2} \log \frac{\gamma_k}{\gamma_{\ell}}$
is the $\ell$-th breakpoint corresponding to $\alpha = \gamma_\ell$.
Now we establish the property of $U(R, \Gamma)$ within each sub-interval by the following lemma.
\begin{lemma}[Local Minimum Property of $U(R, \Gamma)$]\label{lem:sub_convexity_U}
  For any $k \in \{1, 2, \ldots, L\}$ and fixed $\Gamma \succ 0$, any stationary point
  of $U(R, \Gamma)$ in the open interval $R\in (R_k, R_{k+1})$ is a local minimum.
  This local minimum is strict unless all eigenvalues of $\Gamma$ are equal.
\end{lemma}
\begin{IEEEproof}
  See Appendix~\ref{app:sub_convexity_U}.
\end{IEEEproof}

To determine the global maximum of $U(R, \Gamma)$ over $R\geq 0$ for a given
$\Gamma \succ 0$, we can compute the maximum over all sub-intervals as follows:
\begin{align}
   & \max_{R \in \Ico{0}{\infty}} U(R, \Gamma) \nonumber                                                      \\
  \overset{(a)}{=}
   & \max_{\ell \in \{1,2,\ldots,L\}} \left(\max_{R \in [R_{\ell}, R_{\ell+1}]}U(R, \Gamma)\right), \nonumber \\
  \overset{(b)}{=}
   & \max_{\ell \in \{1,2,\ldots,L\}} \max \big\{ U(R_{\ell},\Gamma), U(R_{\ell+1},\Gamma) \big\}, \nonumber  \\
  \overset{(c)}{=}
   & \max_{\ell \in \{1,2, \ldots, L\}} U(R_{\ell},\Gamma), \label{eq:global_max_discrete}
\end{align}
where step $(a)$ comes from the decomposition of the feasible domain
given in Eq.~\eqref{eq:domain_divide}; step $(b)$ follows from continuity and the strict
local minimum property established by Lem.~\ref{lem:sub_convexity_U}, which excludes
any interior maximum on a nonempty open breakpoint interval; and step $(c)$ is due to
$U(R_1,\Gamma)=U(0,\Gamma)=0$ and $\lim_{R\to\infty}U(R,\Gamma)=0$.
Thus, the maximum is attained at one of the finite breakpoints.
By Thm~\ref{thm:gap_upper_bound}, for any $R \ge 0$, we have
\begin{equation}
  \Delta D(R,C,P) \le U(R,\Gamma) \le \max_{\ell\in\{1,2,\ldots,L\}} U(R_\ell,\Gamma).
\end{equation}
This completes the proof.

\section{Proof of Corollary~\ref{cor:universal_gap_bound}}\label{app:universal_gap_bound}

By Thm.~\ref{thm:gap_upper_bound}, the performance gap $\Delta D(R,C,P)$ satisfies
\begin{equation}\label{eq:cor9_gap_to_U}
  0 \leq \Delta D(R,C,P) \leq U(R,\Gamma).
\end{equation}
Hence, to determine the universal upper bound of the gap function $\Delta D(R,C,P)$,
it suffices to determine
\begin{align}
  \text{(P1)} \quad
  \rho^* \triangleq
   & \sup_{\substack{L\in\mathbb N,\,R\geq0,                                             \\ \Gamma\succ 0,\alpha>0,\nu_2>0}}
  \frac{U(R,\Gamma)}{2\operatorname{tr}(\Gamma)}, \label{eq:cor9_rho_def}                \\
   & \qquad \mathrm{s.t.} \quad \eqref{eq:alpha_condition}, \eqref{eq:opt_D_multiplier}.
\end{align}
Directly solving the problem (P1) is quite complicated due to its high-dimensional
non-convex structure and the arbitrary dimension $L$ of the covariance matrix $\Gamma$.
To address this, we first formulate an optimization problem with only two independent
variables and prove that its optimal value coincides with $\rho^*$. We then characterize
the optimal solution using the KKT conditions and rigorously establish its existence
and uniqueness, thereby obtaining the stated universal bound.

\subsection{Preliminary Definitions and Properties}
Firstly, we provide definitions of some auxiliary functions and variables that will be
used later, along with some properties they satisfy.

Let $\bm{\gamma} = (\gamma_1,\gamma_2,\ldots,\gamma_L)$ and $\alpha>0$ represent
the classical water level given in Def.~\ref{def:r_tilde}, then we can
obtain a normalized vector $\bar{\bm{\gamma}} = (\bar{\gamma}_1,\bar{\gamma}_2,\ldots,\bar{\gamma}_L)$
by dividing each element by the water level $\alpha$, i.e.,
$\bar{\gamma}_\ell \triangleq \frac{\gamma_\ell}{\alpha}$ for $\ell=1,2,\ldots,L$
and a normalized multiplier $\bar{\nu}_2 \triangleq \frac{\nu_2}{\alpha}$.
For $M\in\mathbb N$, let $\bm{w} \triangleq (w_1,\ldots,w_M)$ and
$\tilde{\bm{\gamma}}\triangleq (\tilde{\gamma}_1,\ldots,\tilde{\gamma}_M)$ satisfy
$w_i = \frac{N_i}{\sum_{\ell=1}^{L}\bar{\gamma}_\ell}$ and
$\{\tilde{\gamma}_i\}_{i=1}^{M}=\left\{\bar{\gamma}_\ell: \ell\in\{1,\ldots,L\}\right\}$,
respectively, where
$N_i \triangleq \left| \left\{\ell\in\{1,\ldots,L\}:\bar{\gamma}_\ell=\tilde{\gamma}_i \right\} \right|$
represent the multiplicity of each distinct value $\tilde{\gamma}_i$ and
$M \leq L$ is the number of distinct values in $\bar{\bm{\gamma}}$.
Define two auxiliary functions
\begin{align}
  \varphi_1(\nu,\gamma) & \triangleq \sqrt{\gamma^2+\nu^2}-\nu
  -\sqrt{\gamma(\gamma-1)_+}, \label{eq:cor9_phi_def}                                   \\
  \varphi_2(\nu,\gamma) & \triangleq \log\left(1+\sqrt{1+\frac{\gamma^2}{\nu^2}}\right)
  - 1- \log^+\gamma, \label{eq:cor9_psi_def}
\end{align}
for $\nu,\gamma>0$, and define four feasible sets
\begin{align}
  \mathcal{A}_s^{(L)}
   & \triangleq \left\{(\alpha,\nu_2,R,\bm{\gamma}):
  \eqref{eq:alpha_condition}, \eqref{eq:opt_D_multiplier} \right\},        \\
  \mathcal{A}_n^{(L)}
   & \triangleq \left\{(\bar{\nu}_2,\bar{\bm{\gamma}}):
  \sum_{\ell=1}^{L} \varphi_2(\bar{\nu}_2,\bar{\gamma}_\ell) = 0 \right\}, \\
  \mathcal A_m^{(M)}
   & \triangleq \Bigg\{(\bar{\nu}_2,\bm{w},\tilde{\bm{\gamma}}):
  \bar{\nu}_2 >0,w_i>0,\tilde{\gamma}_i>0,\nonumber                        \\
   & \qquad \sum_{i=1}^{M} w_i\tilde{\gamma}_i=1,
  \sum_{i=1}^{M} w_i\varphi_2(\bar{\nu}_2,\tilde{\gamma}_i)=0 \Bigg\},     \\
  \mathcal{D}
   & \triangleq \bigg\{(\bar{\nu}_2,\tilde{\gamma}_1,w_1,w_2):
  \tilde{\gamma}_1> \gamma_0(\bar{\nu}_2)\triangleq
  \frac{4\bar{\nu}_2^2}{4\bar{\nu}_2^2-1}, \nonumber                       \\
   & \qquad \bar{\nu}_2>\frac{1}{2}, 0 < w_1,w_2 < 1 \bigg\}.
\end{align}
Then we can give the following optimization problems
\begin{equation}\label{eq:cor9_rhobar_def}
  \text{(P2)} \quad
  \bar\rho \triangleq \sup_{(\bar{\nu}_2,\bm w,\tilde{\bm\gamma})\in\mathcal A_m^{(2)}}
  \sum_{i=1}^{2}w_i\varphi_1(\bar{\nu}_2,\tilde{\gamma}_i),
\end{equation}
and
\begin{align}
  \text{(P3)} \quad \rho' \triangleq
  \!\sup_{(\bar{\nu}_2\!,\tilde{\gamma}_1\!,w_1\!,w_2)\in\mathcal{D}}\!
   & w_1\varphi_1(\bar{\nu}_2,\tilde{\gamma}_1) \!+\! w_2\varphi_1(\bar{\nu}_2,1), \label{eq:rp_obj1}          \\
  \text{s.t.} \qquad\quad
   & \hspace{-2em} w_1 \varphi_2(\bar{\nu}_2,\tilde{\gamma}_1)+w_2 \varphi_2(\bar{\nu}_2,1)=0,\label{eq:rp_c1} \\
   & \hspace{-2em} w_1\tilde{\gamma}_1+w_2=1.\label{eq:rp_c2}
\end{align}
The problem (P2) can be viewed as a reformulation of problem (P1) that is restricted to source
covariance matrices with at most two distinct eigenvalues, while problem (P3) is a further
reduction of (P2) by fixing $\tilde{\gamma}_2=1$ and specifying the admissible ranges of
the remaining variables.
Now we give a lemma to state some properties that will be used later.
\begin{lemma}\label{lem:related_properties_cor9}
  \renewcommand{\labelenumi}{(\theenumi)} 
  For the functions and variables defined above, we have the following properties:
  \begin{enumerate}
    \item\label{item:cor9_1}
          The supremum of the following two problems coincides
          \begin{align}
             & \sup_{M\in\mathbb N} \sup_{(\bar{\nu}_2,\bm w,\tilde{\bm\gamma})\in\mathcal A_m^{(M)}}
            \sum_{i=1}^{M}w_i\varphi_1(\bar{\nu}_2,\tilde{\gamma}_i) \nonumber                          \\
             & \qquad \qquad \qquad = \sup_{(\bar{\nu}_2,\bm w,\tilde{\bm\gamma})\in\mathcal A_m^{(2)}}
            \sum_{i=1}^{2}w_i\varphi_1(\bar{\nu}_2,\tilde{\gamma}_i). \label{eq:cor9_extreme_reduction}
          \end{align}
    \item\label{item:cor9_2}
          The definitions of $\bar{\nu}_2$, $\bar{\bm{\gamma}}$, $\tilde{\bm{\gamma}}$
          $\bm{w}$ and $\bm{N}$ imply that
          \begin{align}
             & \sum_{\ell=1}^{L}\varphi_2(\bar{\nu}_2,\bar{\gamma}_\ell)=0,\label{eq:cor9_2_1}     \\
             & \frac{U(R,\Gamma)}{2\operatorname{tr}(\Gamma)} =
            \frac{\sum_{\ell=1}^{L} \varphi_1(\bar{\nu}_2,\bar{\gamma}_\ell)}
            {\sum_{\ell=1}^{L}\bar{\gamma}_\ell}, \label{eq:cor9_2_2}                              \\
             & \sum_{i=1}^{M} w_i \varphi_2(\bar{\nu}_2,\tilde{\gamma}_i) = 0, \label{eq:cor9_2_3} \\
             & \sum_{i=1}^{M} w_i\tilde{\gamma}_i = 1, \label{eq:cor9_2_4}                         \\
             & \sum_{\ell=1}^{L} \varphi_1(\bar{\nu}_2,\bar{\gamma}_\ell) =
            \sum_{i=1}^{M} w_i \varphi_1(\bar{\nu}_2,\tilde{\gamma}_i)
            \sum_{\ell=1}^{L}\bar{\gamma}_\ell. \label{eq:cor9_2_5}
          \end{align}
    \item\label{item:cor9_3}
          For $\nu>0$ and $\gamma>1$, the following properties hold:
          \begin{align}
             & \frac{\partial\varphi_2(\nu,\gamma)}{\partial\gamma}<0, \label{eq:cor9_3_1} \\
             & \frac{\partial\varphi_2(\nu,\gamma)}{\partial\nu} = \frac{1}{\nu\ln 2}
            \frac{\partial\varphi_{1}(\nu,\gamma)}{\partial\nu} < 0, \label{eq:cor9_3_2}   \\
             & \frac{\partial}{\partial \gamma} \left[\varphi(\nu,\gamma)-\gamma
              \frac{\partial \varphi(\nu,\gamma)}{\partial \gamma}\right]=-\gamma
            \frac{\partial^2 \varphi(\nu,\gamma)}{\partial \gamma^2}, \label{eq:cor9_3_3}  \\
             & \lim_{\gamma\to\infty} \varphi_1(\bar{\nu}_2,\gamma)
            =\frac{1}{2}-\bar{\nu}_2, \label{eq:cor9_3_4}                                  \\
             & \lim_{\gamma\to\infty}\varphi_2(\bar{\nu}_2,\gamma)
            = -\log(2\bar{\nu}_2). \label{eq:cor9_3_5}
          \end{align}
    \item\label{item:cor9_4}
          The supremum in problem (P3) can be attained at an interior point of $\mathcal{D}$.
  \end{enumerate}
\end{lemma}
\begin{IEEEproof}
  See Appendix~\ref{app:gap_pointwise_bound}.
\end{IEEEproof}

\subsection{\texorpdfstring{$\rho^*=\rho'$}{rho*=rho'}}
In this part, we first establish equality $\rho^*=\bar{\rho}$ by proving
the two inequalities $\rho^* \leq \bar{\rho}$ and $\rho^* \geq \bar{\rho}$, separately.
Then, we determine the optimal values or ranges for the variables in
problem (P2) to show $\bar{\rho}=\rho'$, which implies that $\rho^*=\rho'$.

\subsubsection{\texorpdfstring{$\rho^*\leq\bar{\rho}$}{rho*<=bar rho}}
We first apply a transformation of variables to represent the $L$-dimensional source
covariance matrix $\Gamma$ by a weighted spectrum formula $\tilde{\bm{\gamma}}$
with $M$ distinct eigenvalues, then use the equality constraints to eliminate
the dependent variables. Finally, we restrict the arbitrary dimension $M$ to $2$
without changing the supremum. Specifically,
\begin{align}
  \rho^*
   & = \sup_{L\in \mathbb{N},(\alpha,\nu_2,R,\bm{\gamma}) \in \mathcal{A}_s^{(L)}}
  \frac{U(R,\Gamma)}{2\operatorname{tr}(\Gamma)}, \nonumber                                                      \\
   & \overset{(a)}{\leq} \sup_{L\in \mathbb{N},(\bar{\nu}_2,\bar{\bm{\gamma}}) \in \mathcal{A}_n^{(L)}}
  \frac{\sum_{\ell=1}^{L} \varphi_1(\bar{\nu}_2,\bar{\gamma}_\ell)}
  {\sum_{\ell=1}^{L}\bar{\gamma}_\ell}, \nonumber                                                                \\
   & \overset{(b)}{\leq} \sup_{M\in \mathbb{N},(\bar{\nu}_2,\bm{w},\tilde{\bm{\gamma}}) \in \mathcal{A}_m^{(M)}}
  \sum_{i=1}^{M} w_i \varphi_1(\bar{\nu}_2,\tilde{\gamma}_i),\nonumber                                           \\
   & \overset{(c)}{=} \sup_{(\bar{\nu}_2,\bm{w},\tilde{\bm{\gamma}}) \in \mathcal{A}_m^{(2)}}
  \sum_{i=1}^{2} w_i \varphi_1(\bar{\nu}_2,\tilde{\gamma}_i),\nonumber                                           \\
   & = \bar{\rho}, \label{eq:P1_obj_to_P2_obj}
\end{align}
where step $(a)$ follows by Eqs.~\eqref{eq:cor9_2_1} and \eqref{eq:cor9_2_2};
step $(b)$ comes from Eqs.~\eqref{eq:cor9_2_3}-\eqref{eq:cor9_2_5};
and step $(c)$ applies the Lem.~\ref{lem:related_properties_cor9}\eqref{item:cor9_1}.

\subsubsection{\texorpdfstring{$\rho^* \geq \bar{\rho}$}{rho* >= bar rho}}
The reverse inequality follows because every feasible two-point value of problem
(P2) can be approached by finite-dimensional Gaussian sources with the prescribed
trace in (P1). Specifically, given any feasible solution to (P2), we first
construct a set of feasible solutions to (P1), then show that their objective
values converge to that of the given solution.

For any feasible two-point value in problem (P2),
a one-point solution is represented exactly by a constant normalized sequence,
so consider two distinct eigenvalues with $w_1,w_2>0$.
Let $\tilde{\gamma}_1>\tilde{\gamma}_2$ and set
\begin{equation}\label{eq:cor9_zeta_general}
  \zeta\triangleq\frac{w_1}{w_1+w_2}\in(0,1).
\end{equation}
Choose positive integers $N_{1,n}$ and $N_{2,n}$ such that
\begin{equation}
  \zeta_n\triangleq \frac{N_{1,n}}{N_{1,n}+N_{2,n}}\to\zeta.
\end{equation}
For every $\gamma>0$, $\frac{\partial \varphi_{2}(\nu,\gamma)}{\partial \nu} < 0$.
Hence, the implicit function theorem applied to equality moment constraint
gives a sequence $\bar\nu_{2,n}\to\bar{\nu}_2$ satisfying
\begin{equation}\label{eq:cor9_adjusted_nu}
  \zeta_n\varphi_2(\bar\nu_{2,n},\tilde{\gamma}_1) +(1-\zeta_n)\varphi_2(\bar\nu_{2,n},\tilde{\gamma}_2)=0.
\end{equation}
Let $L_n\triangleq N_{1,n}+N_{2,n}$ and fix the total variance by constructing
\begin{align}
  \alpha_n  & \triangleq \frac{\operatorname{tr}(\Gamma)}
  {N_{1,n}\tilde{\gamma}_1+N_{2,n}\tilde{\gamma}_2},\label{eq:cor9_alpha_n}             \\
  \Gamma_n  & \triangleq\operatorname{diag}
  \big(\underbrace{\alpha_n\tilde{\gamma}_1,\ldots,\alpha_n\tilde{\gamma}_1}_{N_{1,n}},
  \underbrace{\alpha_n\tilde{\gamma}_2,\ldots,\alpha_n\tilde{\gamma}_2}_{N_{2,n}}\big), \\
  R_n       & \triangleq \frac{1}{2} \left( N_{1,n}\log^+\tilde{\gamma}_1+N_{2,n}
  \log^+\tilde{\gamma}_2\right),                                                        \\
  \nu_{2,n} & \triangleq\alpha_n\bar\nu_{2,n}. \label{eq:cor9_nu2_n}
\end{align}
Then Eq.~\eqref{eq:alpha_condition} follows from the definition of $R_n$,
and Eq.~\eqref{eq:cor9_adjusted_nu} is precisely Eq.~\eqref{eq:opt_D_multiplier} after
normalization. Thus, every constructed point is feasible for problem (P1).
Moreover,
\begin{align}
  \frac{U(R_n,\Gamma_n)}{2\operatorname{tr}(\Gamma)}
   & =\frac{\zeta_n\varphi_1(\bar\nu_{2,n},\tilde{\gamma}_1) +(1-\zeta_n)\varphi_1(\bar\nu_{2,n},\tilde{\gamma}_2)}
  {\zeta_n\tilde{\gamma}_1+(1-\zeta_n)\tilde{\gamma}_2} \nonumber                                                                     \\
   & \rightarrow \frac{\zeta\varphi_1(\bar{\nu}_2,\tilde{\gamma}_1) +(1-\zeta)\varphi_1(\bar{\nu}_2,\tilde{\gamma}_2)}
  {\zeta\tilde{\gamma}_1+(1-\zeta)\tilde{\gamma}_2} \nonumber                                                                         \\
   & =w_1\varphi_1(\bar{\nu}_2,\tilde{\gamma}_1) +w_2\varphi_1(\bar{\nu}_2,\tilde{\gamma}_2), \label{eq:cor9_objective_approximation}
\end{align}
where the last equality follows from $w_1\tilde{\gamma}_1+w_2\tilde{\gamma}_2=1$
and Eq.~\eqref{eq:cor9_zeta_general}. Taking the supremum proves $\rho^*\geq\bar\rho$.
If $\tilde{\gamma}_1=\tilde{\gamma}_2$, then $\bar{\rho}=0$, which cannot be optimal.
Hence, without loss of generality, it suffices to consider the case when
$\tilde{\gamma}_1>\tilde{\gamma}_2>0$ in the subsequent analysis.

\subsubsection{\texorpdfstring{$\bar{\rho}=\rho'$}{bar rho = rho'}}
To prove $\bar{\rho}=\rho'$, we first localize the optimal smaller support point
$\tilde{\gamma}_2^*$ at $1$ by the result of Cor.~\ref{cor:property_of_U}, after which
the range of $\bar{\nu}_2$ and $\tilde{\gamma}_1$ can be restricted to $\mathcal D$
by analyzing necessary conditions for the constraints in problem (P2)
that admit $w_1,w_2>0$.

\begin{lemma}[Localization of the Smaller Support Point]\label{lem:cor9_localization}
  The optimal value of the smaller support point
  $\tilde{\gamma}_2$ is $\tilde{\gamma}_2^*=1$.
\end{lemma}
\begin{IEEEproof}
  By Cor.~\ref{cor:property_of_U}, the piecewise function $U(R,\Gamma)$
  attains its maximum over $R$ at a transition point. For any $\Gamma$ with only two
  distinct eigenvalues, the only transition at $R>0$ occurs when the
  classical water level $\alpha$ equals the smaller eigenvalue, allowing us to
  fix the eigenvalue ratio $\tilde\gamma_2=1$ without changing the supremum.
\end{IEEEproof}

For a two-point feasible solution whose lower support point
is $\tilde{\gamma}_2^*=1$, both the admissible parameter region and the
corresponding weights can be characterized explicitly.
Indeed,
\begin{align}
   & (\bar{\nu}_2,w_1,w_2,\tilde{\gamma}_1,\tilde{\gamma}_2^*) \in \mathcal A_m^{(2)}\nonumber                                        \\
   & \qquad \Longrightarrow w_1 \varphi_2(\bar{\nu}_2,\tilde{\gamma}_1) + w_2 \varphi_2(\bar{\nu}_2,\tilde{\gamma}_2^*) = 0 \nonumber \\
   & \qquad \overset{(a)}{\Longleftrightarrow} \exists\,\gamma>1:\ \varphi_2(\bar{\nu}_2,\gamma)<0\nonumber                           \\
   & \qquad \overset{(b)}{\Longleftrightarrow} \lim_{\gamma\to\infty}\varphi_2(\bar{\nu}_2,\gamma) < 0 \nonumber                      \\
   & \qquad \overset{(c)}{\Longleftrightarrow} -\log(2\bar{\nu}_2)<0 \nonumber                                                        \\
   & \qquad \Longleftrightarrow \bar{\nu}_2>\frac{1}{2}.
\end{align}
where step $(a)$ is due to $\varphi_2(\bar{\nu}_2,\tilde{\gamma}_2^*)=\varphi_2(\bar{\nu}_2,1)>0$
and $w_1,w_2>0$; step $(b)$ follows from the strict decrease of $\varphi_2(\bar{\nu}_2,\gamma)$
w.r.t. $\gamma$, as established in Eq.~\eqref{eq:cor9_3_1};
and step $(c)$ is due to Eq.~\eqref{eq:cor9_3_5}.
Moreover, for $\bar{\nu}_2>\frac{1}{2}$, solving the equation
$\varphi_2(\bar{\nu}_2,\gamma)=0$ gives
\begin{align}
  \gamma = \gamma_0(\bar{\nu}_2)\triangleq \frac{4\bar{\nu}_2^2}{4\bar{\nu}_2^2-1}. \label{eq:cor9_gamma0}
\end{align}
Together with the constraint~\eqref{eq:rp_c2}, we have $0<w_1,w_2<1$.
Thus, the admissible parameter region is $(\bar{\nu}_2,\tilde{\gamma}_1,w_1,w_2)\in\mathcal{D}$,
and the original problem satisfies $\rho^* =\bar{\rho} = \rho'$.

Remark that the equality $\rho^* = \rho'$ is an equality of suprema.
It does not assert that an arbitrary irrational two-point weight is represented exactly in a
fixed finite dimension. Therefore, we can solve the
optimal value $\rho'$ defined in problem (P3) alternatively to obtain $\rho^*$.

\subsection{Solution of \texorpdfstring{$\rho'$}{rho'}}
To determine the optimal value $\rho'$ of problem (P3), we derive the
necessary conditions for optimality by the Lagrange multiplier method.
The determinant of the two constraint gradients w.r.t. $(w_1,w_2)$ is
\begin{equation}
  \varphi_2(\bar{\nu}_2,\tilde\gamma_1) -\tilde\gamma_1\varphi_2(\bar{\nu}_2,1)<0,
\end{equation}
so the linear-independence constraint qualification holds. Let $\mu_1$ and $\mu_2$
be the optimal Lagrange multipliers corresponding to the constraint~\eqref{eq:rp_c1}
and~\eqref{eq:rp_c2}, respectively. The KKT conditions at any interior maximizer are therefore
\begin{align}
   & \varphi_1(\bar{\nu}_2^*,\tilde{\gamma}_1^*) - \mu_2\tilde{\gamma}_1^*
  - \mu_1\varphi_2(\bar{\nu}_2^*,\tilde{\gamma}_1^*) =0, \label{eq:cor9_kkt_w1}                     \\
   & \varphi_1(\bar{\nu}_2^*,1) -\mu_2 -\mu_1\varphi_2(\bar{\nu}_2^*,1) =0, \label{eq:cor9_kkt_w2}  \\
   & \varphi_{1\gamma}(\bar{\nu}_2^*,\tilde{\gamma}_1^*) -\mu_2 -\mu_1
  \varphi_{2\gamma}(\bar{\nu}_2^*,\tilde{\gamma}_1^*) =0, \label{eq:cor9_kkt_gamma}                 \\
   & w_1^*\big[ \varphi_{1\nu}(\bar{\nu}_2^*,\tilde{\gamma}_1^*) -\mu_1
  \varphi_{2\nu}(\bar{\nu}_2^*,\tilde{\gamma}_1^*) \big] \nonumber                                  \\
   & \quad + w_2^*\big[ \varphi_{1\nu}(\bar{\nu}_2^*,1) -\mu_1\varphi_{2\nu}(\bar{\nu}_2^*,1) \big]
  = 0, \label{eq:cor9_kkt_nu}
\end{align}
where $\varphi_{1\nu}$, $\varphi_{1\gamma}$, $\varphi_{2\nu}$, and $\varphi_{2\gamma}$ denote the
partial derivatives of $\varphi_1(\nu,\gamma)$ and $\varphi_2(\nu,\gamma)$ w.r.t. $\nu$
and $\gamma$, respectively. The uniqueness of the solution to the KKT conditions is
given by the following lemma.
\begin{lemma}[Uniqueness of the KKT Point]\label{lem:cor9_kkt_uniqueness}
  The system of KKT conditions~\eqref{eq:cor9_kkt_w1}-\eqref{eq:cor9_kkt_nu} is
  equivalent to Eq.~\eqref{eq:universal_gap_constant} and has a unique solution.
\end{lemma}
\begin{IEEEproof}
  See Appendix~\ref{app:cor9_kkt_uniqueness}.
\end{IEEEproof}

As established in Lem.~\ref{lem:related_properties_cor9}\eqref{item:cor9_4}, the global
maximizer exists in the interior and every such maximizer must satisfy this KKT system,
thus this solution is the unique global maximizer of the problem (P3).
Solving Eq.~\eqref{eq:universal_gap_constant} by the numerical
method gives the unique solution $\bar{\nu}_2^* \approx 1.6556$, and
$\tilde{\gamma}_1^* \approx 2.7144$.
Multiplying Eqs.~\eqref{eq:cor9_kkt_w1} and \eqref{eq:cor9_kkt_w2} by $w_1$ and $w_2$,
respectively, and using the two equality constraints~\eqref{eq:rp_c1}-\eqref{eq:rp_c2} gives
\begin{equation}\label{eq:cor9_rho_value}
  \rho^*=\mu_2=\varphi(\bar{\nu}_2^*,1) \approx 0.1448.
\end{equation}
Finally, Eqs.~\eqref{eq:cor9_gap_to_U} and~\eqref{eq:cor9_rho_def} yield
\begin{equation}\label{eq:cor9_final_bound}
  0\leq\Delta D(R,C,P) \leq U(R,\Gamma) \leq 2 \rho^*\operatorname{tr}(\Gamma).
\end{equation}

Applying the construction $(R_n,\Gamma_n)$ in Eqs.~\eqref{eq:cor9_alpha_n}-\eqref{eq:cor9_nu2_n}
to the optimizer of (P3) gives $\operatorname{tr}(\Gamma_n)=\operatorname{tr}(\Gamma)$.
By the tightness statement following Thm.~\ref{thm:gap_upper_bound}, with $P=0$ and
$C\to\infty$, a diagonal choice $C_n\to\infty$ then ensures
$\frac{\Delta D(R_n,C_n,0)}{2\operatorname{tr}(\Gamma)}\to\rho^*$.
Hence, $2\rho^*\operatorname{tr}(\Gamma)$ is an asymptotically achievable universal upper
bound, which completes the proof.

\section{Proof of Theorem~\ref{thm:analytical_ratio}}\label{app:analytical_ratio}
We will first show that the infimum of $\eta(R,C,P)$ over $P\geq 0$ is approached
as $P=0$ by monotonicity, then proving $\eta(R,C,P)\geq \eta(R,\infty,0)$ for any
$C\geq 0$ and $P\geq 0$ with the help of Thm.~\ref{thm:gap_upper_bound}. Finally,
let $L(R,\Gamma) = \eta(R,\infty,0)$ to obtain an analytical lower bound of $\eta(R,C,P)$.

We first state the required monotonicity property.

\begin{lemma}[Monotonicity of $\eta(R,C,P)$ in $P$]\label{lem:monotonicity_P_eta}
  For any given $R>0$ and $C\geq 0$, the ratio $\eta(R,C,P)$ is non-decreasing
  w.r.t. the perception limit $P\geq 0$.
\end{lemma}
\begin{IEEEproof}
  See Appendix~\ref{app:monotonicity_P_eta}.
\end{IEEEproof}

It follows that
\begin{align}
  \eta(R,C,P)
   & \overset{(a)}{\geq}\eta(R,C,0) \nonumber                                 \\
   & =1-\frac{\Delta D(R,C,0)}{\tilde{D}(R,C,0)} \nonumber                    \\
   & \overset{(b)}{\geq}1-\frac{U(R,\Gamma)}{\tilde{D}(R,C,0)} \nonumber      \\
   & \overset{(c)}{\geq}1-\frac{U(R,\Gamma)}{\tilde{D}(R,\infty,0)} \nonumber \\
   & \overset{(d)}{=}\frac{D(R,\infty,0)}{\tilde{D}(R,\infty,0)} \nonumber    \\
   & =\eta(R,\infty,0),
\end{align}
where step $(a)$ is based on the monotonicity established
in Lem.~\ref{lem:monotonicity_P_eta}; step $(b)$ uses the gap bound in
Thm.~\ref{thm:gap_upper_bound}; step $(c)$ comes from $\tilde{D}(R,C,0)$
is non-increasing in $C$, since the greedy allocation $\tilde{\mathbf r}$
is independent of $C$, whereas the feasible set for optimizing $\mathbf r'$
expands as $C$ increases; and step $(d)$ is due to the fact that
$U(R,\Gamma)=\tilde{D}(R,\infty,0)-D(R,\infty,0)$.
Dividing Eq.~\eqref{eq:D_Cinf_P0} by Eq.~\eqref{eq:D_tilde_Cinf_P0}
directly yields the analytical formula for $\eta(R,\infty,0)$, which is
the expression of $L(R,\Gamma)$ in Eq.~\eqref{eq:L_R_definition_ratio}.
Thus, $\eta(R,C,P)\geq L(R,\Gamma)$, which completes the proof.

\section{Proof of Corollary~\ref{cor:property_of_L}}\label{app:property_of_L}

Similar to the proof of Cor.~\ref{cor:property_of_U}, we will show that
$L(R, \Gamma)$ is a piecewise function over $R$ and any stationary point
within the sub-interval divided by the breakpoints is a local maximum.
With this local maximum property, the global minimum of $L(R, \Gamma)$ can
be found only at the finite set of breakpoints, which
are exactly $\{R_1, R_2, \dots, R_L\}$.

If all eigenvalues of $\Gamma$ are equal, $L(R,\Gamma)\equiv1$, and the
conclusion follows immediately from Thm.~\ref{thm:analytical_ratio}.
Otherwise, use the partition in Eq.~\eqref{eq:domain_divide} and fix an
interior stationary point $R_0\in(R_k,R_{k+1})$ of $L(R,\Gamma)$.
We use $x_\ell$ and $S(\nu)$ as defined in Appendix~\ref{app:sub_convexity_U},
with $\alpha_0$ and $\nu_0$ evaluated at $R=R_0$.
Differentiating Eq.~\eqref{eq:L_R_definition_ratio} using
Eqs.~\eqref{eq:alpha_condition} and~\eqref{eq:opt_D_multiplier} gives
\begin{equation}\label{eq:stationary_condition_L}
  \frac{\nu_0}{\alpha_0}
  =\frac{L(R_0,\Gamma)}{2k}\sum_{\ell=1}^k x_\ell
  \le\frac{1}{2k}\sum_{\ell=1}^k x_\ell,
\end{equation}
where $0<L(R_0,\Gamma)\le1$ follows from Eq.~\eqref{eq:L_R_definition_ratio},
Thm.~\ref{thm:greedy_solution} and Thm.~\ref{thm:analytical_ratio}. Consequently,
\begin{align}
  S(\nu_0)\sum_{\ell=1}^k(x_\ell^3+x_\ell)
   & \overset{(a)}{\ge} S\left(\frac{\alpha_0}{2k}\sum_{\ell=1}^k x_\ell\right)
  \sum_{\ell=1}^k(x_\ell^3+x_\ell) \nonumber                                         \\
   & \overset{(b)}{>} 2k\sum_{\ell=1}^k x_\ell, \label{eq:second_derivative_chain_L}
\end{align}
where step $(a)$ follows from Eq.~\eqref{eq:stationary_condition_L} and
$S'(\nu_0)=-\sum_{\ell=1}^L\gamma_\ell^2 (\gamma_\ell^2+\nu_0^2)^{-3/2}<0$;
step $(b)$ applies the Jensen argument in Eq.~\eqref{eq:second_derivative_chain}
with its parameter $\nu_0$ replaced by $\frac{\alpha_0}{2k}\sum_{\ell=1}^k x_\ell$.
Since the denominator in Eq.~\eqref{eq:L_R_definition_ratio} is positive,
the quotient rule at $R_0$ yields
\begin{align}
   & \operatorname{sgn}\left(\left.\frac{\partial^2L(R,\Gamma)}{\partial R^2}\right|_{R=R_0}\right)\nonumber \\
   & \qquad = \operatorname{sgn}
  \left[2k\sum_{\ell=1}^k x_\ell-S(\nu_0)\sum_{\ell=1}^k(x_\ell^3+x_\ell)\right] = -1,
  \label{eq:L_second_derivative_stationary}
\end{align}
where the equality of signs is based on Eq.~\eqref{eq:stationary_condition_L},
and the last equality follows from Eq.~\eqref{eq:second_derivative_chain_L}.
Thus, every interior stationary point is a strict local maximum.

By continuity of $L(R,\Gamma)$ over $R\ge0$ and the fact that
$L(R_1,\Gamma)=L(0,\Gamma) =\lim_{R\to\infty}L(R,\Gamma)=1$,
the same endpoint argument as in the proof of Cor.~\ref{cor:property_of_U}
gives Eq.~\eqref{eq:global_min_L_overR}. Therefore, combined with
Thm.~\ref{thm:analytical_ratio}, we have
\begin{equation}
  \eta(R,C,P)\ge L(R,\Gamma)\ge\min_{\ell\in\{1,2,\ldots,L\}}L(R_\ell,\Gamma).
\end{equation}
This completes the proof.

\section{Proof of Corollary~\ref{cor:universal_ratio_bound}}\label{app:universal_ratio_bound}
By Thm.~\ref{thm:analytical_ratio}, proving $\eta^*\leq \eta(R,C,P)$ is equivalent
to showing that $L(R,\Gamma)\geq \eta^*$ for any $R > 0$ and $\Gamma\succ 0$.
To this end, we first establish a summation inequality and a pointwise bound by
applying the convexity of $\varphi(\nu,\gamma)$.
Then, we construct a sequence of two-tier sources whose performance ratios approach
the proposed constant $\eta^*$, thereby establishing its sharpness.

Since the greedy solution is an upper bound on the optimal solution, $\eta(R,C,P) \leq 1$.
For the extreme cases in Cor.~\ref{cor:tightness_extremes}, the two distortions
coincide and $\eta(R,C,P)=1$. The following proof continues to use the notations
introduced in Appendix~\ref{app:universal_gap_bound}.

\subsection{Universality of the Lower Bound}
To prove $\eta^*$ is a universal lower bound of $\eta(R,C,P)$, we first introduce
a lemma to give some related properties.
\begin{lemma}\label{lem:related_properties_cor11}
  \renewcommand{\labelenumi}{(\theenumi)} 
  For any $\nu>0$ and $\gamma>0$, we have:
  \begin{enumerate}
    \item\label{item:cor11_1}
          The multiplier $\tilde{\nu}_2$ defined as the root of Eq.~\eqref{eq:nu1_star} is unique.
    \item\label{item:cor11_2}
          The constant $\eta^*\in (0, 1)$.
    \item\label{item:cor11_3}
          For $\bar{\nu}_2$, $\tilde{\nu}_2$ and the normalized
          eigenvalues $\bar{\gamma}_\ell$, there is
          \begin{align}
            \sum_{\ell=1}^{L}\varphi(\tilde{\nu}_2,\bar{\gamma}_\ell) \geq
            \sum_{\ell=1}^{L}\varphi_1(\bar{\nu}_2,\bar{\gamma}_\ell).
            \label{eq:ratio_multiplier_comparison}
          \end{align}
    \item\label{item:cor11_4}
          \emph{Pointwise bound}: For any $\gamma>0$, there is
          \begin{equation}\label{eq:ratio_pointwise_bound}
            \varphi(\tilde{\nu}_2,\gamma) \leq (1-\eta^*)\tilde d(\gamma),
          \end{equation}
          where $\tilde d(\gamma) \triangleq \gamma-\sqrt{\gamma(\gamma-1)_+}>0$ for $\gamma>0$.
  \end{enumerate}
\end{lemma}
\begin{IEEEproof}
  See Appendix~\ref{app:ratio_pointwise_bound}.
\end{IEEEproof}

With these properties, we have
\begin{align}
  \eta(R,C,P)
   & \overset{(a)}{\geq}L(R,\Gamma)
  \nonumber                                                                                 \\
   & \overset{(b)}{=} 1-\frac{\sum_{\ell=1}^{L} \varphi_1(\bar{\nu}_2,\bar{\gamma}_\ell)}
  {\sum_{\ell=1}^{L}\tilde d(\bar{\gamma}_\ell)} \nonumber                                  \\
   & \overset{(c)}{\geq} 1-\frac{\sum_{\ell=1}^{L}\varphi(\tilde{\nu}_2,\bar{\gamma}_\ell)}
  {\sum_{\ell=1}^{L}\tilde d(\bar{\gamma}_\ell)} \nonumber                                  \\
   & \overset{(d)}{\geq} 1-\frac{(1-\eta^*)\sum_{\ell=1}^{L}\tilde d(\bar{\gamma}_\ell)}
  {\sum_{\ell=1}^{L}\tilde d(\bar{\gamma}_\ell)} \nonumber                                  \\
   & = \eta^*,\label{eq:ratio_universal_chain}
\end{align}
where step $(a)$ applies Thm.~\ref{thm:analytical_ratio}; step $(b)$ substitutes the
definition of $L(R,\Gamma)$ in Eq.~\eqref{eq:L_R_definition_ratio} with
$\varphi_1(\bar{\nu}_2,\bar{\gamma}_\ell)$ and $\tilde d(\bar{\gamma}_\ell)$;
step $(c)$ is based on Eq.~\eqref{eq:ratio_multiplier_comparison} established in
Lem.~\ref{lem:related_properties_cor11};
and step $(d)$ follows by Eq.~\eqref{eq:ratio_pointwise_bound}.
Solving the unique scalar equation~\eqref{eq:nu1_star} gives
$\tilde{\nu}_2\approx0.8916$ and $\eta^*\approx0.7517$.
This establishes the stated universal lower bound $\eta^*$.

\subsection{Sharpness of the Lower Bound}
We prove that the universal lower bound $\eta^*$ is asymptotically achievable
by constructing a source covariance matrix with only two distinct eigenvalues, whose ratio
tends to infinity. Define
\begin{equation}\label{eq:ratio_asymptotic_fraction}
  \zeta^*\triangleq
  \frac{\varphi_2(\tilde{\nu}_2,1)}
  {\varphi_2(\tilde{\nu}_2,1)+\log(2\tilde{\nu}_2)}
  \in(0,1).
\end{equation}
For sufficiently large integers $n$, set $N_{1,n}=\lfloor n\zeta^*\rfloor$,
$N_{2,n}=n-N_{1,n}$, and $\zeta_n=\frac{N_{1,n}}{n} \to \zeta^*$. Construct
\begin{align}
   & \alpha_n\triangleq\frac{\operatorname{tr}(\Gamma)}{nN_{1,n}+N_{2,n}}, \\
   & \Gamma_n\triangleq\operatorname{diag}\bigl(
  \underbrace{n\alpha_n,\ldots,n\alpha_n}_{N_{1,n}},
  \underbrace{\alpha_n,\ldots,\alpha_n}_{N_{2,n}}\bigr),                   \\
   & R_n \triangleq\frac{N_{1,n}}{2}\log n,
\end{align}
where $\alpha_n$ corresponds to the classical water level for $(R_n,\Gamma_n)$, and
$\operatorname{tr}(\Gamma_n)=(nN_{1,n}+N_{2,n})\alpha_n=\operatorname{tr}(\Gamma)$ for any $n$.
The corresponding normalized multiplier $\bar{\nu}_{2,n}>0$ is the unique solution of
\begin{equation}\label{eq:ratio_sequence_moment}
  \zeta_n\varphi_2(\bar{\nu}_{2,n},n)+(1-\zeta_n)\varphi_2(\bar{\nu}_{2,n},1)=0.
\end{equation}
For each fixed $\nu>0$,
$\zeta_n\varphi_2(\nu,n)+(1-\zeta_n)\varphi_2(\nu,1)\rightarrow-\zeta^*\log(2\nu)+(1-\zeta^*)\varphi_2(\nu,1)$,
whose limit is strictly decreasing in $\nu$ and vanishes at $\nu=\tilde{\nu}_2$ by
Eq.~\eqref{eq:ratio_asymptotic_fraction}. Thus, convergence at $\tilde{\nu}_2-\epsilon$ and
$\tilde{\nu}_2+\epsilon$ for every sufficiently small $\epsilon>0$ gives
$\bar{\nu}_{2,n}\to\tilde{\nu}_2$. It follows that
\begin{align}
  L(R_n,\Gamma_n)
   & =1-\frac{\zeta_n\varphi_1(\bar{\nu}_{2,n},n)+(1-\zeta_n)\varphi_1(\bar{\nu}_{2,n},1)}
  {\zeta_n\tilde d(n)+1-\zeta_n}\nonumber                                                    \\
   & \to 1-\frac{\zeta^*(1/2-\tilde{\nu}_2)+(1-\zeta^*)\varphi_1(\tilde{\nu}_2,1)}
  {\zeta^*/2+1-\zeta^*}\nonumber                                                             \\
   & \overset{(a)}{=} 1-\frac{(1-\eta^*)(\zeta^*/2+1-\zeta^*)}{\zeta^*/2+1-\zeta^*}\nonumber \\
   & =\eta^*, \label{eq:ratio_sharp_limit}
\end{align}
where step $(a)$ uses $\varphi_1(\nu,\gamma)=\varphi(\nu,\gamma)+\nu\ln2\,\varphi_2(\nu,\gamma)$,
Eq.~\eqref{eq:ratio_constant_identity}, Eq.~\eqref{eq:ratio_H_infinity}, and the
equality $-\zeta^*\log(2\tilde{\nu}_2)+(1-\zeta^*)\varphi_2(\tilde{\nu}_2,1)=0$.

Finally, the proof of Thm.~\ref{thm:analytical_ratio} gives
\begin{equation}
  \lim_{C\to\infty}\eta(R_n,C,0)=L(R_n,\Gamma_n),
\end{equation}
for each fixed $n$. Choose a finite $C_n$ large enough that
\begin{equation}
  L(R_n,\Gamma_n)
  \leq\eta(R_n,C_n,0)
  \leq L(R_n,\Gamma_n)+\frac1n.
\end{equation}
Together with Eq.~\eqref{eq:ratio_sharp_limit},
$\eta(R_n,C_n,0)\to\eta^*$ under arbitrary fixed trace of the source covariance matrix.
Hence, the lower bound $\eta^*$ is asymptotically achievable, and
no larger universal lower bound is possible, which completes the proof.

\section{Proof of Lemma~\ref{lem:optimal_solution_properties}}\label{app:existence_uniqueness_w}
The problem~\eqref{eq:prob_re1}-\eqref{eq:prob_re4} is jointly convex in
$(\mathbf d,\mathbf d',T)$, since its constraints are convex as in
Lem.~\ref{lem:convexity_of_fixed_T}, and $(\sqrt T-\sqrt P)^2$ is convex on $T\ge P$.
The finite rate budget $R$ bounds every $d_\ell,d_\ell'$ away from zero, and the
objective diverges as $T\to\infty$, thus an optimal solution exists. Slater's condition
holds by taking $d_\ell=d_\ell'=1-\epsilon$ for sufficiently small $\epsilon>0$ and then
choosing $T$ sufficiently large. Thus optimal multipliers exist, and strong duality holds.
Now we prove the properties in turn.

\subsubsection*{1) Property~(\ref{item:optimal_solution_boundary})}
Firstly, we show that every optimal solution must satisfy $\hat T>P$.
The constraint~\eqref{eq:prob_re2} gives
\begin{align}
  \hat T \geq \sum_{\ell=1}^L W_\ell(\hat r_\ell,\hat r_\ell')
  \geq \sum_{\ell=1}^L \gamma_\ell \hat d_\ell' > 0.
\end{align}
If $\hat T=P>0$, the optimality in $T$ gives $-\mu\ge0$,
hence $\mu=0$. Together with complementary slackness,
the minimization of $\mathcal L$ in $\mathbf d$ reduces to the classical
RD problem~\eqref{eq:opt1}-\eqref{eq:opt1_constraint}, so that
$\hat d_\ell=2^{-2\tilde r_\ell}$ and
\begin{align}
  P=\hat T \geq \sum_{\ell=1}^L W_\ell(\tilde r_\ell,\hat r_\ell')
  \overset{(a)}{\geq} \sum_{\ell=1}^L W_\ell(\tilde r_\ell,\tilde r_\ell') = \tilde P,
\end{align}
where step $(a)$ follows from Lem.~\ref{lem:optimal_of_r_prime}. This contradicts
the assumption that $P<\tilde P$. Hence, the optimal solution $\hat T>P$,
which implies $0<\mu \leq 1$.

A mixed boundary with exactly one of $\hat d_\ell,\hat d_\ell'$
equal to $1$ cannot minimize $\mathcal L$: decreasing that coordinate
by $\epsilon$ decreases the square-root term by order $\sqrt\epsilon$,
while the remaining changes are of order $\epsilon$.
Nor can all pairs equal $(1,1)$, since both rate constraints would
be slack and complementary slackness would give $\nu=\nu'=0$.
Decreasing any one pair to $(1-\epsilon,1-\epsilon)$ would then
change $\mathcal L$ by $-(1+\mu)\gamma_\ell\epsilon<0$.
Thus every pair is either interior or equals $(1,1)$,
and at least one pair is interior.

\subsubsection*{2) Property~(\ref{item:optimal_solution_multipliers}) and~(\ref{item:optimal_solution_active})}
Existence of optimal multipliers has been proven above, now
we focus on the uniqueness. At any interior component, the KKT
conditions~\eqref{eq:partial_3_2}-\eqref{eq:partial_3_3} give
\begin{align}
  \nu  & =\gamma_\ell\hat d_\ell
  \left(1-\mu+\mu\sqrt{\frac{1-\hat d_\ell'}{1-\hat d_\ell}}\right)>0,\label{eq:nu_def}               \\
  \nu' & =\mu\gamma_\ell\hat d_\ell' \sqrt{\frac{1-\hat d_\ell}{1-\hat d_\ell'}}>0.\label{eq:nup_def}
\end{align}
Eq.~\eqref{eq:partial_3_1} also gives $\mu=1$ if and only if $P=0$. To establish
uniqueness, fix any optimal solution $(\hat{\mathbf d},\hat{\mathbf d}',\hat T)$.
By strong duality, it minimizes $\mathcal L$ for every dual optimum.
Since $\hat T>P$ and at least one component is interior,
Eqs.~\eqref{eq:partial_3_1},~\eqref{eq:nu_def}, and~\eqref{eq:nup_def} uniquely determine
$\mu$, $\nu$, and $\nu'$ from this fixed optimal solution.
Hence, the optimal multiplier triple is unique.

All three normalized multipliers are strictly positive, thus the
Complementary slackness gives all three
constraints~\eqref{eq:prob_re2}-\eqref{eq:prob_re4} hold with equality.

\subsubsection*{3) Property~(\ref{item:optimal_solution_root})}
For fixed optimal multipliers, the part of Eq.~\eqref{eq:Lagrangian_RDCP}
depending on $(\mathbf d,\mathbf d')$ is
$\sum_{\ell=1}^L\gamma_\ell\Big[(1-\mu)d_\ell-2\mu\sqrt{(1-d_\ell)(1-d_\ell')}\Big]
  -\nu\sum_{\ell=1}^L\ln d_\ell-\nu'\sum_{\ell=1}^L\ln d_\ell'$,
where the negative-logarithm terms are strictly convex because $\nu,\nu'>0$, and the
remaining terms are convex. Since $\mathcal L$ separates into this strictly convex part
and a function of $T$, every primal optimum has the same allocation
$(\hat{\mathbf d},\hat{\mathbf d}')$.
Then the equality condition of Eq.~\eqref{eq:prob_re2} uniquely
determines $\hat T$, and the bijective rate transformation
gives uniqueness of $(\hat{\mathbf r},\hat{\mathbf r}')$.

At $(d_\ell,d_\ell')=(1,1)$, the necessary and sufficient condition for
optimality can be given by the non-negativity of the directional derivative along
$(d_\ell,d_\ell')=(1-\epsilon x,1-\epsilon y)$ for all $x,y\ge 0$, namely
\begin{align}
   & \qquad [\nu-\gamma_\ell(1-\mu)]x+\nu'y-2\mu\gamma_\ell\sqrt{xy}\ge 0, \nonumber                \\
   & \overset{(a)}{\Longleftrightarrow} [\nu-\gamma_\ell(1-\mu)]\nu'\ge\mu^2\gamma_\ell^2,\nonumber \\
   & \Longleftrightarrow \gamma_\ell \le
  \frac{\sqrt{\nu'^2(1 \!-\! \mu)^2+4\mu^2\nu\nu'}-\nu'(1 \!-\! \mu)}{2\mu^2} = \varpi,
\end{align}
where step $(a)$ follows from the quadratic form in $(\sqrt x,\sqrt y)$.
Therefore, strict convexity and
Lem.~\ref{lem:optimal_solution_properties}\eqref{item:optimal_solution_boundary}
give the stated corner condition $\gamma_\ell\leq \varpi$.

Next, we prove the existence and uniqueness of the root of Eq.~\eqref{eq:w_l} for
each component satisfying $\gamma_\ell>\varpi$. For $\gamma_\ell>\varpi$, define
the residual function of Eq.~\eqref{eq:w_l} as
\begin{align}
  f(w_\ell)
   & = w_\ell^3+\left(1-\mu+\frac{\nu'-\nu}{\gamma_\ell}\right)w_\ell^2 \nonumber     \\
   & \qquad + \left[\frac{(1-\mu)\nu'}{\gamma_\ell}-\mu^2\right]w_\ell -\mu^2(1-\mu).
\end{align}
At the left endpoint,
\begin{equation}\label{eq:f_left_point}
  \begin{cases}
    f(0)=-\mu^2(1-\mu)<0,  & \text{for } 0<\mu<1, \\
    f(0)=0,\quad f'(0)=-1, & \text{for } \mu=1,
  \end{cases}
\end{equation}
so $f(w_\ell)<0$ for all sufficiently small $w_\ell>0$ in both cases.
At the right endpoint,
\begin{align}\label{eq:f_right_point}
  f\!\left(\frac{\mu^2\gamma_\ell}{\nu'}\right) = \left(\frac{\mu^2\gamma_\ell}{\nu'}\right)^2
  \left(\frac{\mu^2\gamma_\ell}{\nu'}+1-\mu-\frac{\nu}{\gamma_\ell}\right)>0,
\end{align}
where the strict inequality follows by $\gamma_\ell>\varpi$.
Then the continuity of $f(w_\ell)$ gives the existence of a root
in $(0,\mu^2\gamma_\ell/\nu')$. Moreover, throughout this interval,
$w_\ell+1-\mu>0$ and $\mu^2 > \nu'w_\ell/\gamma_\ell$, thus
\begin{align}
   & \frac{\mathrm d}{\mathrm d w_\ell} \left(\frac{f(w_\ell)}{w_\ell^2(w_\ell+1-\mu)}\right)= \nonumber \\
   & \qquad \qquad \frac{2\mu^2-\frac{\nu'w_\ell}{\gamma_\ell}}{w_\ell^3}
  + \frac{\nu}{\gamma_\ell(w_\ell+1-\mu)^2} > 0,
\end{align}
This strictly increasing quotient has the same zeros as $f(w_\ell)$,
so the root is unique. This completes the proof.

\section{Proof of Lemma~\ref{lem:monotonicity_P}}\label{app:monotonicity_P}
We prove the monotonicity by evaluating the partial derivative
of $\Delta D(R,C,P)$ w.r.t. $P$ in two distinct regimes.
If $R=0$, then $\mathbf r=\tilde{\mathbf r}=\mathbf0$ and
$\psi(0,r_\ell')=0$, so Eqs.~\eqref{eq:prob5_1} and~\eqref{eq:D_tilde_definition}
give $D(0,C,P)=\tilde D(0,C,P)$. Subsequently, we will concentrate on the case
where $R>0$.

\subsubsection*{1) Inactive Perception Regime}
When $P \ge \tilde{P}$, the perception constraint is inactive, and
both $D(R,C,P)$ and $\tilde{D}(R,C,P)$ collapse to the same distortion
$\underline{D}(R)$, which implies $\Delta D(R,C,P) = 0$.
Thus, $\Delta D(R,C,P)$ is trivially non-increasing w.r.t. $P$ in this regime.

\subsubsection*{2) Active Perception Regime}
For the active perception regime, we first establish the regularity needed
for the derivative argument. For any set of feasible solutions,
there is $0\le T=\sum_{\ell=1}^L W_\ell(r_\ell,r_\ell')\le2\operatorname{tr}(\Gamma)$,
and for $P_1,P_2\ge0$,
\begin{align}
   & \left|\bigl[(\sqrt T-\sqrt{P_1})_+\bigr]^2
  -\bigl[(\sqrt T-\sqrt{P_2})_+\bigr]^2\right| \nonumber \\
   & \qquad \le 2\sqrt{2\operatorname{tr}(\Gamma)}
  \left|\sqrt{P_1}-\sqrt{P_2}\right|, \label{eq:monotonicity_P_uniform_modulus}
\end{align}
where $x\mapsto(\sqrt T-x)_+$ is $1$-Lipschitz and takes values
in $[0,\sqrt T]$ for $x\ge0$.
Since the feasible set in Eqs.~\eqref{eq:prob5_1}-\eqref{eq:prob5_3}
does not depend on $P$, taking the infimum preserves this bound.
Applying the same argument to Eq.~\eqref{eq:D_tilde_definition} and
using the bound $|\sqrt{P_1}-\sqrt{P_2}|\le|P_1-P_2|/(2\sqrt\epsilon)$
whenever $P_1,P_2\ge\epsilon>0$, we can conclude that both $D(R,C,P)$ and
$\tilde D(R,C,P)$ are continuous for $P\ge0$ and locally
Lipschitz for $P>0$.
We now evaluate their derivatives w.r.t. $P$ respectively.

The derivative of the greedy upper bound for $P>0$ can be straightforwardly computed
according to its definition in Eq.~\eqref{eq:D_tilde_definition} as follows:
\begin{equation}\label{eq:D_tilde_derivate_P}
  \frac{\partial \tilde{D}(R,C,P)}{\partial P} = 1 - \sqrt{\frac{\tilde{P}}{P}}.
\end{equation}
Hence, the difficulty lies in determining the derivative of $D(R,C,P)$, which is defined
as the optimal value of a parametric optimization problem.

Fix a point $P_0\in(0,\tilde P)$ at which $D(R,C,P)$ is differentiable.
Let $(\hat{\mathbf r},\hat{\mathbf r}')$ be an optimal allocation at $P_0$, and let
$\hat T= \sum_{\ell=1}^{L}\gamma_\ell\big(2-2^{-2 \hat{r}_\ell}-2\psi(\hat{r}_\ell,\hat{r}_\ell')\big)$
denote its total perception penalty.
By virtue of the optimal-value sensitivity analysis
in~\cite[Sec.~5.6.3]{boyd2004convex}, we have
\begin{align}
    & \left. \frac{\partial D(R,C,P)}{\partial P} \right|_{P=P_0} \nonumber                                  \\
  \overset{(a)}{=}
    & \left. \frac{\partial \mathcal{L}(\mathbf{d},\mathbf{d'},T,\lambda_2,\lambda_3,\lambda_4)}{\partial P}
  \right|_{(\mathbf{r},\mathbf{r}')=(\hat{\mathbf{r}},\hat{\mathbf{r}}')} \nonumber                          \\
  \overset{(b)}{=}
    & \left. \frac{\partial {\big(\sqrt{T}-\sqrt{P}\big)}^2}{\partial P} \right|_{P=P_0,T=\hat{T}} \nonumber \\
  = & 1 - \sqrt{\frac{\hat{T}}{P_0}}, \label{eq:D_derivate_P}
\end{align}
where step $(a)$ comes from the optimal-value sensitivity analysis, which states that
the partial derivative
of the DRCP function $D(R,C,P)$ w.r.t. $P$ can be evaluated by the derivative of the
Lagrangian function given in Eq.~\eqref{eq:Lagrangian_RDCP} at the optimal solution
$(\hat{\mathbf{r}},\hat{\mathbf{r}}')$; and
step $(b)$ follows by taking the Lagrange function in Eq.~\eqref{eq:Lagrangian_RDCP} into
step $(a)$.
Consequently,
\begin{equation}
  \left.\frac{\partial\Delta D(R,C,P)}{\partial P}\right|_{P=P_0}
  =\frac{\sqrt{\hat T}-\sqrt{\tilde P}}{\sqrt{P_0}}.
\end{equation}
By Def.~\ref{def:r_tilde}, the first baseline allocation $\tilde{\mathbf{r}}$
is the optimal rate allocation that minimizes Eq.~\eqref{eq:opt1}, which implies that
\begin{equation}\label{eq:opt_distortion_to_greedy}
  \sum_{\ell=1}^L \gamma_\ell 2^{-2 \tilde{r}_\ell} \le \sum_{\ell=1}^L \gamma_\ell 2^{-2 \hat{r}_\ell}.
\end{equation}
Since it has been established in Thm.~\ref{thm:greedy_solution} that $\tilde{D}(R,C,P) \ge D(R,C,P)$,
taking $P=P_0$ in this inequality yields
\begin{align}
   & \sum_{\ell=1}^L \gamma_\ell 2^{-2 \tilde{r}_\ell}+
  {\left(\sqrt{\tilde{P}}-\sqrt{P_0}\right)}^2 \nonumber              \\
   & \qquad \qquad \ge\sum_{\ell=1}^L \gamma_\ell 2^{-2 \hat{r}_\ell}
  + {\left( \sqrt{\hat{T}} - \sqrt{P_0} \right)}^2.
\end{align}
Combining with Eq.~\eqref{eq:opt_distortion_to_greedy} and $\hat T>P_0$, it follows that
$\hat{T} \le \tilde{P}$. Therefore, for almost every $P_0\in(0,\tilde P)$, we have
\begin{equation}\label{eq:monotonicity_P}
  \left.\frac{\partial \Delta D(R,C,P)}{\partial P}\right|_{P=P_0} \le 0,
\end{equation}
which means $\Delta D(R,C,P)$ is non-increasing when $P\in(0,\tilde P)$.
Continuity from Eq.~\eqref{eq:monotonicity_P_uniform_modulus} extends this
conclusion to $P=0$ and $P=\tilde P$.
Combining with the inactive perception regime proves monotonicity
on the entire perception domain.

\section{Proof of Lemma~\ref{lem:monotonicity_C}}\label{app:monotonicity_C}
We prove the monotonicity by evaluating the first-order derivative of $\Delta D(R,C,P)$
w.r.t. $C$ at the extreme case $P=0$. Assume that $R>0$ and $\gamma_1\ge\cdots\ge\gamma_L>0$.
With perfect perception quality $P=0$, $D(R,C,0)$ can be expressed as
\begin{align}
  D(R,C,0) = \min_{\mathbf{r},\mathbf{r}' \in \mathbb{R}_{+}^L}
   & \sum_{\ell=1}^L 2 \gamma_{\ell} \left( 1 - \psi(r_\ell, r_\ell') \right),\label{eq:obj_P0_1} \\
  \text{s.t.} \quad
   & \sum_{\ell=1}^L r_\ell \leq R, \label{eq:constra_P0_2}                                       \\
   & \sum_{\ell=1}^L r_\ell' \leq R+C,\label{eq:constra_P0_3}
\end{align}
and its greedy upper bound $\tilde{D}(R,C,0)$ is given by
\begin{align}
  \tilde{D}(R,C,0) = \min_{\mathbf{r}'' \in \mathbb{R}_{+}^L}
   & \sum_{\ell=1}^L 2 \gamma_{\ell} \left( 1-\psi(\tilde{r}_\ell, r_\ell'') \right),\label{eq:obj_P0_2} \\
  \text{s.t.} \quad
   & \sum_{\ell=1}^L r_\ell'' \leq R+C,\label{eq:constra_P0_4}
\end{align}
where $\tilde{\mathbf{r}}$ is the fixed baseline allocation strategy given in Def.~\ref{def:r_tilde}.
For fixed $R>0$, both $D$ and $\tilde{D}$ are convex functions of $C$. Thus, they are
locally Lipschitz on $C>0$. By Appendix~\ref{app:existence_uniqueness_w},
both programs satisfy Slater's condition. Therefore, the unique multipliers
characterized in Lem.~\ref{lem:optimal_solution_properties} and
Appendix~\ref{app:optimal_of_r_prime} yield the sensitivity
identity below wherever both objective functions are differentiable.

Inheriting the normalized multipliers $\nu_1'$ and $\beta$ given in Cor.~\ref{cor:optimal_solution_perfect_P}
and Def.~\ref{def:r_prime}, the Lagrange multipliers for the constraint~\eqref{eq:constra_P0_4}
in $\tilde{D}(R,C,0)$ and constraint~\eqref{eq:constra_P0_3} in $D(R,C,0)$ are $(4\ln 2)/\sqrt{\beta}$
and $2\nu_1'\ln 2$, respectively. Based on the optimal value sensitivity analysis,
\begin{align}
  \frac{\partial\Delta D(R,C,0)}{\partial C}
   & = \frac{\partial\tilde D(R,C,0)}{\partial C}
  -\frac{\partial D(R,C,0)}{\partial C} \nonumber          \\
   & = -\frac{4\ln 2}{\sqrt{\beta}}+2\nu_1'\ln 2 \nonumber \\
   & = 2(\ln 2)\left(\nu_1'-\frac{2}{\sqrt{\beta}}\right).
\end{align}
Therefore, to prove the monotonicity of $\Delta D(R,C,P)$, it suffices to show the
non-negativity of $\nu_1'-\frac{2}{\sqrt{\beta}}$ in the subsequent analysis.
Specifically, we exploit the monotonicity and concavity of a constructed auxiliary
function to obtain the desired result.

Firstly, we show the relative magnitudes of $\tilde{r}_\ell$ and $\hat{r}_\ell$
by comparing their cutoff thresholds $\alpha$ and $\varpi_0$ given in Def.~\ref{def:r_tilde}
and Cor.~\ref{cor:optimal_solution_perfect_P}, respectively.
Cor.~\ref{cor:properties_opt} establishes that $\hat r_\ell\le\hat r_\ell'$ on
every active component, which implies $\nu_1>\nu_1'$ when $C>0$. Therefore,
\begin{align}
  \sum_{\ell=1}^L \frac{1}{2}\log^+ \! \Bigl(\frac{\gamma_\ell}{\varpi_0}\Bigr)
  \overset{(a)}{=} R \!+\! \frac{C}{2} \ge R \overset{(b)}{=}
  \sum_{\ell=1}^L\frac{1}{2}\log^+ \! \Bigl(\frac{\gamma_\ell}{\alpha}\Bigr),
  \label{eq:gap_C_cutoff_comparison}
\end{align}
where step $(a)$ comes from the Eq.~\eqref{eq:varpi0_condition},
while step $(b)$ is due to Eq.~\eqref{eq:alpha_condition}.
Applying the monotonicity of the function
$x\mapsto\sum_{\ell=1}^L\frac{1}{2}\log^+\left(\frac{\gamma_\ell}{x}\right)$,
there is $\varpi_0 \le \alpha$, which means the active set of
$\tilde{\mathbf r}$ is contained in that of $\hat{\mathbf r}$.
Let $m_c$ denote the number of active components of $\hat{\mathbf r}$, then
\begin{equation}\label{eq:gap_C_zero_sum}
  \sum_{\ell=1}^{m_c}(\hat r_\ell-\tilde r_\ell) =
  \sum_{\ell=1}^{L}(\hat r_\ell-\tilde r_\ell) = R-R = 0.
\end{equation}
For any component satisfying $\gamma_\ell>\alpha$, Eqs.~\eqref{eq:def_r_tilde}
and~\eqref{eq:opt_perfectP_r} yield
\begin{align}
  \hat r_\ell \!-\! \tilde r_\ell = \frac{1}{2}\! \log \! \left[\frac{\alpha}{2\nu_1}
    \!\left(\sqrt{4+\frac{(\nu_1-\nu_1')^2}{\gamma_\ell^2}}
    \!+\! \frac{\nu_1-\nu_1'}{\gamma_\ell}\right)\!\right]\!,
  \label{eq:gap_C_allocation_difference}
\end{align}
which is non-increasing in $\gamma_\ell$.
For the components satisfying $\varpi_0<\gamma_\ell\le\alpha$, we have
$\hat r_\ell-\tilde r_\ell=\hat r_\ell>0$.
Consequently, there exists $k_c\in\{1,2,\ldots,m_c\}$ such that
\begin{align}
  \begin{cases}
    \hat r_\ell-\tilde r_\ell\le0, & 1\le\ell\le k_c, \\
    \hat r_\ell-\tilde r_\ell\ge0, & k_c<\ell\le m_c.
  \end{cases}\label{eq:gap_C_single_crossing}
\end{align}
If the two allocations $\tilde{\mathbf r}$ and $\hat{\mathbf r}$ coincide, take $k_c=m_c$.

For $v>0$, define an auxiliary function
\begin{equation}\label{eq:gap_C_sum_rate_response}
  S_v(\mathbf r) \triangleq \sum_{\ell=1}^L \frac{1}{2} \!\log\!
  \left(\frac{1+\sqrt{1+4\gamma_\ell^2(1-2^{-2r_\ell})/v^2}}{2}\right)\!,
\end{equation}
which is concave in $\mathbf r$ and strictly decreasing in $v$ for
$R>0$. Thus proving $\nu_1'\ge\frac{2}{\sqrt{\beta}}$ is equivalent to showing
$S_{\nu_1'}(\tilde{\mathbf r})\le S_{\frac{2}{\sqrt{\beta}}}(\tilde{\mathbf r})$.
When $v=\nu_1'$, the KKT condition~\eqref{eq:mult1} gives
\begin{equation}\label{eq:gap_C_response_gradient}
  \partial_{r_\ell}S_{\nu_1'}(\hat{\mathbf r}) = \frac{\nu_1}
  {\sqrt{(\nu_1')^2+4\gamma_\ell^2(1-2^{-2\hat r_\ell})}},
\end{equation}
for $1\le\ell\le m_c$. Since both $\gamma_\ell$ and $\hat r_\ell$ are non-increasing
in $\ell$, the denominator of Eq.~\eqref{eq:gap_C_response_gradient} is non-increasing in $\ell$,
which implies that
\begin{align}
  \partial_{r_1}S_{\nu_1'}(\hat{\mathbf r}) \le \partial_{r_2}S_{\nu_1'}(\hat{\mathbf r})
  \le \cdots \le \partial_{r_{m_c}}S_{\nu_1'}(\hat{\mathbf r}). \label{eq:gap_C_gradient_order}
\end{align}
Therefore,
\begin{align}
  S_{\nu_1'}(\tilde{\mathbf r})
   & \overset{(a)}{\le}S_{\nu_1'}(\hat{\mathbf r}) +\sum_{\ell=1}^{m_c}
  \partial_{r_\ell}S_{\nu_1'}(\hat{\mathbf r})(\tilde r_\ell-\hat r_\ell) \nonumber          \\
   & \overset{(b)}{=}R+C +\sum_{\ell=1}^{m_c}
  \left[\partial_{r_\ell}S_{\nu_1'}(\hat{\mathbf r})
  -\partial_{r_{k_c}}S_{\nu_1'}(\hat{\mathbf r})\right](\tilde r_\ell-\hat r_\ell) \nonumber \\
   & \overset{(c)}{\le}R+C \nonumber                                                         \\
   & \overset{(d)}{=}S_{\frac{2}{\sqrt{\beta}}}(\tilde{\mathbf r}),
  \label{eq:gap_C_response_comparison}
\end{align}
where step $(a)$ comes from the supporting-hyperplane inequality for the concave
function $S_{\nu_1'}(\mathbf r)$ and $\hat{r}_\ell=\tilde{r}_\ell=0$ for $\ell>m_c$;
step $(b)$ follows by the identity $S_{\nu_1'}(\hat{\mathbf r})=\sum_{\ell=1}^L\hat{r}_\ell'=R+C$
and Eq.~\eqref{eq:gap_C_zero_sum}; step $(c)$ is based on
Eqs.~\eqref{eq:gap_C_single_crossing} and~\eqref{eq:gap_C_gradient_order},
which make every product in the preceding sum non-positive;
and step $(d)$ is based on Eqs.~\eqref{eq:def_r_prime} and~\eqref{eq:lambda_condition}.

Therefore, $\nu_1'\ge \frac{2}{\sqrt{\beta}}$, which implies
$\partial\Delta D(R,C,0)/(\partial C)\ge0$ for almost every $C>0$.
Since $\Delta D(R,C,0)$ is continuous on $(0,\infty)$, it is non-decreasing on $C>0$.
Cor.~\ref{cor:tightness_extremes} gives $\Delta D(R,0,0)=0$,
while Thm.~\ref{thm:greedy_solution} gives $\Delta D(R,C,0)\ge0$.
Hence, the non-decreasing property can be extended to $C=0$.
For $R=0$, the gap is identically zero, as shown in
Appendix~\ref{app:monotonicity_P}, which completes the proof.

\section{Proof of Lemma~\ref{lem:sub_convexity_U}}\label{app:sub_convexity_U}
If all eigenvalues are equal, $U(R,\Gamma)\equiv0$, so we concentrate on the case
where the eigenvalues are not all equal. In this case, to prove that only local minima
can exist within the open interval, we
demonstrate that $\left.\frac{\partial^2 U(R,\Gamma)}{\partial R^2}\right|_{R=R_0}>0$
holds for an arbitrary stationary point $R_0\in(R_k,R_{k+1})$.

For any interior stationary point $R_0\in(R_k,R_{k+1})$ satisfying
$\frac{\partial U(R,\Gamma)}{\partial R}\big|_{R=R_0} = 0$, differentiating
Eq.~\eqref{eq:U_R_definition} using Eqs.~\eqref{eq:alpha_condition}
and~\eqref{eq:opt_D_multiplier} yields
\begin{equation}\label{eq:stationary_condition}
  \frac{\nu_0}{\alpha_0} = \frac{1}{2k}\sum_{\ell=1}^k x_\ell,
\end{equation}
where $x_\ell\triangleq\sqrt{\frac{\gamma_\ell}{\gamma_\ell-\alpha_0}}>1$,
and $\alpha_0$, $\nu_0$ denote the values of $\alpha$, $\nu_2$ at $R=R_0$, respectively.
Furthermore, the second-order derivative of $U(R,\Gamma)$ at $R=R_0$ can be expressed as
\begin{equation}\label{eq:U_second_derivative}
  \left.\frac{\partial^2 U(R,\Gamma)}{\partial R^2}\right|_{R=R_0} = 4(\ln2)^2
  \left[\frac{\alpha_0}{2k^2}\sum_{\ell=1}^k(x_\ell^3+x_\ell)-\frac{2\nu_0}{S(\nu_0)}\right],
\end{equation}
where
$S(\nu)\triangleq\sum_{\ell=1}^L\Bigl(1-\frac{\nu}{\sqrt{\gamma_\ell^2+\nu^2}}\Bigr)>0$.
To prove that Eq.~\eqref{eq:U_second_derivative} is positive, it suffices to show that
\begin{equation}
  S(\nu_0)\sum_{\ell=1}^k(x_\ell^3+x_\ell) > \frac{4k^2\nu_0}{\alpha_0}.
\end{equation}

To this end, define an auxiliary function
\begin{equation}
  F(x)\triangleq 1-\frac{\nu_0(x^2-1)}{\sqrt{\alpha_0^2x^4+\nu_0^2(x^2-1)^2}},
\end{equation}
for $x>1$, which is positive and strictly convex on $(1,\infty)$. It leads to
\begin{align}
  S(\nu_0)\sum_{\ell=1}^k(x_\ell^3+x_\ell)
   & \overset{(a)}{\ge}\left(\sum_{\ell=1}^k F(x_\ell)\right)
  \left(\sum_{\ell=1}^k(x_\ell^3+x_\ell)\right) \nonumber                               \\
   & \overset{(b)}{\ge}k^2F\!\left(\frac{2\nu_0}{\alpha_0}\right)
  \left[\left(\frac{2\nu_0}{\alpha_0}\right)^3+\frac{2\nu_0}{\alpha_0}\right] \nonumber \\
   & \overset{(c)}{=} \frac{4k^2\nu_0}{\alpha_0}, \label{eq:second_derivative_chain}
\end{align}
where step $(a)$ follows from $F(x_\ell)=1-\nu_0/\sqrt{\gamma_\ell^2+\nu_0^2}$ for $\ell\le k$
and the positivity of all summands; step $(b)$ applies Jensen's inequality separately to the
positive, strictly convex functions $F(x)$ and $x^3+x$; and step $(c)$ is due to
$F(2\nu_0/\alpha_0)=2\alpha_0^2/(\alpha_0^2+4\nu_0^2)$, where $2\nu_0/\alpha_0>1$ by
Eq.~\eqref{eq:stationary_condition}.
At least one of steps $(a)$ and $(b)$ is strict: if $k<L$, the omitted summands
in $S(\nu_0)$ are strictly positive; if $k=L$, the $x_\ell$ are not all equal
because the eigenvalues are not all equal, so Jensen's inequality is strict.

Therefore, when the eigenvalues are not all equal, every interior stationary point
is a strict local minimum, which means $U(R,\Gamma)$ cannot admit any local
maximum within the open interval $(R_k,R_{k+1})$. This completes the proof.

\section{Proof of Lemma~\ref{lem:related_properties_cor9}}\label{app:gap_pointwise_bound}
\subsubsection*{1) Property~(\ref{item:cor9_1})}
Property~\ref{item:cor9_1} reduces the optimization over arbitrary finite supports $M$
to at most two support points without changing the supremum. This reduction exploits the
linearity of the objective and constraints in the weights $\bm w$, with $\bar\nu_2$ and
$\tilde{\bm\gamma}$ fixed. Specifically, our proof constructs a monotone path of feasible
solutions along which the objective is non-decreasing. Whenever more than two weights are
positive, a perturbation preserving both constraints can eliminate at least one positive
weight without decreasing the objective. Repeating this procedure reduces any finite-support
feasible solution to one supported on at most two points, thereby establishing the
desired result.

We first present an explicit construction of a feasible solution with fewer dimensions.
For any feasible solutions $(\bar{\nu}_2,\bm w,\tilde{\bm\gamma})\in\mathcal A_m^{(M)}$
with $M>2$, we can always find a nonzero vector $\bm a=(a_1,a_2,\ldots,a_M)$ satisfying
\begin{align}\label{eq:cor9_weight_direction1}
  \sum_{i=1}^{M}a_i\tilde{\gamma}_i=\sum_{i=1}^{M}a_i\varphi_2(\bar{\nu}_2,\tilde{\gamma}_i)=0,
\end{align}
and
\begin{align}
  \sum_{i=1}^{M}a_i\varphi_1(\bar{\nu}_2,\tilde{\gamma}_i) \geq 0. \label{eq:cor9_weight_direction2}
\end{align}
Since $\tilde{\gamma}_i>0$, Eq.~\eqref{eq:cor9_weight_direction1} implies $\min_i a_i<0<\max_i a_i$.
Set $t_*\triangleq\min_{i:a_i<0}(\frac{w_i}{-a_i})>0$ and $w_i^+\triangleq w_i+t_*a_i$.
Then,
\begin{align}
   & \min_{1\leq i\leq M}w_i^+=\min_{1\leq i\leq M}(w_i+t_*a_i)\overset{(a)}{=}0,           \\
   & \sum_{i=1}^{M}w_i^+\tilde{\gamma}_i=\sum_{i=1}^{M}w_i\tilde{\gamma}_i
  +t_*\sum_{i=1}^{M}a_i\tilde{\gamma}_i\overset{(b)}{=}1,                                   \\
   & \sum_{i=1}^{M}w_i^+\varphi_2(\bar{\nu}_2,\tilde{\gamma}_i)
  =\sum_{i=1}^{M}w_i\varphi_2(\bar{\nu}_2,\tilde{\gamma}_i) \nonumber                       \\
   & \qquad + t_*\sum_{i=1}^{M}a_i\varphi_2(\bar{\nu}_2,\tilde{\gamma}_i)\overset{(b)}{=}0,
\end{align}
where step $(a)$ follows from the definition of $t_*$ and $w_i^+$, which
ensures that all updated weights are non-negative and at least one
vanishes; and step $(b)$ is based on Eq.~\eqref{eq:cor9_weight_direction1}.
Besides,
\begin{align}
  \sum_{i=1}^{M}w_i^+\varphi_1(\bar{\nu}_2,\tilde{\gamma}_i)
   & =\sum_{i=1}^{M}(w_i+t_*a_i)\varphi_1(\bar{\nu}_2,\tilde{\gamma}_i) \nonumber \\
   & =\sum_{i=1}^{M}w_i\varphi_1(\bar{\nu}_2,\tilde{\gamma}_i)
  +t_*\sum_{i=1}^{M}a_i\varphi_1(\bar{\nu}_2,\tilde{\gamma}_i)\nonumber           \\
   & \overset{(a)}{\geq}\sum_{i=1}^{M}w_i\varphi_1(\bar{\nu}_2,\tilde{\gamma}_i),
  \label{eq:cor9_weight_reduction}
\end{align}
where step $(a)$ follows by $t_*>0$ and Eq.~\eqref{eq:cor9_weight_direction2}.
Thus, deleting the zero-weight terms preserves feasibility and does
not decrease the objective.

Repeating this construction yields a feasible solution with at most two positive
weights after at most $M-2$ iterations. Consequently,
\begin{align}
   & \sup_{\substack{M\in\mathbb N                                  \\
  (\bar{\nu}_2,\bm w,\tilde{\bm\gamma})\in\mathcal A_m^{(M)}}}
  \sum_{i=1}^{M}w_i\varphi_1(\bar{\nu}_2,\tilde{\gamma}_i)\nonumber \\
   & \overset{(a)}{\leq}\sup_{\substack{M\in\{1,2\}                 \\
  (\bar{\nu}_2,\bm w,\tilde{\bm\gamma})\in\mathcal A_m^{(M)}}}
  \sum_{i=1}^{M}w_i\varphi_1(\bar{\nu}_2,\tilde{\gamma}_i)\nonumber \\
   & \overset{(b)}{=}
  \sup_{(\bar{\nu}_2,\bm w,\tilde{\bm\gamma})\in\mathcal A_m^{(2)}}
  \sum_{i=1}^{2}w_i\varphi_1(\bar{\nu}_2,\tilde{\gamma}_i)\nonumber \\
   & \overset{(c)}{\leq}\sup_{M\in\mathbb N}
  \sup_{(\bar{\nu}_2,\bm w,\tilde{\bm\gamma})\in\mathcal A_m^{(M)}}
  \sum_{i=1}^{M}w_i\varphi_1(\bar{\nu}_2,\tilde{\gamma}_i),\label{eq:cor9_finite_two_point_equivalence}
\end{align}
where step $(a)$ is based on the preceding reduction; step $(b)$
follows by allowing the weight $w_i=0$; and step $(c)$ holds trivially
because the outer supremum includes $M=2$. Hence, all quantities
in Eq.~\eqref{eq:cor9_finite_two_point_equivalence} are equal.

\subsubsection*{2) Property~(\ref{item:cor9_2})}
Eqs.~\eqref{eq:cor9_2_1} and \eqref{eq:cor9_2_2} follow by the definition of
$\bar{\nu}_2$ and $\bar{\bm{\gamma}}$, which implies that
\begin{align}
   & \quad \sum_{\ell=1}^{L}\varphi_2(\bar{\nu}_2,\bar{\gamma}_\ell)\nonumber                   \\
   & = \sum_{\ell=1}^{L} \log \left(1+\sqrt{1+\frac{\bar{\gamma}_\ell^2}{\bar{\nu}_2^2}}\right)
  - \sum_{\ell=1}^{L} 1 -\sum_{\ell=1}^{L}\log^{+}\bar{\gamma}_\ell \nonumber                   \\
   & =\sum_{\ell=1}^{L} \log \left(1+\sqrt{1+\frac{\bar{\gamma}_\ell^2}{\bar{\nu}_2^2}}\right)
  -L-2R \nonumber                                                                               \\
   & =0,
\end{align}
and
\begin{align}
  \frac{U(R,\Gamma)}{2\operatorname{tr}(\Gamma)}
   & =\frac{\alpha\sum_{\ell=1}^{L} \left[ \sqrt{\bar{\gamma}_\ell^2+\bar{\nu}_2^2}
      -\bar{\nu}_2 -\sqrt{\bar{\gamma}_\ell (\bar{\gamma}_\ell-1)_+} \right]}
  {\alpha\sum_{\ell=1}^{L}\bar{\gamma}_\ell} \nonumber                              \\
   & =\frac{\sum_{\ell=1}^{L} \varphi_1(\bar{\nu}_2,\bar{\gamma}_\ell)}
  {\sum_{\ell=1}^{L}\bar{\gamma}_\ell}.
\end{align}
Eqs.~\eqref{eq:cor9_2_3}-\eqref{eq:cor9_2_5} come from the definition of
$\tilde{\bm{\gamma}}$, $\bm{w}$ and $\bm{N}$, which implies that
\begin{align}
   & \sum_{i=1}^{M} w_i \varphi_2(\bar{\nu}_2,\tilde{\gamma}_i)
  = \frac{\sum_{i=1}^{M} N_i\varphi_2(\bar{\nu}_2,\tilde{\gamma}_i)}
  {\sum_{\ell=1}^{L}\bar{\gamma}_\ell}
  = \frac{\sum_{\ell=1}^{L} \varphi_2(\bar{\nu}_2,\bar{\gamma}_\ell)}
  {\sum_{\ell=1}^{L}\bar{\gamma}_\ell} \nonumber                                     \\
   & \qquad \qquad \qquad \quad = 0.                                                 \\
   & \sum_{i=1}^{M} w_i\tilde{\gamma}_i
  =\frac{\sum_{i=1}^{M} N_i\tilde{\gamma}_i}{\sum_{\ell=1}^{L}\bar{\gamma}_\ell}
  =\frac{\sum_{\ell=1}^{L} \bar{\gamma}_\ell}{\sum_{\ell=1}^{L}\bar{\gamma}_\ell}=1. \\
   & \sum_{\ell=1}^{L} \varphi_1(\bar{\nu}_2,\bar{\gamma}_\ell) =
  \sum_{i=1}^{M} N_i \varphi_1(\bar{\nu}_2,\tilde{\gamma}_i)  = \sum_{i=1}^{M} w_i
  \varphi_1(\bar{\nu}_2,\tilde{\gamma}_i) \sum_{\ell=1}^{L}\bar{\gamma}_\ell.
\end{align}

\subsubsection*{3) Property~(\ref{item:cor9_3})}
When $\nu>0$ and $\gamma>1$, Eq.~\eqref{eq:cor9_3_1} comes from the fact that
\begin{equation}
  \frac{\partial\varphi_2(\nu,\gamma)}{\partial\gamma}
  = -\frac{\nu}{\gamma\sqrt{\gamma^2+\nu^2}\ln 2} < 0,
\end{equation}
and Eqs.~\eqref{eq:cor9_3_2}-\eqref{eq:cor9_3_5} follow trivially by direct
computation of the derivatives and limits.

\subsubsection*{4) Property~(\ref{item:cor9_4})}
In this part, we show that the supremum in problem (P3) is attained at an interior point
of $\mathcal{D}$. Specifically, we construct a nonempty positive superlevel set which
is bounded and stays away from the boundary of $\mathcal{D}$. The maximization
can therefore be restricted to a compact subset of the interior, where continuity
guarantees attainment.

For every feasible $(\bar{\nu}_2,\tilde{\gamma}_1,w_1,w_2)$ of (P3), the two equality
constraints in problem (P3) uniquely determine
\begin{align}
  w_1 & = \frac{\varphi_2(\bar{\nu}_2,1)}{\tilde{\gamma}_1\varphi_2(\bar{\nu}_2,1)
  -\varphi_2(\bar{\nu}_2,\tilde{\gamma}_1)}, \label{eq:cor9_w1}                                    \\
  w_2 & = \frac{-\varphi_2(\bar{\nu}_2,\tilde{\gamma}_1)}{\tilde{\gamma}_1\varphi_2(\bar{\nu}_2,1)
    -\varphi_2(\bar{\nu}_2,\tilde{\gamma}_1)}. \label{eq:cor9_w2}
\end{align}
Substituting these expressions into the objective~\eqref{eq:rp_obj1} gives
\begin{align}
   & \quad w_1\varphi_1(\bar{\nu}_2,\tilde{\gamma}_1) +w_2\varphi_1(\bar{\nu}_2,1) \nonumber   \\
   & = \frac{\varphi_2(\bar{\nu}_2,1)\varphi_1(\bar{\nu}_2,\tilde{\gamma}_1)
    -\varphi_2(\bar{\nu}_2,\tilde{\gamma}_1)\varphi_1(\bar{\nu}_2,1)}
  {\tilde{\gamma}_1\varphi_2(\bar{\nu}_2,1)-\varphi_2(\bar{\nu}_2,\tilde{\gamma}_1)} \nonumber \\
   & \triangleq \mathcal Q(\bar{\nu}_2,\tilde{\gamma}_1). \label{eq:cor9_Q_def}
\end{align}

The parameter pair $(1,3)$, with the weights determined by
Eqs.~\eqref{eq:cor9_w1}-\eqref{eq:cor9_w2}, is feasible for (P3), and
$\mathcal Q(1,3)\approx0.1047>0$. Therefore, it suffices to show that
the positive superlevel set
\begin{align}\label{eq:cor9_positive_superlevel}
  \Big\{(\bar\nu_2,\tilde\gamma_1,w_1,w_2)
   & \in\mathcal{D}:\eqref{eq:rp_c1}, \eqref{eq:rp_c2}, \nonumber                            \\
   & \qquad \quad \mathcal Q(\bar\nu_2,\tilde\gamma_1)\geq \frac{1}{2} \mathcal Q(1,3)\Big\}
\end{align}
is compact and contained in the interior of $\mathcal D$.
For every feasible point of (P3), we have
\begin{align}
  \mathcal Q(\bar\nu_2,\tilde\gamma_1)
   & \overset{(a)}{\leq} w_2\varphi_1(\bar\nu_2,1) \leq \varphi_1(\bar\nu_2,1) \nonumber     \\
   & \qquad = \frac{1}{\sqrt{1+\bar\nu_2^2}+\bar\nu_2} \xrightarrow[\bar\nu_2\to\infty]{} 0,
  \label{eq:cor9_large_nu_bound}
\end{align}
and
\begin{align}
  \mathcal Q(\bar\nu_2,\tilde\gamma_1)
   & \le w_2\varphi_1(\bar\nu_2,1) \overset{(b)}{<} \frac{\log(2\bar\nu_2)\varphi_1(\bar\nu_2,1)}
  {\tilde\gamma_1\varphi_2(\bar\nu_2,1)}\nonumber                                                 \\
   & < \frac{\log(2\bar\nu_2)\varphi_1(\bar\nu_2,1)}{\gamma_0(\bar\nu_2)\varphi_2(\bar\nu_2,1)}
  \xrightarrow[\bar\nu_2\downarrow1/2]{} 0, \label{eq:cor9_small_nu_bound}
\end{align}
where step $(a)$ is due to $\varphi_1(\bar\nu_2,\tilde\gamma_1)<0$ for
$\tilde\gamma_1>\gamma_0(\bar\nu_2)$ by Eq.~\eqref{eq:cor9_phi_def}, and step $(b)$ substitutes
Eq.~\eqref{eq:cor9_w2} to eliminate $w_2$ and uses the inequality
$0<-\varphi_2(\bar\nu_2,\tilde\gamma_1)<\log(2\bar\nu_2)$ following
from Eqs.~\eqref{eq:cor9_3_1} and~\eqref{eq:cor9_3_5}. Both bounds are uniform over all
admissible $\tilde\gamma_1$, so Eq.~\eqref{eq:cor9_positive_superlevel} requires
$\bar\nu_2\in[a,b]$ for some $1/2<a<b<\infty$. On this interval, the same estimate gives
\begin{equation}
  \mathcal Q(\bar\nu_2,\tilde\gamma_1)
  \le\frac{\log(2b)\varphi_1(a,1)}
  {\tilde\gamma_1\varphi_2(b,1)}
  \xrightarrow[\tilde\gamma_1\to\infty]{}0,
\end{equation}
where the bound is uniform in $\bar\nu_2\in[a,b]$, since
$\varphi_1(\nu,1)$ and $\varphi_2(\nu,1)$ are positive and decreasing in $\nu$.
Together with Eq.~\eqref{eq:rp_c2}, these bounds show that the
set~\eqref{eq:cor9_positive_superlevel} is bounded.
At any limit point of the set~\eqref{eq:cor9_positive_superlevel}, the denominator
\begin{align}
  \tilde\gamma_1\varphi_2(\bar\nu_2,1)-\varphi_2(\bar\nu_2,\tilde\gamma_1)\ge\varphi_2(b,1)>0,
\end{align}
thus $\mathcal Q(\bar\nu_2,\tilde\gamma_1)$ is continuous in the limit. Besides, the equality
$\varphi_1(\bar\nu_2,\gamma_0(\bar\nu_2))=\varphi_2(\bar\nu_2,\gamma_0(\bar\nu_2))=0$ gives
\begin{align}
  \mathcal Q(\bar\nu_2,\gamma_0(\bar\nu_2))=0<\tfrac12\mathcal Q(1,3)
  \leq \mathcal Q(\bar\nu_2,\tilde\gamma_1).
\end{align}
Hence, $\tilde\gamma_1>\gamma_0(\bar\nu_2)$ at every limit point. The equality constraints
are preserved by continuity, and Eqs.~\eqref{eq:cor9_w1}-\eqref{eq:cor9_w2} give $0<w_1,w_2<1$.
Every limit point remains in the set~\eqref{eq:cor9_positive_superlevel} and in the interior
of $\mathcal D$, establishing the closedness of this bounded set and hence its compactness.
Restricting (P3) to this set preserves the supremum, and the continuity guarantees that the
supremum can be attained. This completes the proof.

\section{Proof of Lemma~\ref{lem:cor9_kkt_uniqueness}}\label{app:cor9_kkt_uniqueness}
In this part, we will show that the system~\eqref{eq:cor9_kkt_w1}-\eqref{eq:cor9_kkt_nu}
has exactly one solution on the feasible domain $\mathcal{D}$ by analyzing the
monotonicity and endpoint values of the relative functions.

We first reduce the full KKT system to an equivalent pair of scalar equations in
$(\bar{\nu}_2,\tilde{\gamma}_1)$ by eliminating the weights and Lagrange multipliers,
which are uniquely recoverable from these two variables within the admissible domain.
From the KKT condition~\eqref{eq:cor9_kkt_nu} and  Eq.~\eqref{eq:cor9_3_2}, we have
\begin{align}
  0 & = w_1^*\big[ \varphi_{1\nu}(\bar{\nu}_2^*,\tilde{\gamma}_1^*) -\mu_1
  \varphi_{2\nu}(\bar{\nu}_2^*,\tilde{\gamma}_1^*) \big]  \nonumber        \\
    & \qquad + w_2^*\big[\varphi_{1\nu}(\bar{\nu}_2^*,1)
  -\mu_1\varphi_{2\nu}(\bar{\nu}_2^*,1) \big] \nonumber                    \\
    & = (\bar{\nu}_2^*\ln 2-\mu_1)
  \big[w_1^*\varphi_{2\nu}(\bar{\nu}_2^*,\tilde{\gamma}_1^*)
    + w_2^*\varphi_{2\nu}(\bar{\nu}_2^*,1) \big].
\end{align}
Since
$w_1^*\varphi_{2\nu}(\bar{\nu}_2^*,\tilde{\gamma}_1^*)+w_2^*\varphi_{2\nu}(\bar{\nu}_2^*,1) < 0$,
we have $\mu_1 = \bar{\nu}_2^*\ln 2$.
By the auxiliary function $\varphi(\nu,\gamma)$ defined in Eq.~\eqref{eq:H_nu_gamma_def},
the KKT conditions~\eqref{eq:cor9_kkt_w1}-\eqref{eq:cor9_kkt_gamma} can be reformulated as
\begin{align}
   & \varphi(\bar{\nu}_2^*,1)=\mu_2, \label{eq:cor9_H_tangency_1}                           \\
   & \varphi(\bar{\nu}_2^*,\tilde{\gamma}_1^*) =\mu_2\tilde{\gamma}_1^*,                    \\
   & \varphi_{\gamma}(\bar{\nu}_2^*,\tilde{\gamma}_1^*)=\mu_2. \label{eq:cor9_H_tangency_3}
\end{align}
Thus, every interior maximizer must satisfy~\eqref{eq:universal_gap_constant},
which can be equivalently expressed as
\begin{align}
  \varphi(\bar{\nu}_2^*,\tilde{\gamma}_1^*) -\tilde{\gamma}_1^* \varphi_{\gamma}(\bar{\nu}_2^*,\tilde{\gamma}_1^*)
   & =0, \label{eq:cor9_first_scalar}  \\
  \varphi_{\gamma}(\bar{\nu}_2^*,\tilde{\gamma}_1^*) -\varphi(\bar{\nu}_2^*,1)
   & =0. \label{eq:cor9_second_scalar}
\end{align}
Remark that the Eqs.~\eqref{eq:cor9_H_tangency_1}-\eqref{eq:cor9_H_tangency_3} have
a clear geometric interpretation, which states that the line $y=\mu_2\gamma$ passes
through the point $(1,\varphi(\bar{\nu}_2^*,1))$ and is tangent to
$\varphi(\bar{\nu}_2^*,\gamma)$ at $\gamma=\tilde{\gamma}_1^*$.

To show the existence and uniqueness of the KKT point, it suffices to show that the
system of~\eqref{eq:cor9_first_scalar}-\eqref{eq:cor9_second_scalar} has a unique solution
$(\bar{\nu}_2^*,\tilde{\gamma}_1^*)$ in the admissible domain $\mathcal{D}$.

We first show that, for each fixed $\bar{\nu}_2>1/2$, Eq.~\eqref{eq:cor9_first_scalar}
has exactly one root strictly larger than $\gamma_0(\bar{\nu}_2)$.
Let $\tilde\gamma_1'>\gamma_0(\bar\nu_2)$ denote the unique solution of
\begin{equation}
  \frac{\sqrt{\tilde\gamma_1(\tilde\gamma_1^2+\bar\nu_2^2)}}{4(\tilde\gamma_1-1)^{3/2}}
  = \bar\nu_2^2.
\end{equation}
Its existence and uniqueness follow because the left-hand side is strictly decreasing
from $\bar\nu_2^2(4\bar\nu_2^2+1)/2>\bar\nu_2^2$ at $\gamma_0(\bar\nu_2)$ to
$1/4<\bar\nu_2^2$ as $\tilde\gamma_1\to\infty$. Consequently, for every
$\tilde\gamma_1>\gamma_0(\bar\nu_2)$,
\begin{align}
   & \quad \frac{\partial}{\partial\tilde\gamma_1}\bigl[\varphi(\bar\nu_2,\tilde\gamma_1)-
  \tilde\gamma_1\varphi_\gamma(\bar\nu_2,\tilde\gamma_1)\bigr]>0 \nonumber                     \\
   & \overset{(a)}{\iff} \varphi_{\gamma\gamma}(\bar\nu_2,\tilde\gamma_1)<0 \nonumber          \\
   & \iff\frac{\sqrt{\tilde\gamma_1(\tilde\gamma_1^2+\bar\nu_2^2)}}
  {4(\tilde\gamma_1-1)^{3/2}} < \bar\nu_2^2 \nonumber                                          \\
   & \overset{(b)}{\iff} \tilde\gamma_1>\tilde\gamma_1', \label{eq:cor9_first_scalar_monotone}
\end{align}
where step $(a)$ is based on the Eq.~\eqref{eq:cor9_3_3}, while step $(b)$ follows from
strict monotonicity of the left-hand side and the definition of $\tilde\gamma_1'$.
On the other hand, the endpoints of the left-hand side
of Eq.~\eqref{eq:cor9_first_scalar} satisfy
\begin{align}
   & \lim_{\tilde{\gamma}_1 \downarrow \gamma_0(\bar{\nu}_2)} \big[\varphi(\bar{\nu}_2,\tilde{\gamma}_1)
    - \tilde{\gamma}_1 \varphi_{\gamma}(\bar{\nu}_2,\tilde{\gamma}_1)\big]
  = 0, \label{eq:cor9_first_scalar_endpoints}                                                            \\
   & \lim_{\tilde{\gamma}_1 \to \infty} \big[ \varphi(\bar{\nu}_2,\tilde{\gamma}_1)
    - \tilde{\gamma}_1 \varphi_{\gamma}(\bar{\nu}_2,\tilde{\gamma}_1) \big]
  = \frac{1}{2} -\bar{\nu}_2 \nonumber                                                                   \\
   & \qquad + \bar{\nu}_2\ln 2\,\log(2\bar{\nu}_2) > 0.
  \label{eq:cor9_first_scalar_endpoints2}
\end{align}
Therefore, Eqs.~\eqref{eq:cor9_first_scalar_monotone}, \eqref{eq:cor9_first_scalar_endpoints}
and \eqref{eq:cor9_first_scalar_endpoints2}
imply that the left-hand side of~\eqref{eq:cor9_first_scalar} first decreases from
zero, then increases to a positive limit, and hence has exactly one root $\tilde{\gamma}_1^*$
when $\tilde{\gamma}_1 \in (\gamma_0(\bar{\nu}_2),\infty)$ for any fixed $\bar{\nu}_2>1/2$.

It remains to show the uniqueness of the root for Eq.~\eqref{eq:cor9_second_scalar}
when $\tilde{\gamma}_1$ takes the value $\tilde{\gamma}_1^*$.
Notice that the derivative of the left-hand side of
Eq.~\eqref{eq:cor9_second_scalar} satisfies
\begin{align}
   & \quad \frac{\mathrm{d}}{\mathrm{d}\bar{\nu}_2}
  \big[ \varphi_{\gamma}(\bar{\nu}_2,\tilde{\gamma}_1^*) -\varphi(\bar{\nu}_2,1) \big]\nonumber      \\
   & =\varphi_{\gamma\nu}(\bar{\nu}_2,\tilde{\gamma}_1^*)
  +\varphi_{\gamma\gamma}(\bar{\nu}_2,\tilde{\gamma}_1^*)
  \frac{\mathrm{d}\tilde{\gamma}_1^*}{\mathrm{d}\bar{\nu}_2} -\varphi_{\nu}(\bar{\nu}_2,1) \nonumber \\
   & \overset{(a)}{=}\frac{ \varphi_{\nu}(\bar{\nu}_2,\tilde{\gamma}_1^*)}{\tilde{\gamma}_1^*}
  - \varphi_{\nu}(\bar{\nu}_2,1) \nonumber                                                           \\
   & =\ln 2 \left[ \varphi_2(\bar{\nu}_2,1)
  -\frac{\varphi_2(\bar{\nu}_2,\tilde{\gamma}_1^*)}{\tilde{\gamma}_1^*} \right]\nonumber             \\
   & \overset{(b)}{>}0, \label{eq:cor9_second_scalar_monotone}
\end{align}
where step $(a)$ comes from
$\frac{\mathrm{d}}{\mathrm{d}\bar{\nu}_2} \big[ \varphi(\bar{\nu}_2,\tilde{\gamma}_1^*)
    -\tilde{\gamma}_1^* \varphi_{\gamma}(\bar{\nu}_2,\tilde{\gamma}_1^*) \big] = 0$,
and step $(b)$ follows from $\varphi_2(\bar{\nu}_2,1)>0> \varphi_2(\bar{\nu}_2,\tilde{\gamma}_1^*)$.
On the other hand, the endpoints of the left-hand side of
Eq.~\eqref{eq:cor9_second_scalar} satisfy
\begin{align}
   & \lim_{\bar{\nu}_2\downarrow 1/2} \big[\varphi_{\gamma}(\bar{\nu}_2,\tilde{\gamma}_1^*)
  -\varphi(\bar{\nu}_2,1) \big]  \nonumber                                                                                          \\
   & \qquad \qquad =\frac{1-\sqrt{5}}{2}+\frac{\ln 2}{2} \log\bigg(\frac{1+\sqrt{5}}{2}\bigg)<0, \label{eq:cor9_second_scalar_left}
\end{align}
and
\begin{align}
   & \quad \lim_{\bar{\nu}_2\to\infty} \big[\varphi_{\gamma}(\bar{\nu}_2,\tilde{\gamma}_1^*)
  - \varphi(\bar{\nu}_2,1) \big] \nonumber                                                   \\
   & = \lim_{\bar{\nu}_2\to\infty} \left(
  \frac{\sqrt{\tilde{\gamma}_1^{*2}+\bar{\nu}_2^{2}}}{\tilde{\gamma}_1^*} -
  \frac{2\tilde{\gamma}_1^*-1}{2\sqrt{\tilde{\gamma}_1^*(\tilde{\gamma}_1^*-1)}}
  -\frac{1}{4\bar{\nu}_2} \right) \nonumber                                                  \\
   & \geq \lim_{\bar{\nu}_2\to\infty} \left(\frac{\bar{\nu}_2}{3} -\frac{3}{2\sqrt{2}} -
  \frac{1}{4\bar{\nu}_2}\right) \to \infty,\label{eq:cor9_second_scalar_right}
\end{align}
where the last inequality follows by the fact that $2<\tilde{\gamma}_1^*<3$
when $\bar{\nu}_2 \to \infty$, which comes from the uniqueness of
$\tilde{\gamma}_1^*$ and
\begin{align}
  \lim_{\bar{\nu}_2\to\infty}
  \frac{\varphi(\bar{\nu}_2,2)-2\varphi_{\gamma}(\bar{\nu}_2,2)}{\bar{\nu}_2} & = \ln 2-1<0,       \\
  \lim_{\bar{\nu}_2\to\infty}
  \frac{\varphi(\bar{\nu}_2,3)-3\varphi_{\gamma}(\bar{\nu}_2,3)}{\bar{\nu}_2} & = \ln 2\log 3-1>0.
\end{align}
Therefore, Eqs.~\eqref{eq:cor9_second_scalar_monotone}-\eqref{eq:cor9_second_scalar_right}
ensure that~\eqref{eq:cor9_second_scalar} has exactly
one solution $\bar{\nu}_2^*\in(1/2,\infty)$ when $\tilde{\gamma}_1=\tilde{\gamma}_1^*$.
Hence, the KKT system has exactly a unique feasible solution.

\section{Proof of Lemma~\ref{lem:monotonicity_P_eta}}\label{app:monotonicity_P_eta}
Fix $R>0$ and $C\geq 0$, so that $\tilde P$ is independent of $P$.
According to Appendix~\ref{app:monotonicity_P}, when $P\geq\tilde P$, both $D(R,C,P)$
and $\tilde D(R,C,P)$ equal $\underline D(R)$, hence $\eta(R,C,P)=1$. Therefore, we will
focus on evaluating the partial derivative of $\eta(R,C,P)$ w.r.t. $P$ on $0<P<\tilde P$
in the following analysis, then extend the conclusion to the endpoints by continuity.

To simplify the notation, let
\begin{equation}
  g(\mathbf r)=\sum_{\ell=1}^L\gamma_\ell 2^{-2r_\ell},
\end{equation}
and abbreviate $D(R,C,P)$ and $\tilde D(R,C,P)$ as $D$ and $\tilde D$,
respectively. In the active perception regime,
\begin{align}
  D        & =g(\hat{\mathbf r})+\bigl(\sqrt{\hat T}-\sqrt P\bigr)^2,     \\
  \tilde D & =g(\tilde{\mathbf r})+\bigl(\sqrt{\tilde P}-\sqrt P\bigr)^2.
\end{align}
By Eqs.~\eqref{eq:D_tilde_derivate_P} and~\eqref{eq:D_derivate_P},
at every point in this interval where $D$ and $\tilde D$ are differentiable,
\begin{equation}\label{eq:eta_derivative_P}
  \frac{\partial\eta(R,C,P)}{\partial P} = \frac{(\sqrt{\tilde P}-\sqrt P)D
    -(\sqrt{\hat T}-\sqrt P)\tilde D}{\sqrt P\,\tilde D^2}.
\end{equation}
Using the preceding distortion representations, we have
\begin{align}
   & \quad \bigl(\sqrt{\tilde P}-\sqrt P\bigr)D
  -\bigl(\sqrt{\hat T}-\sqrt P\bigr)\tilde D \nonumber            \\
   & =\bigl(\sqrt{\tilde P}-\sqrt{\hat T}\bigr)
  \Bigl[g(\tilde{\mathbf r}) -\bigl(\sqrt{\tilde P}-\sqrt P\bigr)
  \bigl(\sqrt{\hat T}-\sqrt P\bigr)\Bigr] \nonumber               \\
   & \qquad+\bigl(\sqrt{\tilde P}-\sqrt P\bigr)
  \bigl(g(\hat{\mathbf r})-g(\tilde{\mathbf r})\bigr) \nonumber   \\
   & \overset{(a)}{\geq}\bigl(\sqrt{\tilde P}-\sqrt{\hat T}\bigr)
  \Bigl[g(\tilde{\mathbf r})-\bigl(\sqrt{\tilde P}-\sqrt P\bigr)
  \bigl(\sqrt{\hat T}-\sqrt P\bigr)\Bigr] \nonumber               \\
   & \overset{(b)}{\geq}\bigl(\sqrt{\tilde P}-\sqrt P\bigr)
  \bigl(\sqrt{\tilde P}-\sqrt{\hat T}\bigr)^2 \nonumber           \\
   & \geq 0, \label{eq:eta_derivative_P_nonnegativity}
\end{align}
where step $(a)$ follows from $g(\hat{\mathbf r})\geq g(\tilde{\mathbf r})$
in Eq.~\eqref{eq:opt_distortion_to_greedy}; step $(b)$ uses
$\hat T\leq\tilde P$ established in Appendix~\ref{app:monotonicity_P} and
the inequality
\begin{align}
  g(\tilde{\mathbf r})
   & \overset{(c)}{=}\sum_{\ell=1}^L\gamma_\ell
  \bigl[2-2^{-2\tilde r_\ell}-2\psi(\tilde r_\ell,\tilde r_\ell)\bigr] \nonumber  \\
   & \overset{(d)}{\geq}\sum_{\ell=1}^L\gamma_\ell
  \bigl[2-2^{-2\tilde r_\ell}-2\psi(\tilde r_\ell,\tilde r'_\ell)\bigr] \nonumber \\
   & =\tilde P \nonumber                                                          \\
   & \geq\bigl(\sqrt{\tilde P}-\sqrt P\bigr)^2,
\end{align}
step $(c)$ substitutes the definition of $\psi(r_\ell,r'_\ell)$ in
Prop.~\ref{prop:original_problem}; and step $(d)$ follows from
$\tilde r_\ell\leq\tilde r'_\ell$ in Cor.~\ref{cor:properties_greedy}
and the monotonicity of $\psi(r_\ell,r'_\ell)$ in its second argument.

By Appendix~\ref{app:monotonicity_P}, $D$ and $\tilde D$ are continuous for $P\ge0$
and locally Lipschitz for $P>0$. Since $\tilde D\ge g(\tilde{\mathbf r})=\underline D(R)>0$,
the ratio is also locally Lipschitz and absolutely continuous on $(0,\tilde P)$.
Eqs.~\eqref{eq:eta_derivative_P} and~\eqref{eq:eta_derivative_P_nonnegativity}
give its derivative non-negative almost everywhere, so the ratio is
non-decreasing on this interval. Continuity extends the conclusion to $[0,\tilde P]$.
Together with $\eta(R,C,P)=1$ for $P\ge\tilde P$, this completes the proof.

\section{Proof of Lemma~\ref{lem:related_properties_cor11}}\label{app:ratio_pointwise_bound}
\subsubsection*{1) Property~(\ref{item:cor11_1})}
To verify the existence and uniqueness of $\tilde{\nu}_2$
defined as the root of Eq.~\eqref{eq:nu1_star}, we analyze the monotonicity
and endpoint behavior of the residual function to Eq.~\eqref{eq:nu1_star}. Let
\begin{align}
  F(\nu) \triangleq \nu \ln\Big[2 \nu \Big(\nu+\sqrt{1+\nu^2}\Big)\Big]
  \!-\! \sqrt{1+\nu^2} \!-\! \nu \!+\! 1. \label{eq:ratio_scalar_residual}
\end{align}
For $\nu>1/2$,
\begin{align}
   & F'(\nu)=\ln\left(2\nu\left(\nu+\sqrt{1+\nu^2}\right)\right)>0,                   \\
   & F\left(\frac{1}{2}\right)=\frac{1}{2}\ln\frac{1+\sqrt5}{2}-\frac{\sqrt5-1}{2}<0, \\
   & \lim_{\nu\to\infty}F(\nu) = +\infty.
\end{align}
Therefore, $F(\nu)=0$ has exactly one root $\tilde{\nu}_2 \in (\frac{1}{2},\infty)$,
which establishes the existence and uniqueness of $\tilde{\nu}_2$.

\subsubsection*{2) Property~(\ref{item:cor11_2})}
We characterize the range of the proposed constant $\eta^*$ in this part.
Notice that
\begin{align}
  \eta^*
   & = 2\tilde{\nu}_2\bigl(1-\ln(2\tilde{\nu}_2)\bigr)\nonumber                             \\
   & \overset{(a)}{=}1+\tilde{\nu}_2-\sqrt{1+\tilde{\nu}_2^{2}}
  +\tilde{\nu}_2\ln\frac{\tilde{\nu}_2+\sqrt{1+\tilde{\nu}_2^{2}}}{2\tilde{\nu}_2}\nonumber \\
   & =1-\varphi(\tilde{\nu}_2,1),\label{eq:ratio_constant_identity}
\end{align}
where step $(a)$ is a rearrangement of $F(\tilde{\nu}_2)=0$ and implies the strict
positivity of $\eta^*$. Moreover, due to $2\nu(1-\ln(2\nu))<1$ for $\nu>1/2$, there is
\begin{align}
  \lim_{\gamma\to\infty}\varphi(\tilde{\nu}_2,\gamma)
  =\frac{1}{2}-\tilde{\nu}_2+\tilde{\nu}_2\ln(2\tilde{\nu}_2)
  =\frac{1-\eta^*}{2}>0, \label{eq:ratio_H_infinity}
\end{align}
thus $0<\eta^*<1$.

\subsubsection*{3) Property~(\ref{item:cor11_3})}
For every $\nu,\gamma>0$,
\begin{align}
  \sum_{\ell=1}^{L}\varphi(\tilde{\nu}_2,\bar{\gamma}_\ell)
   & \overset{(a)}{\geq} \sum_{\ell=1}^{L}\varphi(\bar{\nu}_2,\bar{\gamma}_\ell)
  +(\tilde{\nu}_2-\bar{\nu}_2)\sum_{\ell=1}^{L}\varphi_\nu(\bar{\nu}_2,\bar{\gamma}_\ell)\nonumber \\
   & \overset{(b)}{=}\sum_{\ell=1}^{L}\varphi(\bar{\nu}_2,\bar{\gamma}_\ell)\nonumber              \\
   & \overset{(c)}{=}\sum_{\ell=1}^{L}\varphi_1(\bar{\nu}_2,\bar{\gamma}_\ell),
\end{align}
where step $(a)$ is due to the convexity of $\varphi(\nu,\gamma)$ w.r.t. $\nu$, which is
guaranteed by $\varphi_{\nu\nu}(\nu,\gamma)>0$; step $(b)$ follows by
$\varphi_\nu(\nu,\gamma) =-\ln2\, \varphi_2(\nu,\gamma)$, Eq.~\eqref{eq:alpha_condition}
and Eq.~\eqref{eq:opt_D_multiplier}, which yield
\begin{align}
  \sum_{\ell=1}^{L}\varphi_2(\bar{\nu}_2,\bar{\gamma}_\ell) = \sum_{\ell=1}^{L}\log
  \left(1+\sqrt{1+\frac{\bar{\gamma}_\ell^2}{\bar{\nu}_2^2}}\right)
   & - \sum_{\ell=1}^{L}\log^+\bar{\gamma}_\ell \nonumber \\
   & - L = 0, \label{eq:ratio_normalized_moment}
\end{align}
and step $(c)$ comes from the definition of $\varphi(\nu,\gamma)$ in
Eq.~\eqref{eq:H_nu_gamma_def}, which implies that
$\varphi(\nu,\gamma)=\varphi_1(\nu,\gamma)-\nu\ln2\,\varphi_2(\nu,\gamma)$.

\subsubsection*{4) Property~(\ref{item:cor11_4})}
To prove the pointwise bound in Eq.~\eqref{eq:ratio_pointwise_bound}, we first express
$\varphi(\tilde{\nu}_2,\gamma)$ as a convex function of $\tilde d(\gamma)$
on each of the intervals $0<\gamma\leq1$ and $\gamma>1$.
The endpoint identities place the endpoints of both branches on the
same line through the origin with slope $1-\eta^*$. Since a convex function
lies below the chord joining its endpoints, this line bounds both branches from
above, thereby yielding the desired result.

\emph{1) $\gamma\in \Ioc{0}{1}$}:
For $0<\gamma\leq1$, we have $\tilde d(\gamma)=\gamma$ and
\begin{align}
  \frac{\partial^2\varphi(\tilde{\nu}_2,\gamma)}{\partial\gamma^2} =
  \frac{\partial}{\partial\gamma}
  \left(\frac{\gamma}{\tilde{\nu}_2+\sqrt{\tilde{\nu}_2^2+\gamma^2}}\right)>0,
\end{align}
when $\gamma \in (0,1)$. Thus, $\varphi(\tilde{\nu}_2,\gamma)$ is convex
on $[0,1]$ under the continuous extension $\varphi(\tilde{\nu}_2,0)=0$, and
\begin{align}
  \varphi(\tilde{\nu}_2,\gamma)
   & \overset{(a)}{\leq} (1-\gamma)\varphi(\tilde{\nu}_2,0)+\gamma\varphi(\tilde{\nu}_2,1)\nonumber \\
   & \overset{(b)}{=} (1-\eta^*)\gamma \nonumber                                                    \\
   & = (1-\eta^*)\tilde d(\gamma), \label{eq:ratio_pointwise_lower_branch}
\end{align}
where step $(a)$ follows from the convexity, and step $(b)$ uses
$\varphi(\tilde{\nu}_2,0)=0$ and Eq.~\eqref{eq:ratio_constant_identity}.

\emph{2) $\gamma\in (1,\infty)$}: We first establish convexity of
$\varphi(\tilde{\nu}_2,\gamma)$ as a function of $\tilde d(\gamma)$.
For $\gamma>1$, there is
\begin{equation}\label{eq:ratio_pointwise_d_derivative}
  \tilde d'(\gamma) = -\frac{\tilde d(\gamma)^2}{2\gamma\sqrt{\gamma(\gamma-1)}}<0.
\end{equation}
Hence, $\tilde d(\gamma)$ decreases strictly from $1$ to $1/2$ and can be
used as the independent variable on this branch. With $\tilde{\nu}_2$
fixed, differentiation w.r.t. $\tilde d(\gamma)$ gives
\begin{align}
  \frac{\mathrm d\varphi(\tilde{\nu}_2,\gamma)}{\mathrm d\tilde d(\gamma)}
   & \triangleq \frac{\partial\varphi(\tilde{\nu}_2,\gamma)/\partial\gamma}{\tilde d'(\gamma)}\nonumber \\
   & \overset{(a)}{=} 1+\frac{\sqrt{1+\tilde{\nu}_2^2/\gamma^2}-1}{\tilde d'(\gamma)}\nonumber          \\
   & \overset{(b)}{=} 1-\frac{2\tilde{\nu}_2^2\sqrt{1-1/\gamma}}{\tilde d(\gamma)^2
    \bigl(1+\sqrt{1+\tilde{\nu}_2^2/\gamma^2}\bigr)}, \label{eq:ratio_pointwise_derivative}
\end{align}
where step $(a)$ follows from
\begin{equation}
  \frac{\partial}{\partial\gamma}\bigl(\varphi(\tilde{\nu}_2,\gamma)-\tilde d(\gamma)\bigr)
  = \sqrt{1+\frac{\tilde{\nu}_2^2}{\gamma^2}}-1,
\end{equation}
and step $(b)$ substitutes Eq.~\eqref{eq:ratio_pointwise_d_derivative}.
The fraction on the last line of Eq.~\eqref{eq:ratio_pointwise_derivative} increases
strictly with $\gamma$, since its numerator is strictly increasing and its positive
denominator is strictly decreasing. Consequently,
$\mathrm d\varphi(\tilde{\nu}_2,\gamma)/\mathrm d\tilde d(\gamma)$
decreases strictly with $\gamma$. As $\tilde d(\gamma)$ also decreases
strictly with $\gamma$, this derivative increases strictly with
$\tilde d(\gamma)$, proving convexity on this branch.

By Eqs.~\eqref{eq:ratio_constant_identity} and~\eqref{eq:ratio_H_infinity},
the graph of $\varphi(\tilde{\nu}_2,\gamma)$ as a function of
$\tilde d(\gamma)$ extends continuously to $[\frac{1}{2},1]$, with endpoints
$(\frac{1}{2},\frac{1-\eta^*}{2})$ and $(1,1-\eta^*)$. It follows that
\begin{align}
  \varphi(\tilde{\nu}_2,\gamma)
   & \overset{(a)}{\leq}\bigl(2-2\tilde d(\gamma)\bigr)\frac{1-\eta^*}{2}
  +\bigl(2\tilde d(\gamma)-1\bigr)(1-\eta^*)\nonumber                     \\
   & =(1-\eta^*)\tilde d(\gamma), \label{eq:ratio_pointwise_upper_branch}
\end{align}
where step $(a)$ comes from the convexity given above and
$\frac{1}{2} < \tilde d(\gamma)< 1$.
Combining Eqs.~\eqref{eq:ratio_pointwise_lower_branch} and~\eqref{eq:ratio_pointwise_upper_branch}
proves $\varphi(\tilde{\nu}_2,\gamma)\leq(1-\eta^*)\tilde d(\gamma)$ for all
$\gamma>0$, which completes the proof.

}
\bibliographystyle{IEEEtran}
\bibliography{IEEEabrv,ref}







\end{document}